\documentclass[12pt, a4paper]{article}

\usepackage{blindtext}
\usepackage [a4paper, width=170mm,top=25mm,bottom=25mm]{geometry}
\title{\textbf{MeV Dark Matter with MeV Dark Photons\\ in Abelian Kinetic Mixing Theories}}
\date{\today}
\author{Federico Compagnin \\ Dipartimento di Fisica, Università di Torino, Torino, Italy
\and Stefano Profumo, \\ Department of Physics and Santa Cruz Institute for Particle Physics, \\ University of California, Santa Cruz, CA 95064, USA
\and Nicolao Fornengo, \\ Dipartimento di Fisica, Università di Torino and INFN, Torino, Italy}

\usepackage[utf8]{inputenc}
\usepackage{graphicx}
\usepackage{amssymb}
\usepackage{amsmath}
\usepackage{hyperref}
\usepackage{achemso}
\usepackage{parskip}
\usepackage[sort&compress]{natbib}
\usepackage{comment}
\usepackage{mathrsfs}
\usepackage[nottoc]{tocbibind}
\usepackage[toc,page]{appendix}
\usepackage{lipsum}
\usepackage{outlines}
\usepackage{stackengine}
\usepackage{esvect}
\usepackage{graphicx}
\usepackage{subcaption}
\usepackage{nccmath}
\usepackage[version=4]{mhchem}
\usepackage{graphicx}
\usepackage{float}
\usepackage{subfiles}
\usepackage{textcomp, gensymb}
\usepackage{todonotes}
\usepackage{xcolor}
\usepackage{slashed}
\usepackage{physics}
\usepackage{amsmath,latexsym}
\usepackage{eucal}
\graphicspath{{graphics/}{../graphics/}}
\hypersetup{
    colorlinks,
    citecolor=red,
    linkcolor=red
}

\begin{document}
\setcitestyle{numbers,square}
\maketitle
\section*{\hfil Abstract\hfil}
%The dark sector paradigm has recently gained attention in the particle dark matter studies. There are numerous ways for populating this new sector, some of which make use of the kinetic mixing paradigm. 
We consider the cosmology and phenomenology of a dark photon portal to a simple dark sector consisting of a single, light, fermionic dark matter particle species with mass in the MeV range. We entertain three possible kinetic mixing structures of a new Abelian gauge group U(1)$_{\text{dark}}$ with the visible sector through U(1)$_{\text{e.m.}}$,  U(1)$_{\text{Y}}$ and T[SU(2)$_{\text{L}}$]. We assume the  dark photon to be massive and around the MeV scale, thus close to the mass scale of the dark matter candidate.  We compute the dark matter relic density via freeze-out and freeze-in, entertaining the additional possibilities of (i) a late inflationary period that could dilute the dark matter yield of heavy candidates, and (ii) additional production modes, for models with under-abundant thermal production. We explore the parameter space compatible with a variety of experimental and astrophysical bounds, and discuss prospects for discovery with new CMB probes and MeV gamma-ray telescopes.

%Substantial differences between the theories are present and as such experimental bounds allow for different dark matter candidates.
\newpage

\tableofcontents\setcounter{section}{0}
\newpage

%Begin Introduction ------------------
\section{Introduction}\label{sec:introduction}
The presence of dark matter (DM) has been established in numerous ways with cosmological and astrophysical observations throughout the past eighty years; it is now commonly agreed upon that the DM contributes around $24 \%$ of the Universe critical density at late times \cite{2020planck}. It is additionally agreed upon that this DM does not fit in the Standard Model (SM) of particle physics \cite{thooft71}, and that it consists of a new particle species (or some heavy composite or compact object) \cite{PDGreview}.  %As in every aspect of research, the particle nature of dark matter wasn't the first assumption proposed to account for all the well known discrepancies measured ever since the 30's by Fritz Zwicky \cite{zwicky33,smith36,holmberg37,ambartsumian58,bergh61,finzi63,vera70,freeman70,hawking71,cowsik72,emerson73}, but it gained terrain also due to the completion of the Standard Model of Particle Physics (SM) in 1971 \cite{thooft71} and the apparent impossibility to reconcile the former with the latter. 

The origin and production modes of the cosmological DM remain as mysterious as its particle nature. However, one paradigm has gained significant traction: that the DM consists of a {\em thermal relic}. %There's a quite natural way to interpret the presence of DM in our Universe: 
In the thermal relic paradigm at early times the DM was in thermal equilibrium with the hot primordial plasma, where it interacted with other particles; upon ``freezing out'' of  thermal equilibrium, while possibly non-relativistic, the DM would then free-stream and eventually collapse in the halos observed in the late universe, triggering the formation of structure in the universe. % (hence unveiling the sensitivity to others interactions besides gravity) for then decouple from the plasma and produce the cosmological relic density we measure today. 
Such  cold dark matter (CDM) paradigm is able to reproduce statistical properties of the large scale structure (LSS) of the Universe, and predicts galaxy formations patterns  consistent with observation \cite{Springel:2008zz}. A prototypical CDM candidate is a  weakly interacting (being able to decouple with the measured relic abundance)  massive (hence non-relativistic at the moment of decoupling) particle, or WIMP \cite{arcadi}. %This model has been label as the "WIMP miracle" and it's clear how, relying on this, neutrinos have been immediately discarded as DM candidates \cite{doroshkevich80_1,doroshkevich80_2,white83}. \\ 
%Given this historical genesis of the idea, if one could hope DM to be in fact interacting through different channels besides gravity, it may be explained as charged under new symmetries such as discrete symmetries, e.g. $\mathbb{Z}_2$ or gauge symmetries, e.g. U(1)$_{\text{dark}}$ \cite{bernal16,hambye19} with rich experimental signatures to be searched \cite{jaeckel10,essig13,alexander16}.
While the search for WIMPs via direct and indirect detection has produced no conclusive positive results, the possibility that the DM interact with the visible, Standard Model (SM) sector via some ``portal'' remains extremely well motivated \cite{jaeckel10,essig13,alexander16}.

Here we consider a ``portal'' to a minimal dark sector, consisting of a Dirac fermion DM candidate, blind to all known interactions leaving aside gravity, but charged under a new ``dark'' force, whose quantized nature is represented by a ``Dark Photon" (DP) (for a review see e.g. \cite{Fabbrichesi_2021}). The latter is (weakly) coupled to charges in the SM, via kinetic mixing. Here, we will consider the following three possible kinetic mixing structures: U(1)$_{\text{dark}}\times $U(1)$_{\text{e.m.}}$, U(1)$_{\text{dark}}\times $U(1)$_{\text{Y}}$ and U(1)$_{\text{dark}}\times $T[SU(2)$_{\text{L}}]$ mixing (where ``T'' stands for ``maximal torus'' of a given group).

The structure of our study is as follows. In the following Sec.~\ref{sec:thermal_equilibrium} we  present our choice of kinetic mixing theories, describe their structure, and derive all  couplings needed. This is explicitly done by rotating back the initial Lagrangian to the mass eigenstates where the kinetic terms are  diagonal. The guiding idea is to introduce the minimum number of extra gauge bosons, in this case only one, the DP;  for this additional gauge boson to interact with SM fields, one can only couple it with another neutral gauge boson within the SM, hence our choice of using the electromagnetic field $A_\mu$, the hypercharge field $B_\mu$ and the SU(2)$_\text{L}$ - Cartan subalgebra generator $W^3_\mu$. 

The following Sec.~\ref{sec:production} discusses DM production from the primordial plasma: DPs are unstable particles and as such they may or may not be present at the beginning of DM production depending on whether they are still in equilibrium or have already decoupled and decayed. Here, we discuss these possibilities in detail, and find that the presence of DPs plays  an important role in the solution to the relevant Boltzmann equations in the two production mechanisms discussed in the next section. We outline there the region where one may not assume $n_{A'} \neq 0$ at the moment of ignition of DM production ($n_{A'}$ being the number density of the DP).

 Section \ref{sec:production}  entertains two ``thermal'' DM  production mechanisms, freeze-out and freeze-in, for DM candidates of mass 10~MeV and 100~MeV. %Starting with the freeze-out process where, in a nutshell, DM is assumed to be in thermal equilibrium within the primordial plasma before decoupling from it and producing its so called ``fossil abundance'' measured today, we simplify the Boltzmann equations noticing that they decouple, for then we just need to solve the one for DM. In this way, recalling the dependency of the density parameter, for instance measured by Planck, by our model parameters, 
We derive contour plots outlining the theory parameters combinations that yield the right amount of DM, and the regions where,  for the 100 MeV candidate, a late-time inflationary period dilutes the standard DM abundance and opens up otherwise excluded parameter space. %This will produce some important physical differences between the two candidates also due to our previous study upon DP equilibrium. In fact the heavier DM candidate, being the one that decouples before from the plasma, will come across, in some parameters configurations, to DPs still in equilibrium, which in turn will modify our approximated Boltzmann equations.\\ 
Finally we consider the freeze-in mechanism, where the DM is produced out of equilibrium  starting from a null initial population. In this paradigm, DM is a ``feebly'' interacting massive particle (FIMP), for it is required not to reach thermal equilibrium throughout the entire thermal history of the early Universe \cite{Asaka_2007, Gopalakrishna_2006, Hall_2010, Dvorkin_2021}. In this case, the presence of DPs in the thermal bath will play a critical role in shaping the open parameter space and the ensuing observational and experimental bounds. In fact, when DPs are absent at early times, the only channel left for DM production is an $s$-wave process with  DP exchange between SM particles and DM. In this case, the relevant interactions scale as $\Gamma \sim (g_\text{DM}^{\text{DP}})^2(g_\text{SM}^{\text{DP}})^2T$, while the Hubble rate $H \sim T^2/M_P$, where $M_P$ is Planck's mass, meaning that the leading contributes are those of lower temperatures, i.e. DM will be independent from UV a priori unknown physics.

In Sec.~\ref{sec:future} we briefly review direct and indirect DM searches, as well as constraints on DM from CMB measurements in light of upcoming experiments such as NA64$^{++}$ \cite{Gninenko:2300189}, the MeV telescope GECCO \cite{subGeV1,subGeV2} and the Simons Array \cite{Simons-Array}.
The final Sec.~\ref{sec:conclusions} concludes. \\ We collect relevant analytical results in the appendices.
%: if we assume DPs to be in equilibrium from the very beginning of DM production, generically they will bring DM in equilibrium for it then to be controlled by freeze-out in its decoupling production. Whilst assuming no DPs and no DM, i.e. an empty DS, will allow us to derive some interesting predictions which will be unfortunately too far from current experimental bounds to be constraint by any of them. We'll then be briefly discussing, for this aim, our best experimental bound coming from SN1987A measurements. 

%End Introduction --------------------

%Begin equilibrium --------------------

\section{The dark photon}\label{sec:thermal_equilibrium}
In this Section, we present a concise description of three different channels that may lead to the production of a new massive gauge boson that kinetically mixes with an Abelian SM gauge group. In particular, we  give the terms one needs to add to the SM Lagrangian to obtain the operators stemming from the DP mixing.

\subsection{Kinetic mixing theories}
Kinetic mixing is a well-known  mechanism to introduce DPs  as portals between the SM and a dark sector assumed  to contain the DM candidate. Here we explicitly derive all the needed couplings between the DS and the SM starting from the Lagrangian density where the mixing is explicit. Moreover, we  address some of the differences between the theories presented in the next Section, where phenomenological predictions will be made.

We start by looking at a U(1)$_{\text{e.m.}} \times$ U(1)$_{\text{dark}}$ theory. The  Lagrangian density under consideration is
\begin{equation}\label{darkcrossemlagrangian}
    \begin{split}
        \mathcal{L} = -\dfrac{1}{4} \tilde{F}_{\mu\nu}\tilde{F}^{\mu\nu} -\dfrac{1}{4}\tilde{F}'_{\mu\nu}\tilde{F}'^{\mu\nu}-\dfrac{\varepsilon}{2}\tilde{F}_{\mu\nu}\tilde{F}'^{\mu\nu} + e J_\mu \tilde{A}^\mu + g_D J'_\mu \tilde{A}'^\mu,
    \end{split}
\end{equation}
where $\varepsilon$ is the kinetic mixing parameter while $e$ and $g_D$ are the electric and ``dark" charges. $J'_\mu = \Bar{\chi} \gamma_\mu \chi$ is the ``dark current'',  $\chi$ the DM candidate and $J_\mu$ is the usual electromagnetic current. $\tilde{F}_{\mu\nu}$ and $\tilde{F}'_{\mu\nu}$ are the field strengths respectively associated to the gauge fields $\tilde{A}_\mu $ and $\tilde{A}'_\mu $. Rotation into mass eigenstates is given in App.~ \ref{em_calculations}.

 Another well motivated possibility, relevant to extending the model before the electroweak phase transition, is to make use of the other two neutral gauge bosons within the SM gauge group, namely the hypercharge boson $B_\mu$ and $W^3_\mu$. We indicate the former as the U(1)$_{\text{Y}} \times $ U(1)$_{\text{dark}}$ theory, obtained by rotating 
\begin{equation}\label{lagrangian_mixing_Yxdark}
\begin{split}
    \mathcal{L} = &  -\dfrac{1}{4}B_{\mu\nu}B^{\mu\nu} - \dfrac{1}{4}W^3_{\mu\nu}W^{3,\mu\nu} - \dfrac{1}{4}a'_{\mu\nu}a'^{\mu\nu}- \dfrac{\varepsilon}{2}B_{\mu\nu}a'^{\mu\nu} + \\
    & + (\mathcal{D}_\mu \Phi)^\dagger(\mathcal{D}^\mu \Phi) + \dfrac{1}{2}m_{a'}^2a'_\mu a'^\mu + i \Bar{L}\slashed{\mathcal{D}}L + i\Bar{\ell}_\text{R}\slashed{\mathcal{D}}\ell_\text{R} + i \Bar{Q}\slashed{\mathcal{D}}Q + \\
    & + i\Bar{u}_\text{R}\slashed{\mathcal{D}}u_\text{R} + i\Bar{d}_\text{R}\slashed{\mathcal{D}}d_\text{R} + i \Bar{\chi} \slashed{\mathcal{D}}\chi
\end{split}
\end{equation}
into the mass eigenstates. Notice that in this case, we have to introduce a massive boson $a'_\mu$ which will be rotated into a new DP after symmetry breaking (see App.~\ref{Y/L_calculations}). In what follows, we will refer to the dark photon mass as $m_{A'}$, distinguishing it from $m_{a'}$ (see $m_{A'}(m_{a'})$ below). 
Also notice that the resulting $\hat{Z}_\mu$ boson  inherits the kinetic mixing residual as a slight modification of the SM $Z_\mu$ due to its mass,  see again App.~\ref{Y/L_calculations} for details.  

The  T[SU(2)$_\text{L}$] $\times $ U(1)$_\text{dark}$ case is obtained similarly by substituting the kinetic mixing term
\begin{equation}
    \mathcal{L}_{\text{mix}} = -\dfrac{\varepsilon}{2}W^3_{\mu\nu}a'^{\mu\nu} 
\end{equation}
in place of the $B_\mu$-mixing in \eqref{lagrangian_mixing_Yxdark}.

Before proceeding, we point out that all  vertices containing a DP are of order $\mathcal{O}(\varepsilon)$; moreover in the two theories describing physics prior to the electroweak phase transition, the vertices comprising the $\hat{Z}_\mu$ boson correct SM interactions by terms of order $\mathcal{O}(\varepsilon^2)$. In these latter theories, neutrinos are the only ``anomalous" particles interacting with DPs, with couplings of order $\mathcal{O}(\varepsilon \delta)$, where $\delta = (m_{A'}/m_{Z})^2$.

\subsection{Thermal equilibrium}\label{sc:thermalequilibrium}
We now examine the question of whether dynamically tracking the dark photon abundance in computing the dark matter relic density is necessary. %vanishes Turning to an intermediate point before proceeding, we'll be considering here how does a extremely useful approximation for Boltzmann equations, i.e. $n_{A'} = 0$ should be treated once discussing the role of DP within the primordial plasma.\\
%As said, to simplify Boltzmann equations one could assume the following: by the time the universe reached $T_{\text{f.o.}}$ or $T_{\text{f.i.}}$ (i.e. freeze-out and freeze-in temperatures, see Section III, which in a nutshell are the "ignition" temperatures at which DM starts being produced), depending on the production mechanism one considers, all the $A'$ have already disappeared through decay, i.e. $n_{A'} \ll 1$.\\
First, we point out that DP-SM interactions (such as $ff \longleftrightarrow A'A'$, $f A' \longleftrightarrow f A'$ and $\gamma f \longleftrightarrow A' f$) are greatly suppressed compared to dark sector processes. As a result, 
%
%From this moment on, although the complete set of interactions the DP is subjected to comprises also of $ff \longleftrightarrow A'A'$, $f A' \longleftrightarrow f A'$ and $\gamma f \longleftrightarrow A' f$, we will discard them since these are negligible if compared to the ones regarding the only dark sector. As another interesting subtlety we'd like also to mention that the type of thermal equilibrium we're discussing here is the one of a separate plasma with respect to the one where SM particles are present, in principle characterized by a different temperature $T'$ which shared a common evolution with the SM "traditional" plasma, up to the moment DP decoupled from (for example in the dark $\times$ e.m.) electrically charged fermions. \\
$A'$ is in thermal equilibrium when the rates $\Gamma$ for dark sector processes satisfy 
 $\Gamma_{\chi\chi\to A'A'}/H > 1$ and $\Gamma_{A'A'\to \chi\chi}/H > 1$, or $\Gamma_{A'\chi\to A'\chi}/H > 1$ or both simultaneously. If one of the first two conditions is violated, the third one will prevent the decay, but when both are violated, then the particle is no longer in equilibrium.
 Notice also that we must require annihilation and production to be in equilibrium simultaneously: if one of the two is violated, then both are to be considered no longer available interactions to keep $A'$ in equilibrium.
 Finally if the interactions are slower than the decay rate, then $A'$ will decay, regardless of the equilibrium argument given above.
 
 %These observations are therefore to be taken into consideration when discussing the presence of $A'$ at the moment of DM decoupling (freeze-out) or at beginning of its production (freeze-in): 
 We computed the temperature at which $A'$ starts decaying 
 %because of any of the mentioned reasons
 ($T_D$) 
 and  compared it with the temperatures of freeze-out $T_{\chi,\text{f.o.}}$ and freeze-in $T_{\chi,\text{f.i.}}$, whichever appropriate: if for example $T_D > T_{\chi,\text{f.o.}}$,  DPs  decouple before the beginning of DM production (through the freeze-out mechanism in this case) and hence they may be considered already decayed away. On the other hand if $T_D < T_{\chi,\text{f.o.}}$ then DPs have still to decouple, being therefore present at the moment DM starts being produced; in this case we label the corresponding region in parameter space with $n_{A'} \neq 0$ and the DP-DM Boltzmann equations must be solved as coupled differential equations.
 
\begin{figure}[H]
    \centering
    \subfloat{{\includegraphics[width=7.5cm]{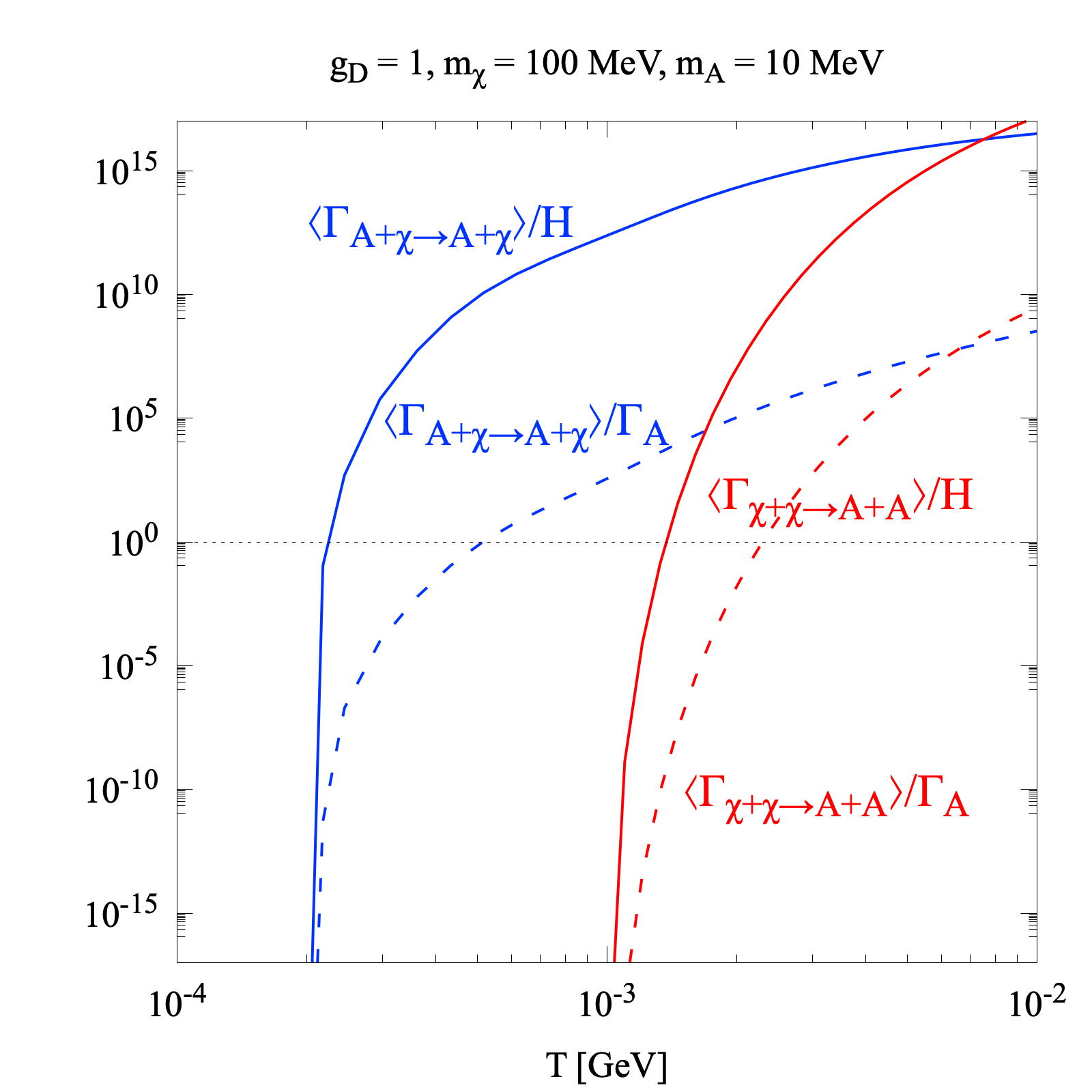} }}%
     \qquad
    \subfloat{{\includegraphics[width=7.5cm]{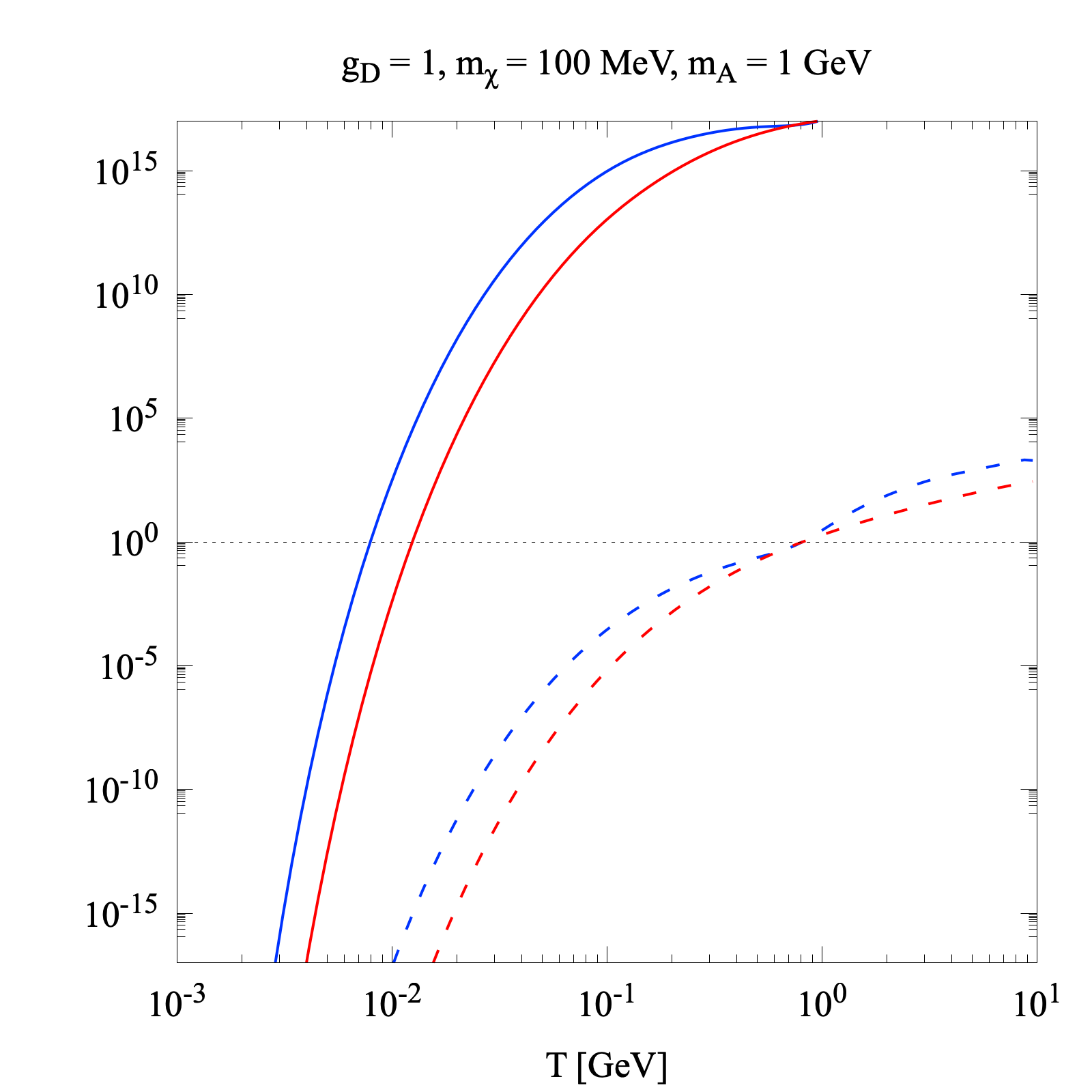} }}%
     \qquad
    \subfloat{{\includegraphics[width=7.5cm]{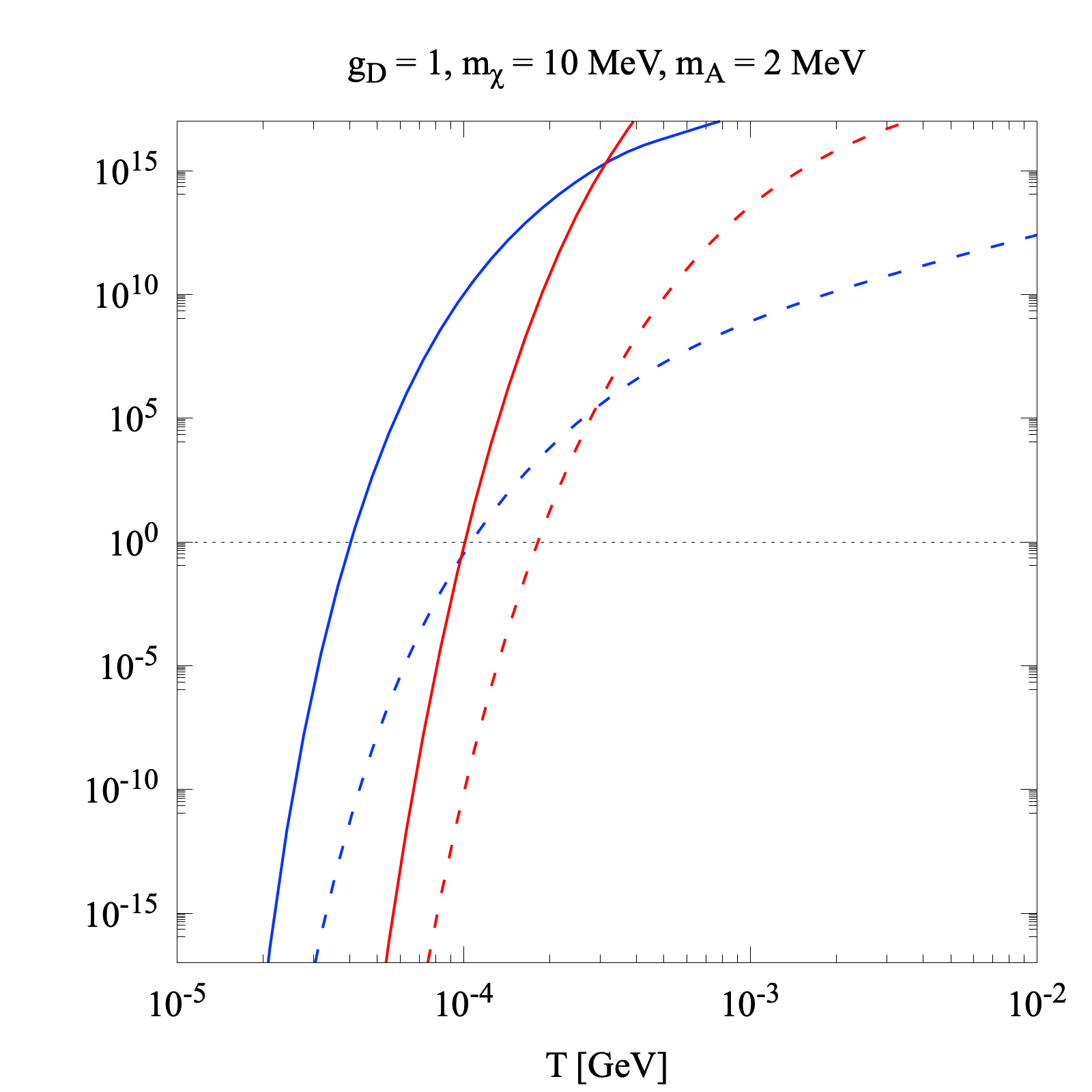} }}%
     \qquad
    \subfloat{{\includegraphics[width=7.5cm]{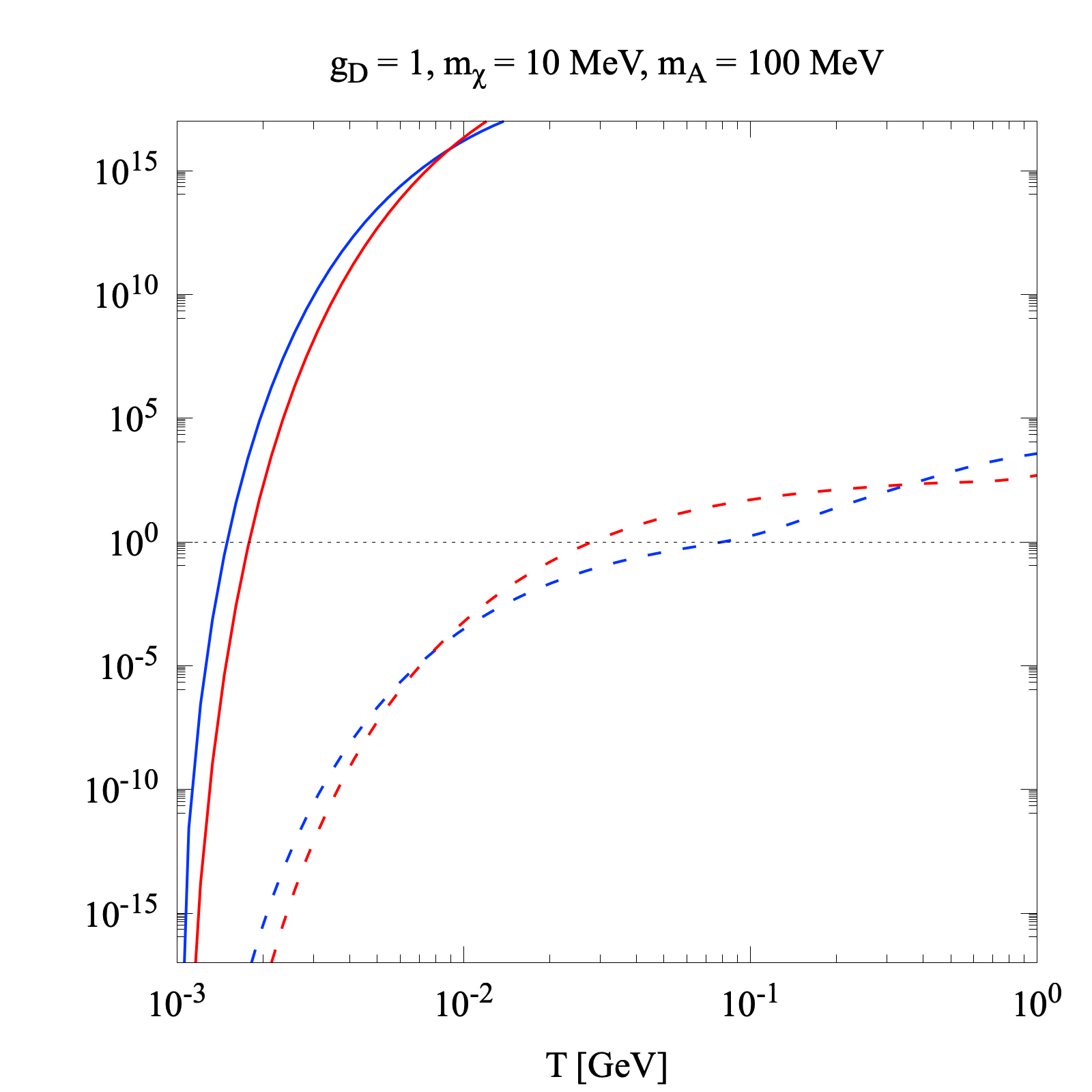} }}%
    \caption{Pair-production and elastic scattering rates, relative to the Hubble rate or to the $A$ boson decay rate, as a function of temperature for by $g_D = 1$, $m_\chi = 10$ MeV and $m_\chi = 100$ MeV, for a few values of $m_{A'}$.}%
    \label{fig:thermalequilibriumplot}%
\end{figure}

 We conclude this section illustrating the interplay of rates discussed above, normalized to the Hubble rate and to the DP decay rate as a function of temperature for a few choices of DM and DP masses. In Fig.~\ref{fig:thermalequilibriumplot} we consider DM masses $m_\chi = 10$ MeV and $m_\chi = 100$ MeV, with $m_{A'} = 2$ MeV, $m_{A'} = 10$ MeV, $m_{A'} = 100$ MeV  and $m_{A'} = 1$ GeV in the dark $\times$ electromagnetic theory. 
In the figure we show four ratios: $\Gamma_{\chi A' \to \chi A'}/H$ , $\Gamma_{\chi A' \to \chi A'}/\Gamma_{A'}$, $\Gamma_{\chi \chi \to A' A'}/H$ and $\Gamma_{\chi \chi \to A' A'}/\Gamma_{A'}$. Notice that when the ratios $\Gamma / H$ are greater than unity,  interactions are rapid enough to keep the DP in equilibrium %(where for elastic interactions only one interaction is needed, due to the fact that $\chi A' \to \chi A'$ is equal to its opposite, while for inelastic ones both are needed for if one of the two is violated the whole annihilation-production process violates the equilibrium requirement), 
as long as in the meantime  the condition  $\Gamma / \Gamma_{A'}>1$ is also satisfied. For example, in the $m_\chi = 100$ MeV and $m_{A'} = 1$ GeV section, even though $A'$ could be in equilibrium thanks to elastic scatterings and inelastic interactions (continuous lines), the decay amplitude takes control and brings it to decay (dashed lines): leading interactions are always the ones we  refer to, hence the dashed interactions ``decouple" at higher temperatures i.e. a little sooner in the thermal history of the plasma. More precisely in the considered case we see that the ``decoupling" temperature is about $1$ GeV while the DM decoupling temperature is $T\sim m_\chi / 20 \sim 5$ MeV; as a result, we are allowed to assume no DP is present at the moment of DM decoupling. Vice-versa, in the upper left plot, we see that the DM decouples at $T \sim 5\times 10^{-4}$ GeV, when the DPs are still present in the plasma. In this case, the DPs abundance must be tracked to compute the final DM abundance.

The freeze-out condition is shown in Figure \ref{f.o.10MeVeps(mA)}, where the light blue region indicates the subspace where $n_{A'}\ll 1 $ cannot be assumed. In the case of freeze-in, instead, one must  distinguish among the different scenarios, as they are relevant at different cosmological times, prior or subsequent to Electroweak Symmetry Breaking (EWSB); in the dark $\times$ electromagnetic theory, since DM may be produced from the very beginning of the thermal history of the universe, but the model is only valid up to EWSB,  $T_{\chi,\text{f.i.}} \sim T_{\text{EWSB}}$ (which is very high) and we get that the presence of DP in this case can never be neglected. The last argument applies also to Y $\times$ dark and L $\times$ dark mixing theories, although in these we considered a generic post-inflationary high temperature $T \sim 200$ GeV (the precise value of the temperature is irrelevant, as we discuss below). Lastly, no matter which theory one considers, DP processes $\chi\chi \longleftrightarrow A'A'$ (and $\chi A' \longleftrightarrow \chi A'$) are always very significant in the ``strong dark'' regime we consider below, one where the coupling $g_D \lesssim \mathcal{O}(1)$.

%their contribute is immediately extremely intense within the "strong dark" regime we'll be interested into - for us, $g_D \lesssim \mathcal{O}(1)$, hence DPs produce immediately a great amount of DM and bring it to equilibrium. \\

%End equilibrium ----------------------

% Begin production -------------------
\section{Production}\label{sec:production}
We consider a standard homogeneous and isotropic FLRW universe, where the collisional Boltzmann equations take the form 
\begin{equation}\label{Boltzmann_equation}
    \dfrac{dY_i}{dT} = -\dfrac{\tilde{g}}{s H T}\left( \dfrac{g_i}{(2 \pi)^3}\int \dfrac{\mathop{d^3p}}{E}\mathcal{C}_{\text{inel}}[f_i;...]\right),
\end{equation}
where $Y_i = n_i/s$ is the $i$-th species abundance, $s = (2\pi^2/45)g_{\star s}T^3$ is total entropy density \cite{kolbturner}, $H$ the Hubble rate, $g_i$ the degrees of freedom (d.o.f.) of $i$, $\mathcal{C}_{\text{inel}}$ the inelastic Collisional operator and $f_i$ the phase space distribution function of $i$. $\tilde{g} = \left( 1+ T d\log g_{\star s} / dT\right)\sim 1$ for $g_{\star s}$ only changes when a d.o.f. becomes non-relativistic in the plasma.\\ 
Integrating \eqref{Boltzmann_equation} we get the current value of $Y_{\chi, 0}$ which we use for estimating the relic abundance
\begin{equation}\label{omegah2}
    \Omega h^2 = \dfrac{m_\chi s_0}{\rho_{\text{crit},0}/h^2}Y_{\chi, 0}(m_\chi,m_{A'},\varepsilon,g_D),
\end{equation}
known to be close to $0.12$ from Planck data \cite{planck16}.

\subsection{Freeze-out}
Producing DM through a freeze-out scenario requires it to be in equilibrium with the plasma at the moment of decoupling, so we may take $Y_\chi = Y_\chi^{\text{eq}}$ with $ n_\chi^{\text{eq}} = (g_i /2 \pi^2) m_i^2 T K_2(m_i/T)$ \cite{gondologel90}, where $K_2$ is the modified Bessel function of the second kind.\\
In the electromagnetic kinetic mixing scenario we have the following system of equations:
\begin{equation}
    \begin{split}
    \begin{cases}
    \dfrac{{d} Y_\chi}{{d}T} = & -\dfrac{1}{s H T}\bigg[ (n_f^{\text{eq}})^2 \langle \sigma v_{ff\to\chi\chi}\rangle - n_\chi^2 \langle \sigma v_{\chi\chi\to ff}\rangle + n_{A'}^2 \langle \sigma v_{A'A'\to\chi\chi}\rangle  \\[2mm] 
    & - n_\chi^2 \langle \sigma v_{\chi\chi\to A'A'}\rangle 
     + 2 n_{A'}\Gamma_{A'\to\chi\chi} \bigg] \\[2mm]
    \dfrac{{d}Y_{A'}}{{d}T} = & -\dfrac{1}{s H T}\bigg[ (n_f^{\text{eq}})^2 \langle \sigma v_{ff\to A'A'}\rangle - n_{A'}^2\langle \sigma v_{A'A'\to ff}\rangle + n_\chi^2 \langle \sigma v_{\chi\chi\to A'A'}\rangle \\[2mm]
    & - n_{A'}^2 \langle \sigma v_{A'A'\to\chi\chi}\rangle - n_{A'}\Gamma_{A'\to\chi\chi} + n_f^{\text{eq}} n_\gamma^\text{eq} \langle \sigma v_{f\gamma\to fA'}\rangle - n_f^{\text{eq}} n_{A'} \langle \sigma v_{fA'\to f\gamma}\rangle \bigg]
    \end{cases}
    \end{split}
\end{equation}
whilst in the hypercharge and $W^3$-mixing theories more terms are needed to account for processes where $\hat{Z}$ participate to the interactions (see App.~\ref{Y/L_calculations} for its definition). On the other hand, as mentioned in the previous section, $\hat{Z}$ is quite heavy if compared to DP and its presence may be neglected in a first approximation. We thus take $n_{A'}$ to be zero at the moment of DM freeze-out, in such a way we are allowed to solve the first equation independently of the second one:
\begin{equation}
\begin{split}
    \dfrac{{d} Y_\chi}{{d}T} = & \dfrac{1}{HT}\langle \sigma v_{\text{ann, tot}}\rangle \left(Y_\chi^2-Y_{\text{eq,}\chi}^2\right) \stackrel{\text{cold relic}}{\sim} \dfrac{1}{HT}\langle \sigma v_{\text{ann, tot}}\rangle Y_\chi^2,
\end{split}
\end{equation}
where we take $Y_{\text{f.o.},\chi} \equiv Y_\chi(T_{\text{f.o.}}) = Y_{\text{eq,}\chi}(T_{\text{f.o.}})\left(1+\delta\right) $ with $\delta = 1.5$  \cite{gondologel90}. In this way, the present-day  relic abundance can be computed as 
\begin{equation}
    Y_{\chi,0} = \dfrac{Y_{\text{f.o.},\chi}}{1+\mathcal{I}\,Y_{\text{f.o.},\chi}},
\end{equation}
with 
\begin{equation}
    \mathcal{I} \equiv \sqrt{\dfrac{\pi}{45}}M_P\int_{T_0}^{T_{\text{f.o.}}} \mathop{dT} \dfrac{g_{\star s}}{\sqrt{g_\star}}\langle \sigma v_{\text{ann, tot}}\rangle.
\end{equation}
We have the following situation for the three theories under consideration:
\begin{align*}
        \text{dark} \times \text{e.m.}&: \langle \sigma v_{\text{ann, tot}}\rangle \sim \langle \sigma v_{\chi\chi\to e^+e^-}\rangle +  \langle \sigma v_{\chi\chi\to A'A'}\rangle \\
    \text{dark} \times \text{Y}&: 
    \langle \sigma v_{\text{ann, tot}}\rangle \sim \langle \sigma v_{\chi\chi\to e_L^+e_L^-} \rangle + \langle \sigma v_{\chi\chi\to e_R^+e_R^-} \rangle  + \sum_{\ell\,=\, e,\,\mu,\,\tau}\langle \sigma v_{\chi\chi\to \nu_{\ell_\text{L}}\nu_{\ell_\text{L}}} \rangle + \langle \sigma v_{\chi\chi\to A'A'} \rangle \\
    \text{dark} \times \text{L}&: 
    \langle \sigma v_{\text{ann, tot}}\rangle \sim \langle \sigma v_{\chi\chi\to e_L^+e_L^-} \rangle + \langle \sigma v_{\chi\chi\to e_R^+e_R^-} \rangle  + \sum_{\ell\,=\, e,\,\mu,\,\tau}\langle \sigma v_{\chi\chi\to \nu_{\ell_\text{L}}\nu_{\ell_\text{L}}} \rangle + \langle \sigma v_{\chi\chi\to A'A'} \rangle
\end{align*}
Notice that we only considered $\mathcal{O}(\varepsilon)$ processes consisting of electron-positron pairs. This has been numerically checked and it turned out that the total contribution of particles heavier than electrons sums up to around $1\%$ of the integral $\mathcal{I}$, and were neglected in what follows.

Finally, the theoretical abundance parameter $\Omega h^2$ can be computed as \eqref{omegah2}.

\subsubsection[10 MeV DM]{10 MeV DM}
\begin{figure}[H]
    \centering
    \subfloat{{\includegraphics[width=7.5cm]{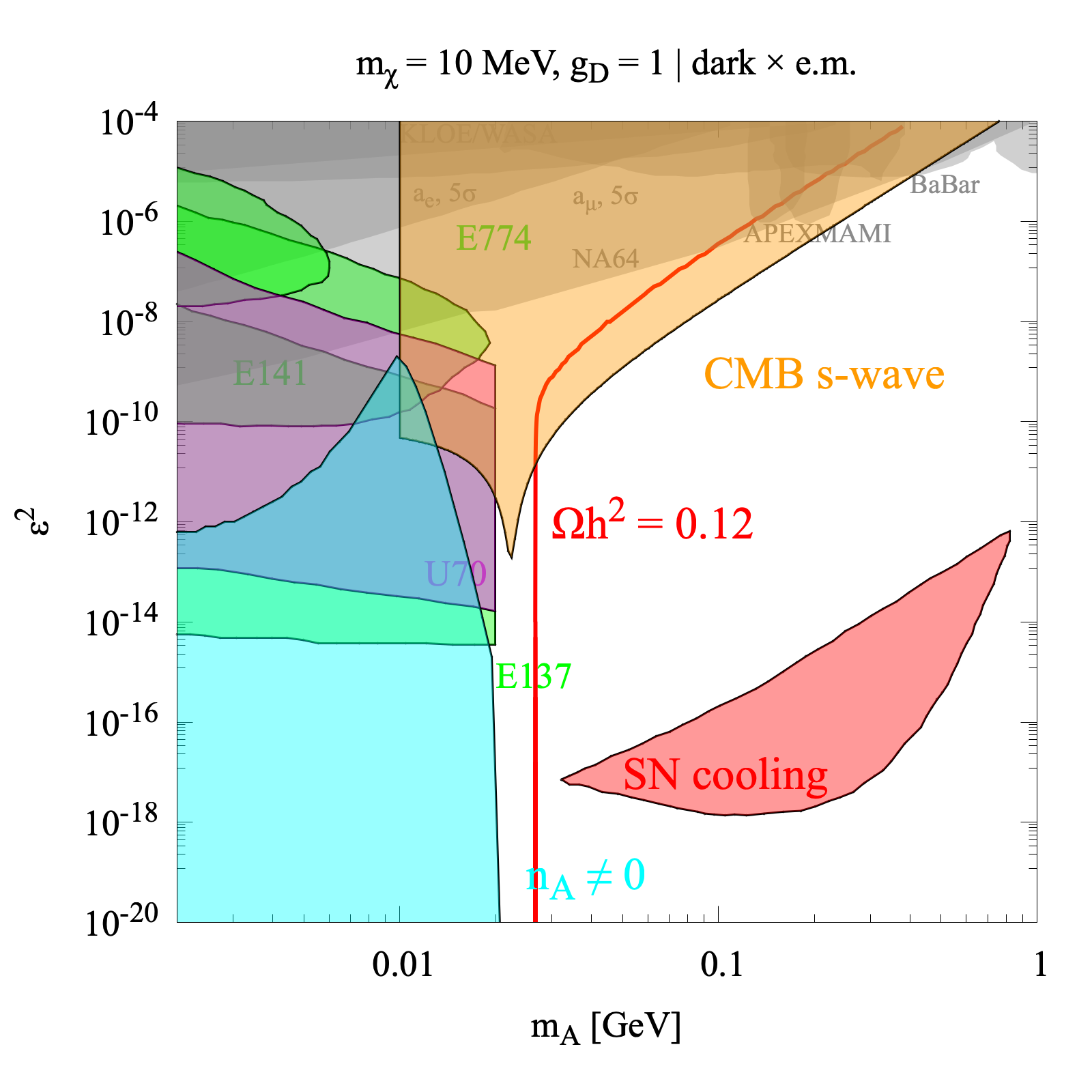} }}%
    \qquad
    \subfloat{{\includegraphics[width=7.5cm]{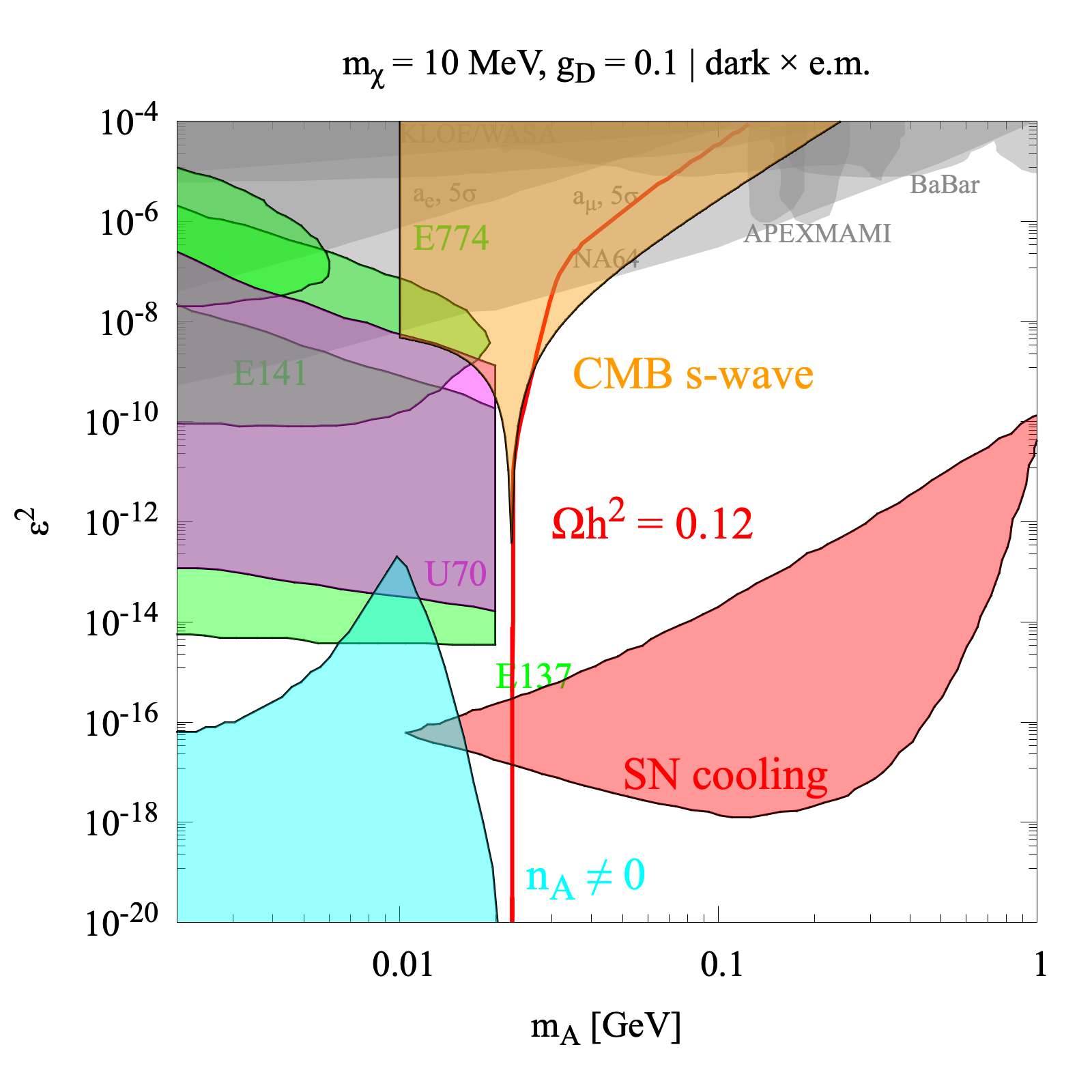} }}%
    \qquad
    \subfloat{{\includegraphics[width=7.5cm]{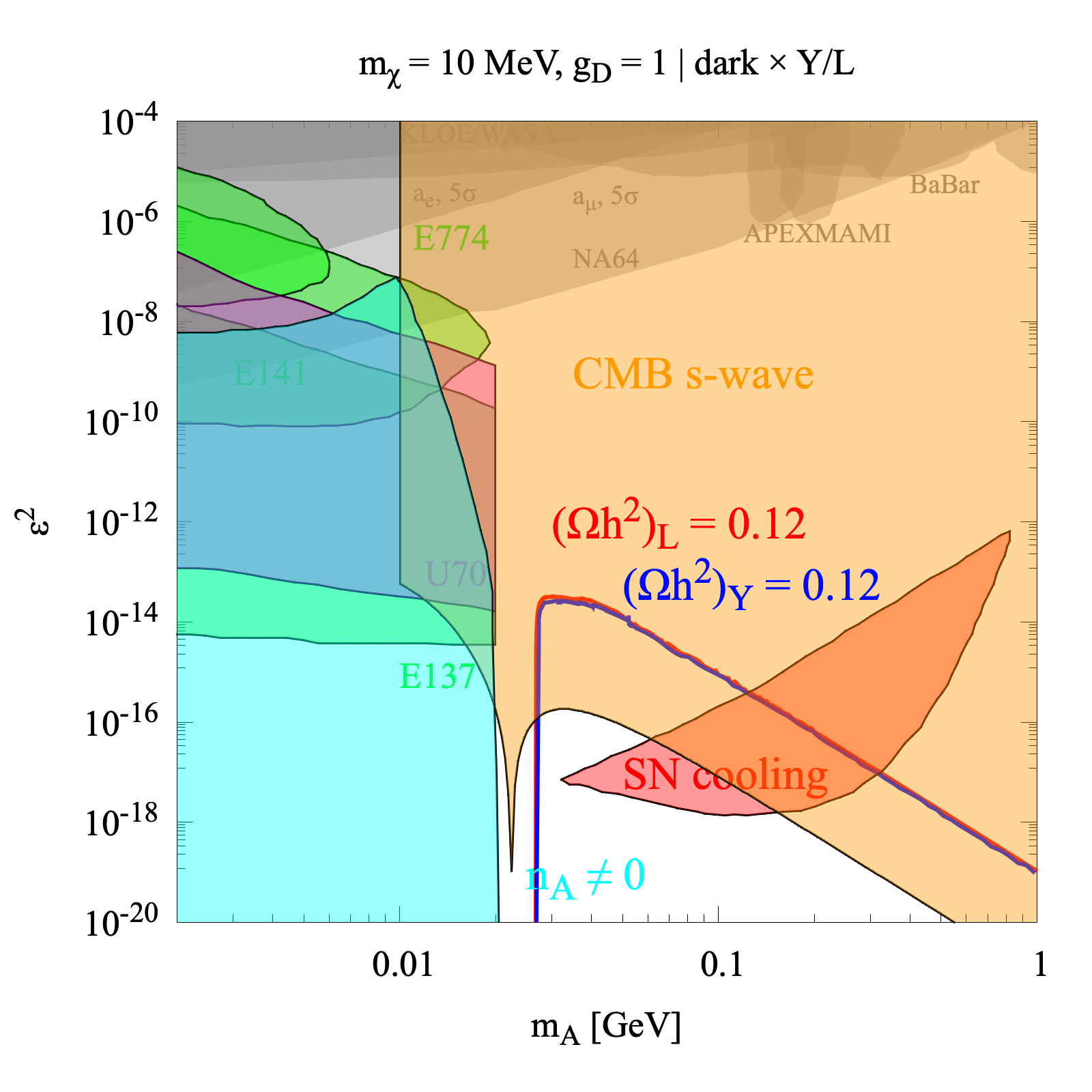} }}%
    \qquad
    \subfloat{{\includegraphics[width=7.5cm]{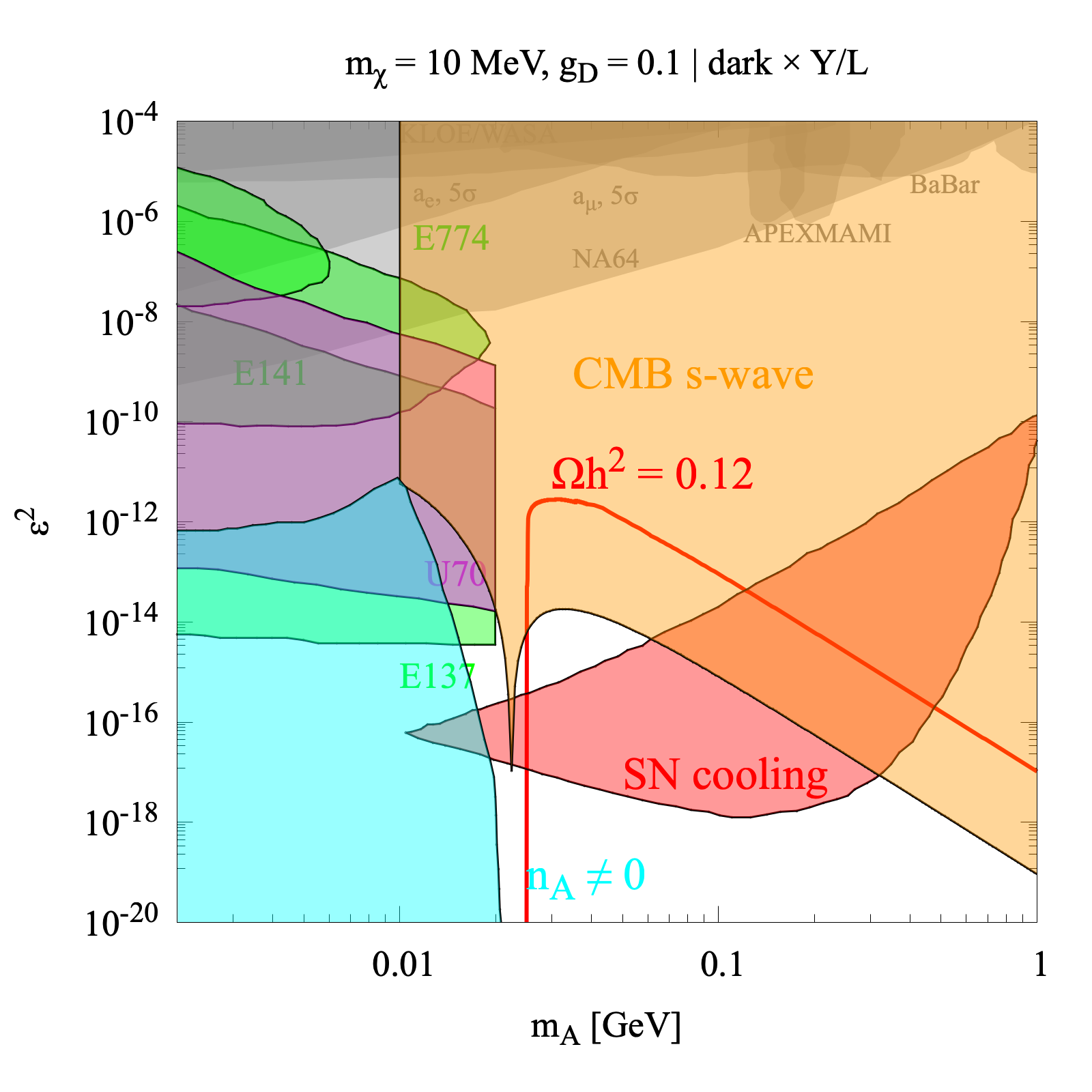} }}%
    \caption{The relevant parameter space for dark $\times$ electromagnetic (above) and dark $\times$ Y/L (below).
We show constraints from beam dump experiments at SLAC: E137, E141 and E774 \cite{andreas12}, beam
dump U70 experiment \cite{mlein11}, fixed target experiments like MAMI \cite{merkel11} and APEX \cite{abrahamyan11}, electron-positron
colliding experiments as BaBar \cite{bjorken09, collabBaBar} and KLOE \cite{archilli12}. Finally, constraints from CMB
assuming s-wave annihilation \cite{slatyer16} and SN-cooling \cite{dreiner14}. The light blue region refers to Section \ref{sc:thermalequilibrium}.}
\label{f.o.10MeVeps(mA)}%
\end{figure}

We start by considering a $m_\chi = 10$ MeV DM candidate. Results are presented starting from Figure (\ref{f.o.10MeVeps(mA)}) as contour plots of $\Omega h^2= 0.12$.%, obtained numerically through a double integration and requiring a relative error of $1\%$ with respect to Planck measurement $\Omega h^2 = 0.12$. \\
    
    The filled regions are constrained by experimental bounds, including beam dump experiments at SLAC such as E137 and E141, E774 at Fermilab \cite{andreas12}, NA64 at CERN, beam dump U70 experiment  \cite{mlein11}, fixed target experiments like MAMI \cite{merkel11} and APEX \cite{abrahamyan11}, $e^+ e^-$ collider experiments such as BaBar \cite{bjorken09}, \cite{collabBaBar} and KLOE \cite{archilli12}. Finally, we show astrophysical constraints from CMB assuming s-wave annihilation \cite{slatyer16} and SN-cooling \cite{dreiner14}. The light blue region refers to the region where
%     where one isn't allowed to assume
$n_{A'} \neq 0 $ and DP are still in equilibrium after DM decoupled.

    First, let us comment on how DM interactions play an important role in defining the correct value of the  relic abundance. For simplicity we'll be referring our general discussion to the dark $\times$ electromagnetic theory, adjusting the argument whenever needed to include the other theories.
    
    When $m_\chi \lesssim m_{A'}$, the  leading interactions are $\chi\chi\to e^+e^- \sim e \varepsilon g_D$ and the weak scale is reached in the coupling constants product for the abundance to be produced correctly. On the other hand, when $m_\chi \gtrsim m_{A'}$ channels $t$ and $u$ open up, making the annihilation into two DPs $\chi\chi\to A'A' \sim g_D^2$ the leading interaction. In our $g_D$ range, the latter is too strong to reach the weak scale, hence DM stays in equilibrium much longer than before, leading to its final relic abundance being extremely small due to Boltzmann suppression.
    
    A very finely-tuned parameters configuration is obtained in the neighborhood of $m_\chi \sim m_{A'}$ in the $\varepsilon^2(m_{A'})$-plane: for dark $\times$ electromagnetic, $\varepsilon^2 \lesssim 10^{-14}$ when $(m_{A'} = 26.6~\text{MeV}, g_D = 1)$, $(m_{A'} = 22.5~\text{MeV}, g_D = 10^{-1})$ whilst for dark $\times$ Y, $\varepsilon^2 \lesssim 10^{-18}$ when $(m_{A'} = 21.6~\text{MeV}, g_D = 1)$ and $\varepsilon^2 \lesssim 10^{-17}$ when $(m_{A'} = 22.1~\text{MeV}, g_D = 10^{-1})$.
    
    As mentioned in Section \ref{sec:production}, the difference between dark $\times$ Y and dark $\times$ L is of order $\mathcal{O}(\varepsilon^2)$: we therefore show, in the third plot in Figure (\ref{f.o.10MeVeps(mA)}) both predictions, but in what follows we will make use only of dark $\times$ Y theory as representative of both.
    
\begin{figure}[H]\label{f.o.10MeV}
    \centering
    \subfloat{{\includegraphics[width=7cm]{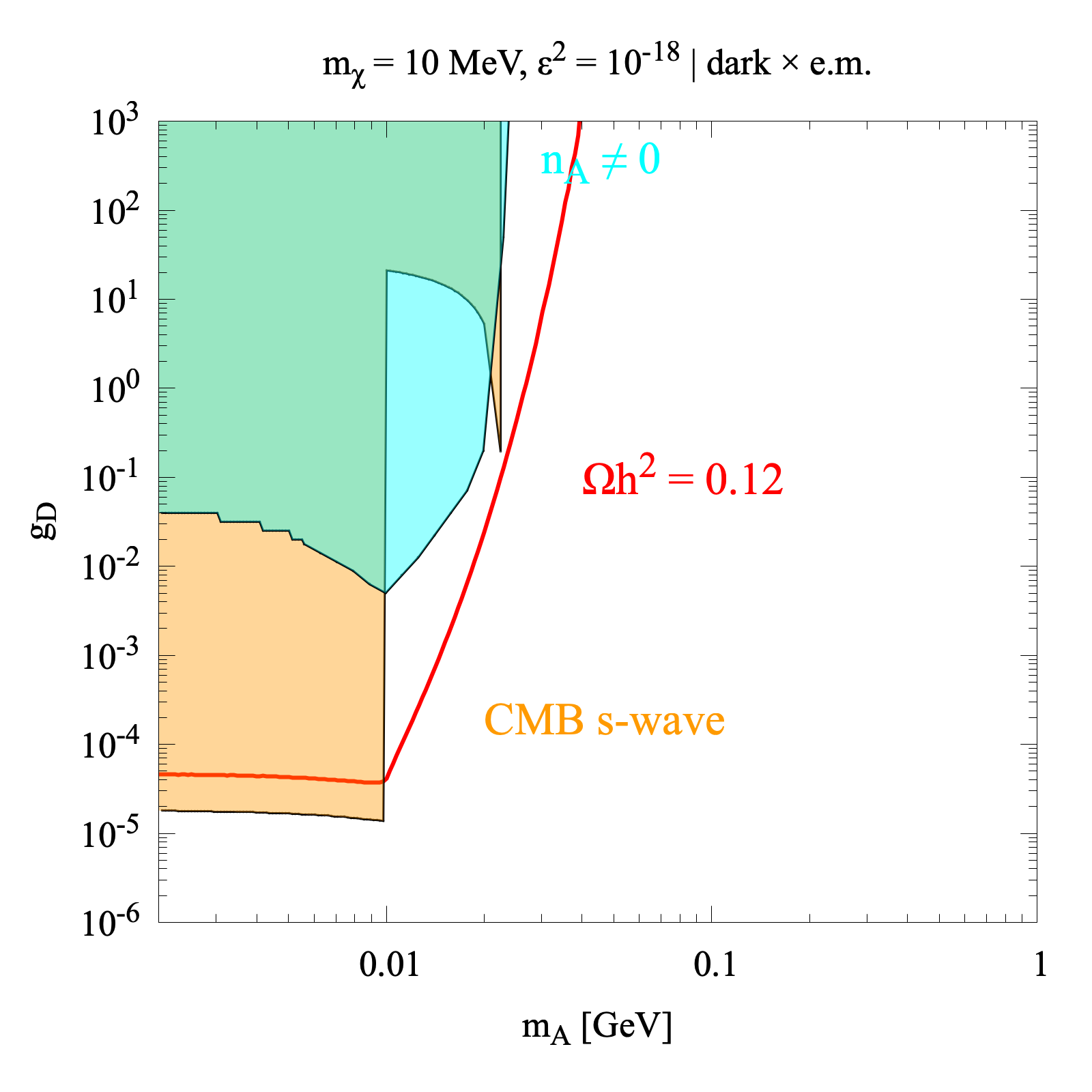} }}%
    \qquad
    \subfloat{{\includegraphics[width=7cm]{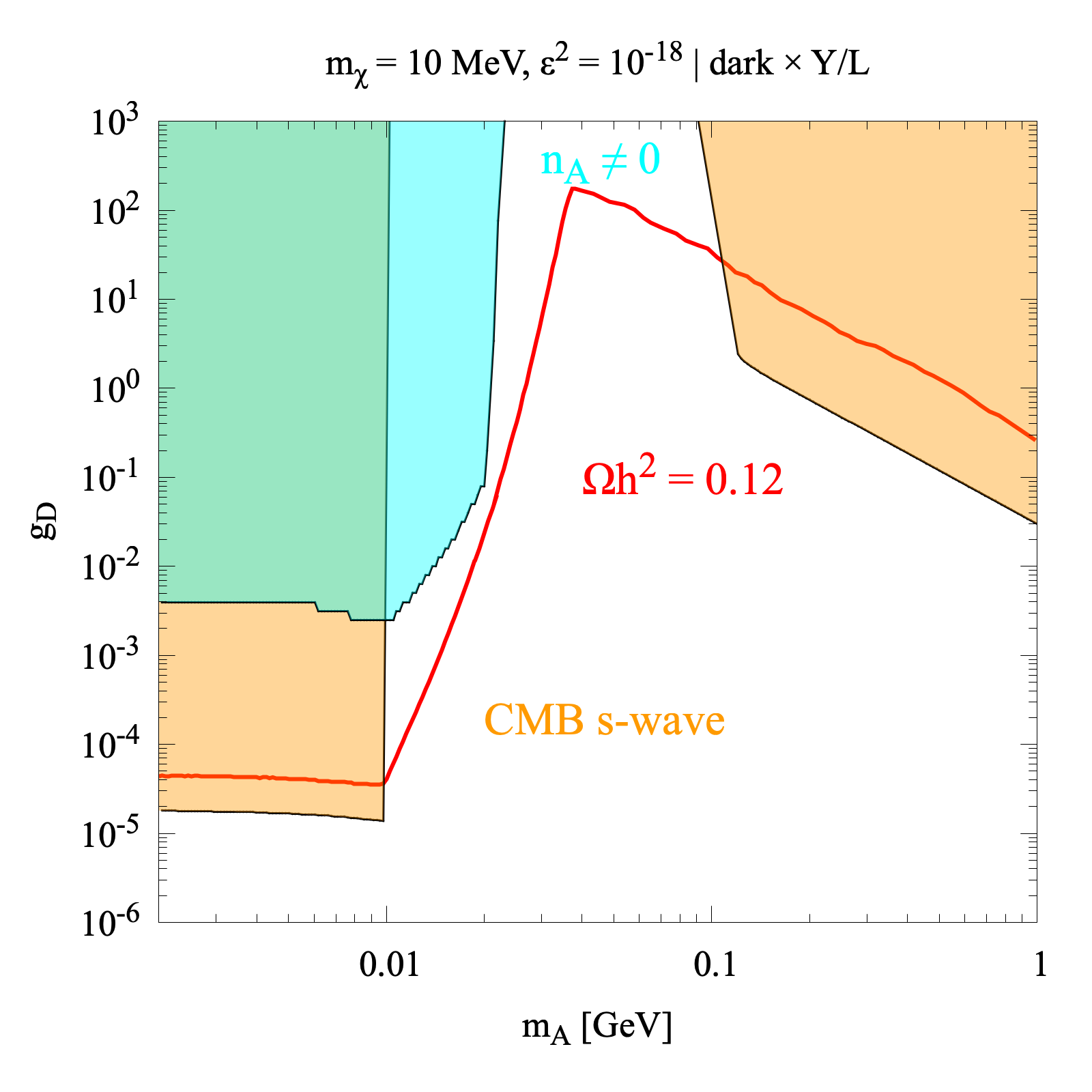} }}%
    
    \caption{The dark $\times$ electromagnetic (left) and dark $\times$ Y (right) (which is similar for the L kinetic mixing theory as well). Notice the clear contribute of leading interactions $\chi\chi\to A'A'$ up to $m_\chi \sim m_{A'}$ forcing $g_D \sim$ const configurations, while  turning to leading $s$-wave annihilation when $m_\chi \lesssim m_{A'}$, and thus to a ``dark strong" regime.}%
    \label{f.o.10MeVgD(mA)}%
\end{figure}

    Another interesting aspect of these results is the  behaviour of $\Omega h^2 = 0.12$ upon changes in $g_D$: while on the right part of the plot the product $q_f e \varepsilon g_D$  decreases, requiring the DP to take up lighter and lighter masses, on the left side the singular behaviour moves slightly to the left, indicating an $\varepsilon$-independent configuration. This can clearly be seen in Figure \ref{f.o.10MeVgD(mA)}, where we  specifically focus on that slice of parameter  sub-space. 
    When $m_\chi \gtrsim m_{A'}$ on-shell DPs production is the leading interaction for greater $g_D$, bringing the annihilation cross section close to threshold value from CMB $s$-wave. Notice that in this case, when considering the dark $\times$ Y theory, we get two regions where CMB bounds are effective: this may be understood looking at this $g_D(m_{A'})$ plane as the $\varepsilon^2 = 10^{-18}$ section of the previous $\varepsilon^2(m_{A'})$ plane for the same theory.
    Finally, the relic density requirement forces a $g_D \sim \; \text{const}$ for  $m_{A'} \lesssim m_\chi$, interestingly already excluded by CMB bounds.
    However, for larger masses, the relic density forces larger values of $g_D$, in a region we indicate as ``dark strong" interacting DM.

%SP: reviewed till here

\subsubsection[Inflated DM]{Inflated DM} 
Here we entertain the possibility that the dark matter candidate be heavier, $m_\chi = 100$ MeV, and that a late entropy injection episode occurs between the dark matter freeze-out temperature and Big Bang Nucleosynthesis (BBN) so that a larger-than-expected thermal relic density can be reconciled with observations. 

%In this subsection we will merge two analysis: the first one, in which we'll be considering the very same aspects of before but concerning a heavier candidate of an indicative mass $m_\chi = 100$MeV, but also we'll make use of the fact that this candidate is representative of a class of DM particles which decouples before Big Bang Nucleosynthesis (BBN) and through them a new phenomenological paradigm is accessible to explore a wider parameters space. 

The freeze-out temperature for a $m_\chi = 100$ MeV Dirac fermion, which decouples as a cold relic is approximately given by $x_{\text{f.o.}} \sim 15 + 3\log(m_\chi / \text{GeV}) \sim 8$ \cite{kolbturner} hence $T_{\text{f.o.}} \sim 13$ MeV. Here, we consider a period of ``Late  Inflation'', by which we mean a model where an entropy dilution episode occurs at times close, but preceding, Big Bang Nucleosynthesis (BBN), i.e. $T_{\text{BBN}} \sim 1~\text{MeV}$. Such episode would dilute the DM relic abundance, and offset its late-time asymptotic value.

As concrete example we consider the model outlined in Ref.~\cite{davoudiasl16} which comprises a real scalar, coupled to fermions as well as self coupled through a suitable potential (bounded from below as a result of a  $\mathbb{Z}_2$ symmetry)
\begin{equation}
    \mathcal{L} = \dfrac{1}{2}\partial_\mu \phi \partial^\mu \phi -V[\phi] - \dfrac{1}{2}\mu_\chi \phi^2 + \sum_f y_f f\Bar{f}\phi
\end{equation}
Requiring the DM to have frozen-out before the late-time ``inflationary period'', by which we indicate the period when $\phi$ decays out of equilibrium increasing the total entropy density of the Universe, leads to a DM relic abundance which is diluted by a factor  $\Delta$ which may be very large, depending on the couplings $y_f$. As a result, the thermal relic abundance is related to its value in absence of the $\phi$ decay by
\begin{equation}
    (\Omega h^2)_\text{after} = \dfrac{ (\Omega h^2)_\text{before}}{\Delta}.
\end{equation}
For this to hold, however, one must ascertain that the DM species decouple {\em before} $\phi$'s decay, thus one needs to consider DM with mass at least $m_{\chi} \gtrsim 20~\text{MeV}$; as a concrete example, here we pick $m_\chi = 100$ MeV.

\begin{figure}[H]
    \centering
    \subfloat{{\includegraphics[width=7.5cm]{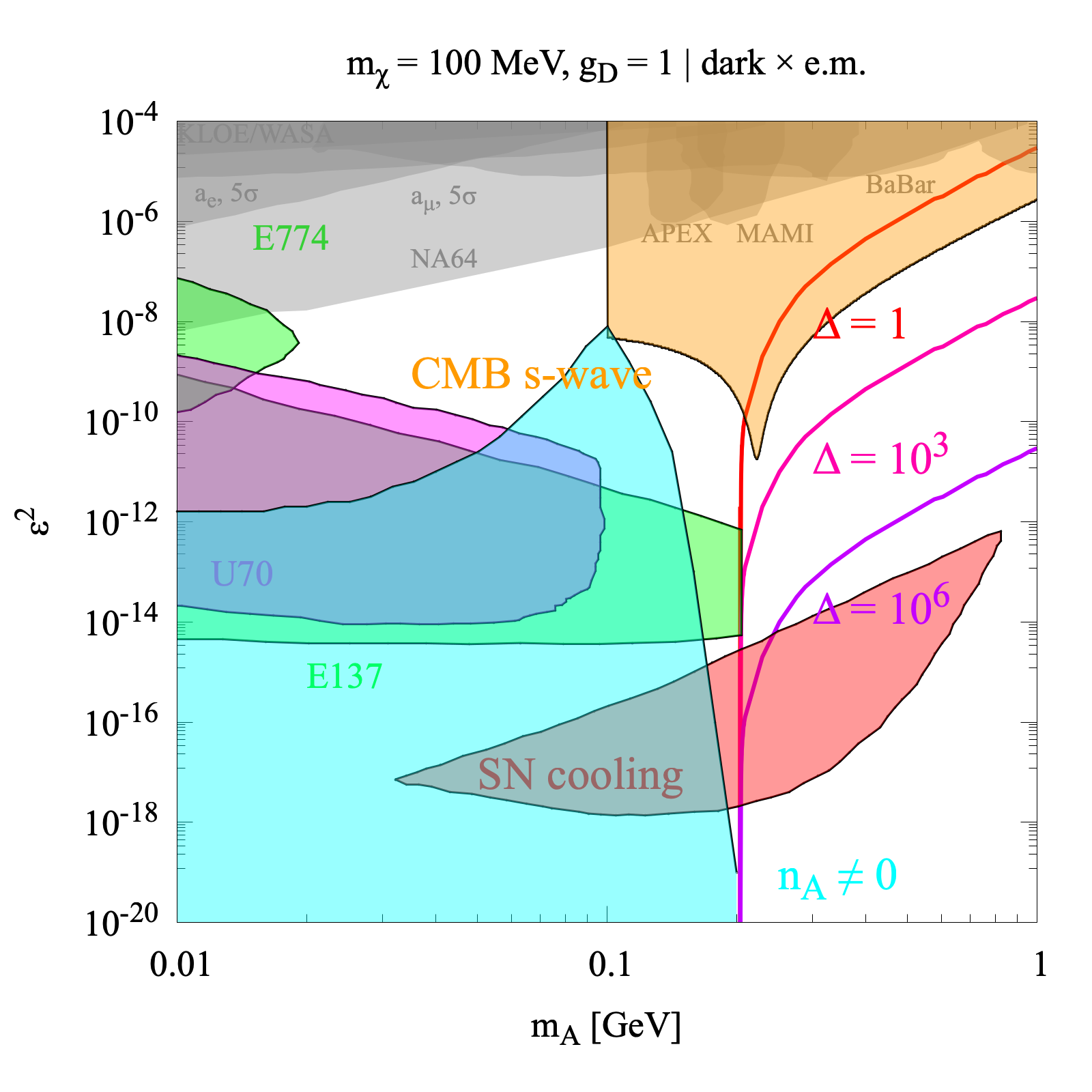} }}%
    \qquad
    \subfloat{{\includegraphics[width=7.5cm]{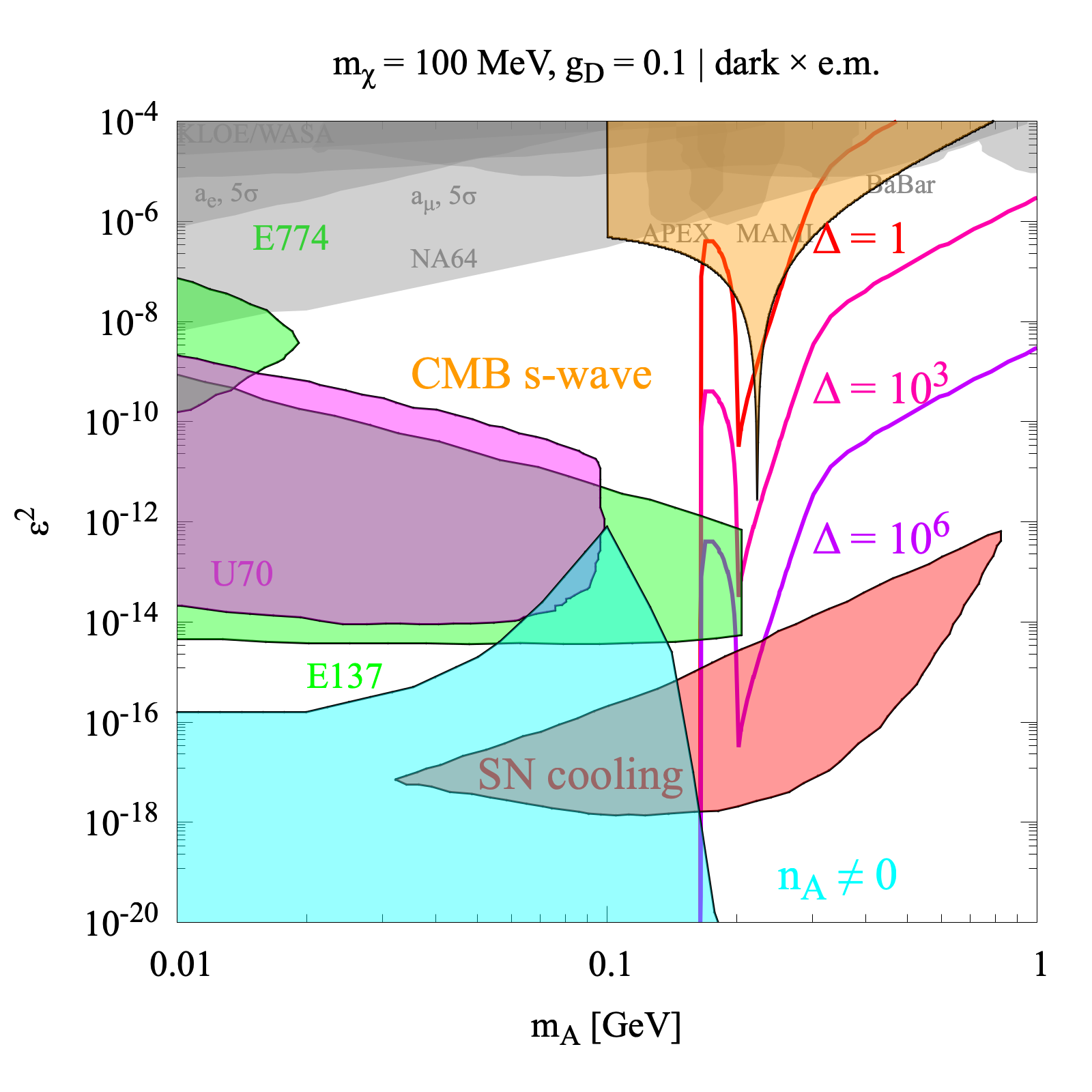} }}%
    \qquad
    \subfloat{{\includegraphics[width=7.5cm]{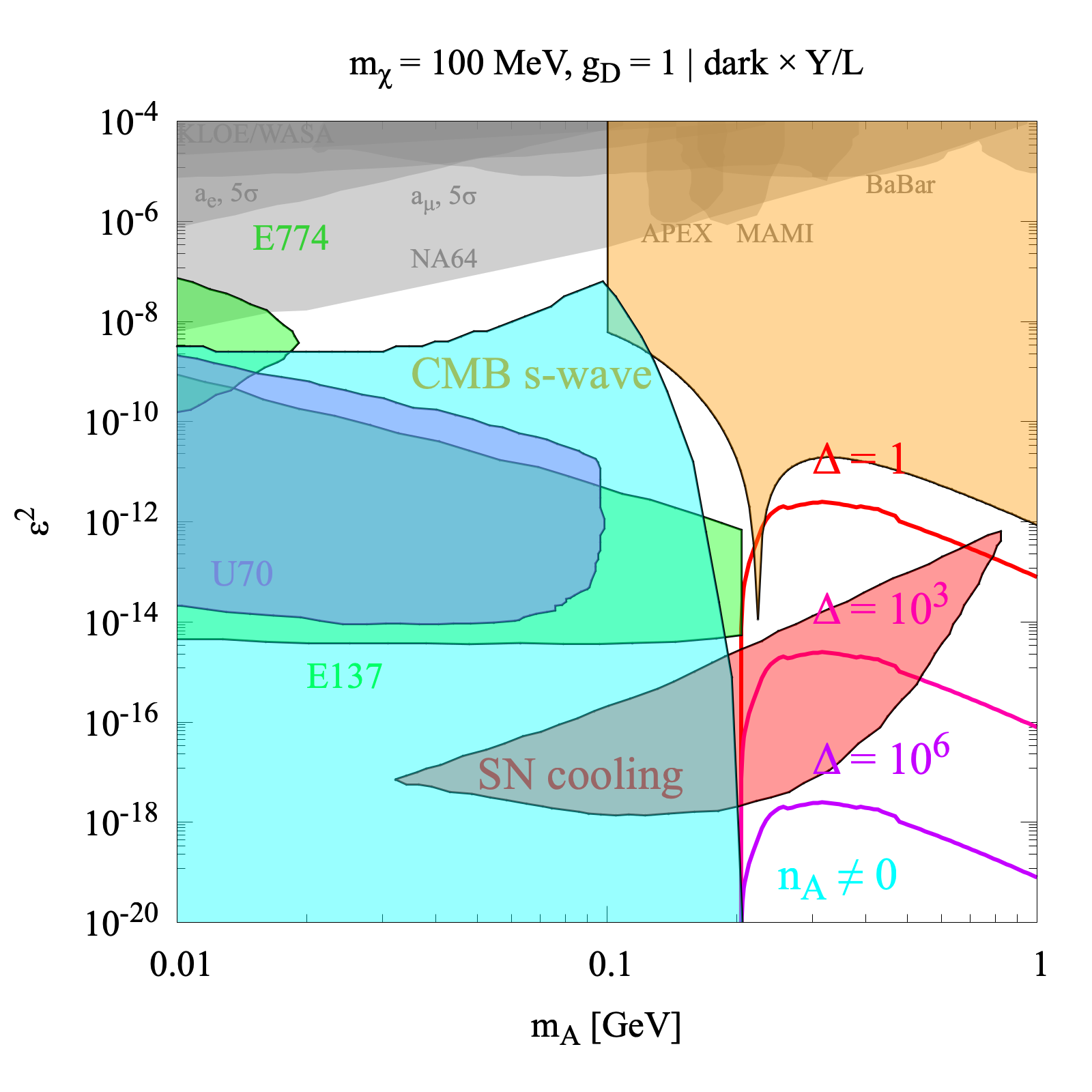} }}%
    \qquad
    \subfloat{{\includegraphics[width=7.5cm]{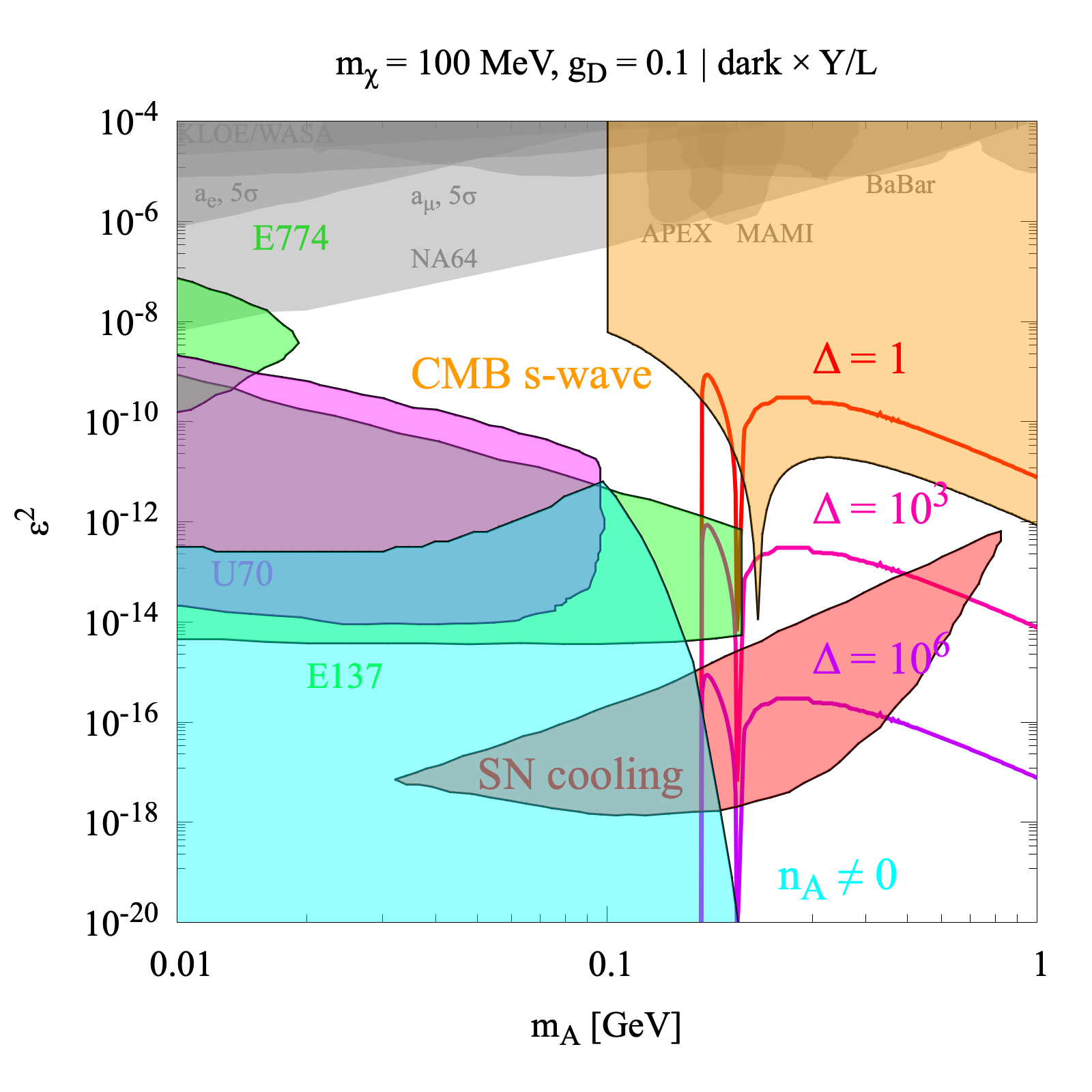} }}%
    \caption{The dark $\times$ electromagnetic (top panels) and dark $\times$ Y (bottom panels) for $m_\chi=100$ MeV and various values of $g_D$ and of the dilution factor $\Delta$. In particular, $\Delta = 1$ refers to standard cosmology (i.e. no second inflationary epoch), while $\Delta = 10^3$ and $\Delta = 10^6$ refer to dilution factors produced by a late-time entropy injection episode.
    }
    \label{f.o.100MeVeps(mA)}%
\end{figure}

\begin{figure}[H]\label{f.o.100MeV}
    \centering
    \subfloat{{\includegraphics[width=7.5cm]{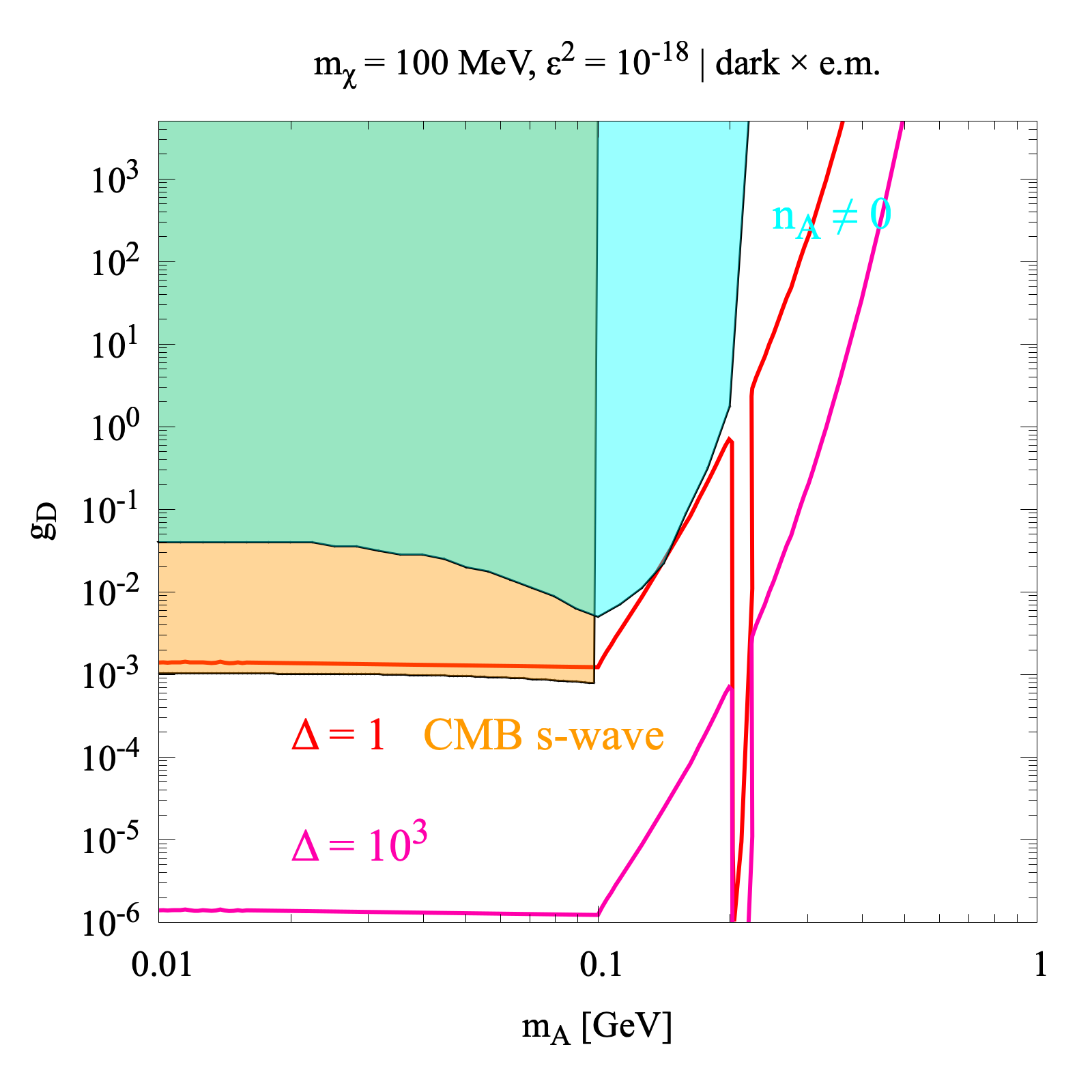} }}%
    \qquad
    \subfloat{{\includegraphics[width=7.5cm]{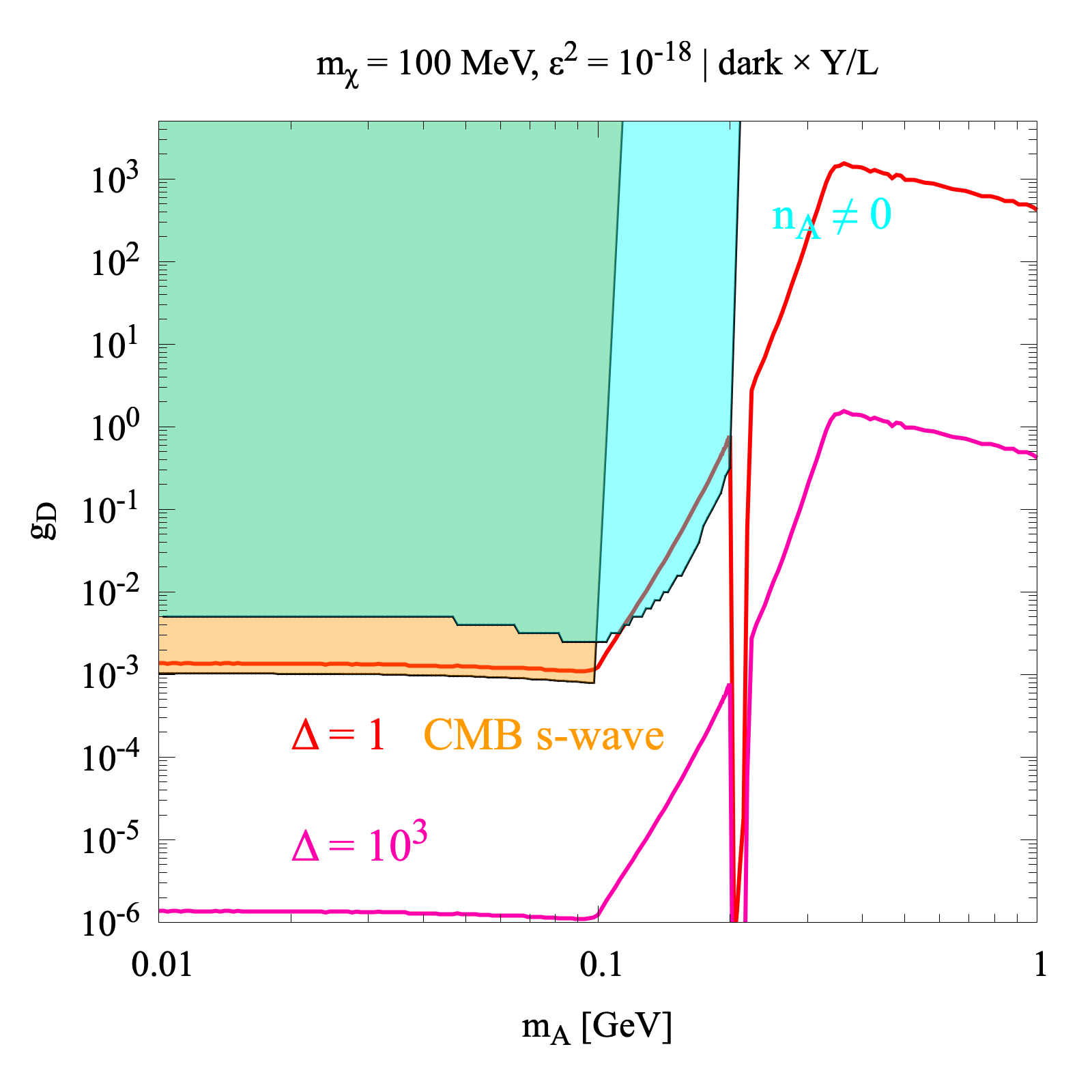} }}%
    
    \caption{%Focus on the
    The $(m_{A}, g_D)$ parameter space for $m_\chi=100$ MeV  for dark $\times$ electromagnetic (left) and dark $\times$ Y (right). Again, notice how a little dilution parameter $\Delta$ is sufficient to evade CMB constraints.}%
    \label{f.o.100MeVgD(mA)}%
\end{figure}

We show slices of the relevant parameter space in Figures \ref{f.o.100MeVeps(mA)} and \ref{f.o.100MeVgD(mA)}.
The two top panels focus on the dark $\times$ electromagnetic case, while in the bottom panels we consider the dark $\times$ Y/L case. The plots show predictions for a DM candidate of $m_\chi = 100$ MeV in a standard cosmology ($\Delta = 1$, red continuous line) and for two dilution factors, $\Delta = 10^3$ (magenta) and $\Delta = 10^6$ (purple). Notice that the latter two open up  new regions of parameter space as the DM yield must be over-abundance to get diluted into the right amount. In turn, this forces the interaction rates (in this region, dominantly $\chi\chi\to e^+e^-$) to be suppressed, with respect to the $\Delta = 1$ case, for the DM to decouple sooner from the primordial plasma and hence to get overproduced by the right amount that later gets diluted away.

In Figure \ref{f.o.100MeVgD(mA)} we focus on a single value for $\varepsilon$, and continue to consider $m_\chi = 100$ MeV.  Here, even a small $\Delta$ allows to evade the CMB constraints, and opens up the parameter space at small dark photon masses $m_A\ll m_\chi$. For $m_A>m_\chi$ the non-trivial dependence of $g_D(m_{A'})$ sets in again. % dilution to take place guaranteed us a spectrum of candidates from about $\Delta \gtrsim 1$ being the non-dilution case very close to experimental bounds' edge. 
%In particular, $\Delta = 10^3$, for instance,  $g_D = 1.39\times 10^{-6}$ up to $m_{A'} = m_\chi$ for then turn to a variable $g_D(m_{A'})$ dependence.\\
%
We conclude that as long as  $T_{\text{f.o.}} > T_{\text{BBN}}$, dilution factors generically enable consideration of a significantly wider portion of the theory parameter space.% us to explore new parameter space regions.

\subsection{Freeze-in}\label{f.i.}
Let's now turn to the freeze-in mechanism. Here one assumes the DM to be absent from the early Universe thermal bath, and DM production to result from out-of-equilibrium processes instead, with the DM remains permanently out of equilibrium. %during its thermal history, although without ever reaching an equilibrium situation. 
In this case, we define an ``initial" temperature of freeze-in production to be the highest temperature compatible with our model, i.e. $T_{\text{f.i.}} = T_{\text{EWSB}}$ for the dark $\times$ electromagnetic model and a generic post-inflationary temperature, which we fix to $T_{\text{f.i.}} = 200~\text{GeV}$ for the dark $\times$ Y/L models. Even though in principle one should consider new physics to uniquely determine such temperatures, we will argue below that the dependence on this $T_{\text{f.i.}}$ value is extremely suppressed and only extremely strong interactions may produce non-negligible production at high temperatures. In a nutshell though, it is sufficient to notice that relevant interactions are faster than the Hubble rate. As mentioned in the introduction, this happens when temperatures are low, from where we deduce that thermal freeze-in is an IR-dominated mechanism. For complete analytical expressions of interaction rates we refer the reader to Appendix \ref{sec:thermal_averaged_cross_section}.

Note that if we assumed DPs to be present during DM freeze-in, due to  strong interactions $g_D \gg \varepsilon$, they would bring DM to equilibrium at least within the DS plasma, resulting in production via freeze-out as discussed above. Assuming then $n_{A'} = 0$, the  Boltzmann equation for DM reads:
\begin{equation}\label{f.i.BE}
    \begin{split}
        \dfrac{\mathop{dY_\chi}}{dT} \sim -\sqrt{\dfrac{\pi}{45}}M_P \dfrac{g_{\star s}}{\sqrt{g_\star}}Y_{\text{eq,}f}^2\langle \sigma v_{\text{prod}}\rangle,
    \end{split}
\end{equation}
where
\begin{align}\label{sigmavproduction}
        \text{dark} \times \text{e.m.}&: \langle \sigma v_{\text{prod}}\rangle \sim \langle \sigma v_{e^+e^-\to\chi\chi}\rangle \\\label{sigmavproduction1}
    \text{dark} \times \text{Y}&: 
    \langle \sigma v_{\text{prod}}\rangle \sim \langle \sigma v_{e_L^+e_L^-\to\chi\chi} \rangle + \langle \sigma v_{e_R^+e_R^-\to\chi\chi } \rangle  + \sum_{\ell\,=\, e,\,\mu,\,\tau}\langle \sigma v_{\nu_{\ell_\text{L}}\nu_{\ell_\text{L}}\to\chi\chi } \rangle \\\label{sigmavproduction2}
    \text{dark} \times \text{L}&: 
     \langle \sigma v_{\text{prod}}\rangle \sim \langle \sigma v_{e_L^+e_L^-\to\chi\chi} \rangle + \langle \sigma v_{e_R^+e_R^-\to\chi\chi } \rangle  + \sum_{\ell\,=\, e,\,\mu,\,\tau}\langle \sigma v_{\nu_{\ell_\text{L}}\nu_{\ell_\text{L}}\to\chi\chi } \rangle
\end{align}
Another possible complication arises in connection with the presence of light quarks participating in DM production (and hence dividing the integration domain into $(T_{\text{EWSB}},\Lambda_{\text{QCD}})$ and $(\Lambda_{\text{QCD}},T_0)$); however, we find that the quark contribution, even after summing over all possible light degrees of freedom, is on the order of $1\%$ of the total relic density. In fact, light quarks only contribute between $(T_{\text{EWSB}},\Lambda_{\text{QCD}})$, i.e. when the high temperature suppression is still active. Electrons on the other hand contribute during the entire thermal history of the early Universe, resulting as the leading production initial states. Hadronic states contributions may come from light hadrons such as pions producing DM in the final state (see e.g. \cite{subGeV2}), but these would be inevitably strongly Boltzmann-suppressed due to the hadron  masses. We note that an exception might consist of purely hadrophilic DM, where it has been shown pions can in fact give a substantial contribute in the case of a scalar mediator \cite{Bhattiprolu_22}.

\begin{figure}[H]
    \centering
    \subfloat{{\includegraphics[width=7.5cm]{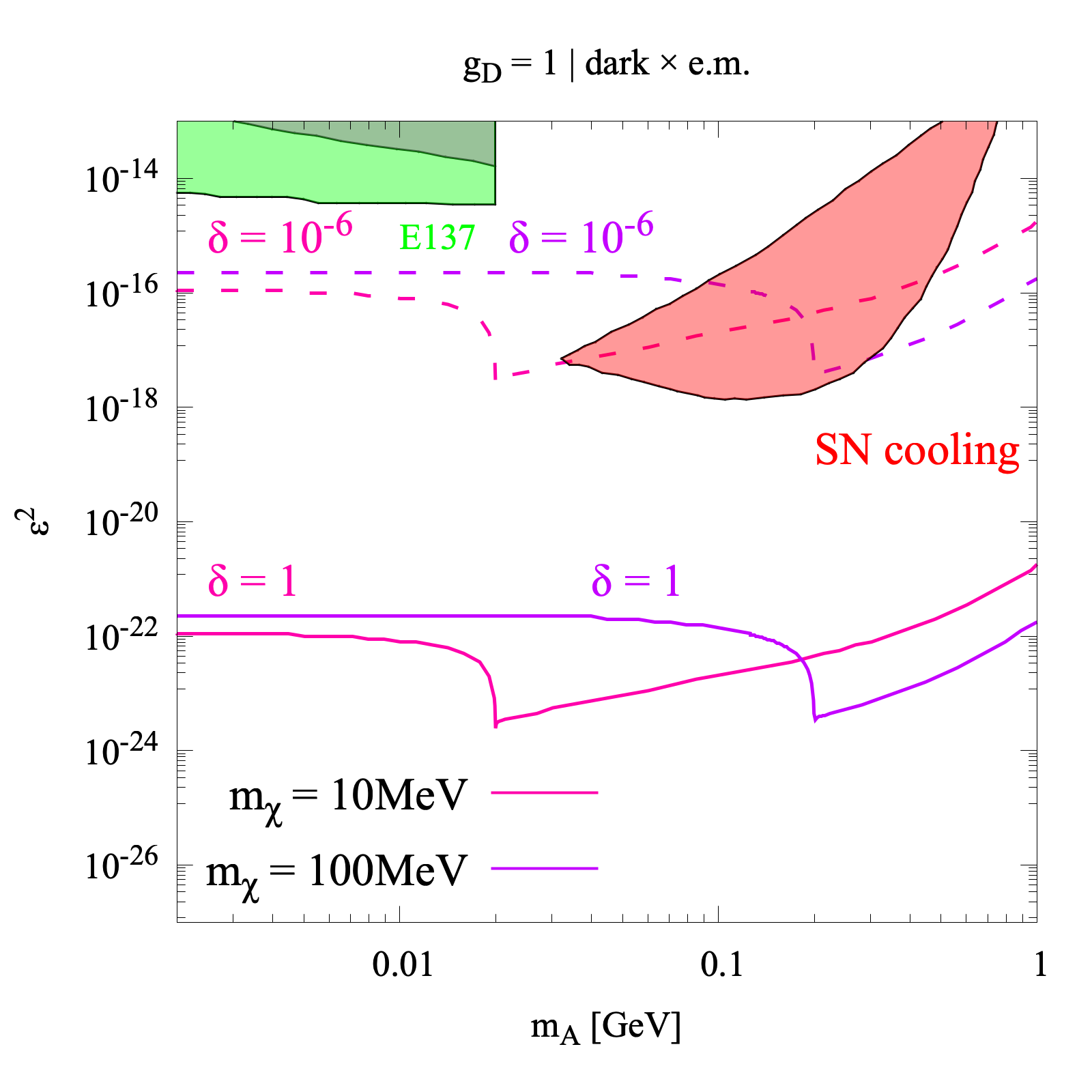} }}%
    \qquad
    \subfloat{{\includegraphics[width=7.5cm]{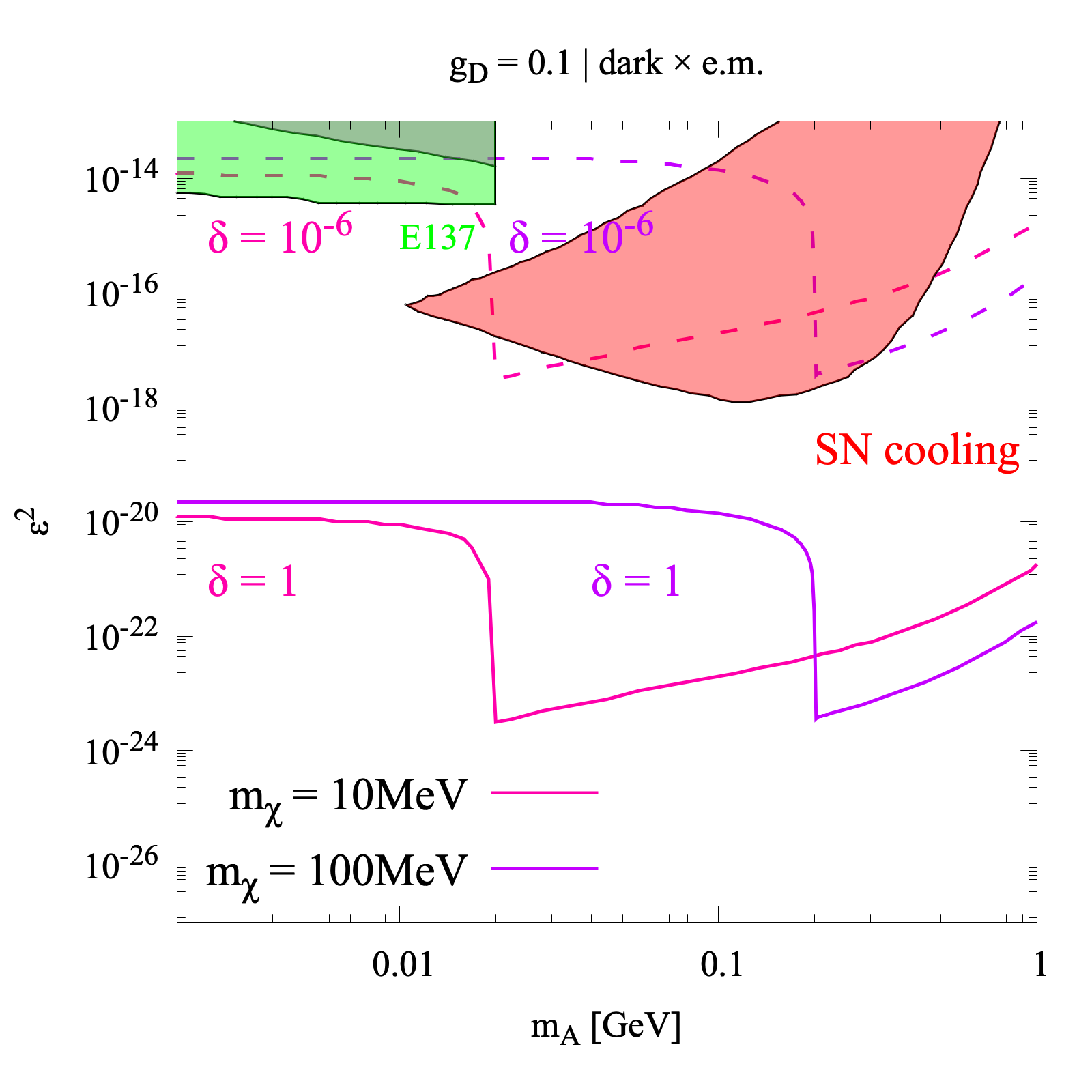} }}%
    \qquad
    \subfloat{{\includegraphics[width=7.5cm]{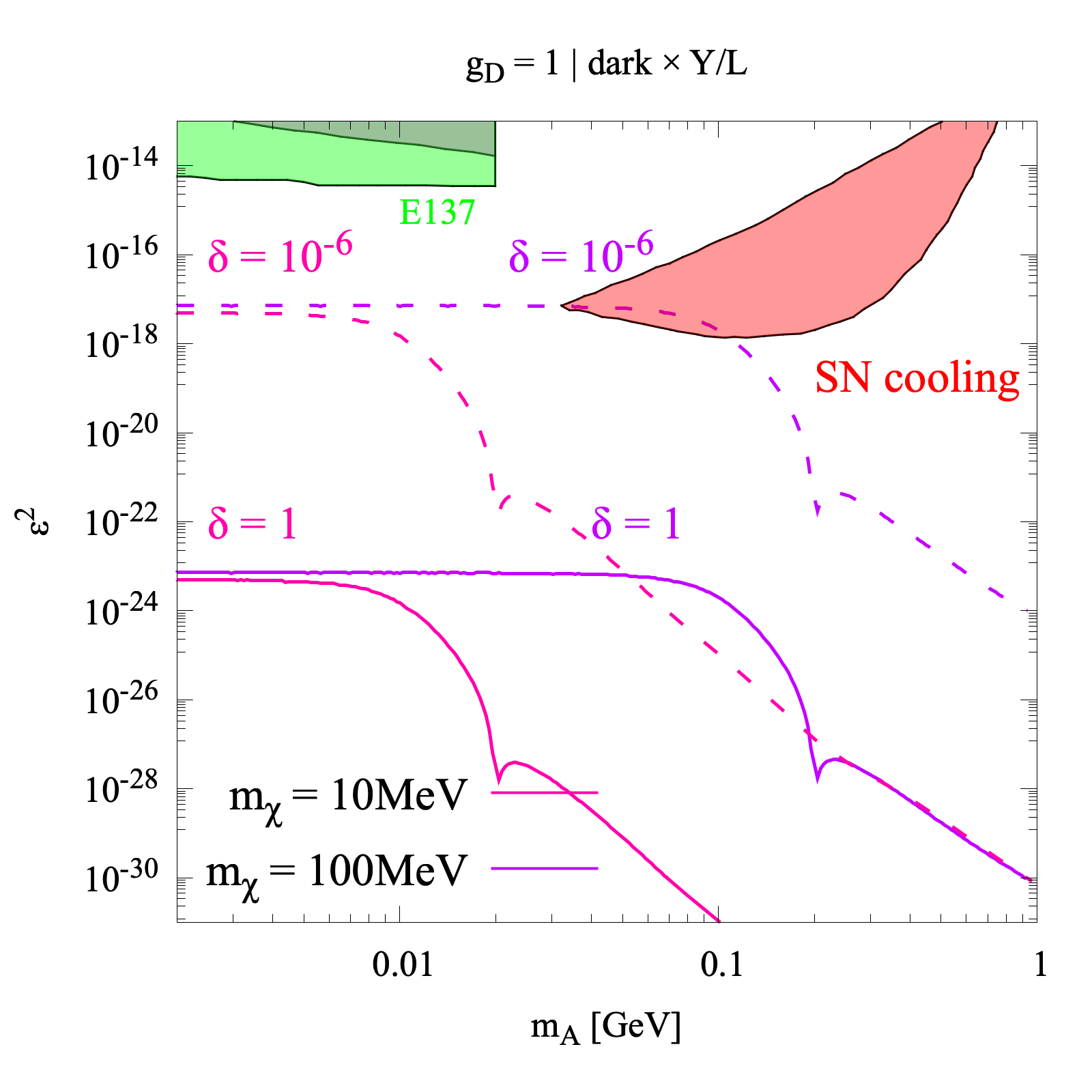} }}%
    \qquad
    \subfloat{{\includegraphics[width=7.5cm]{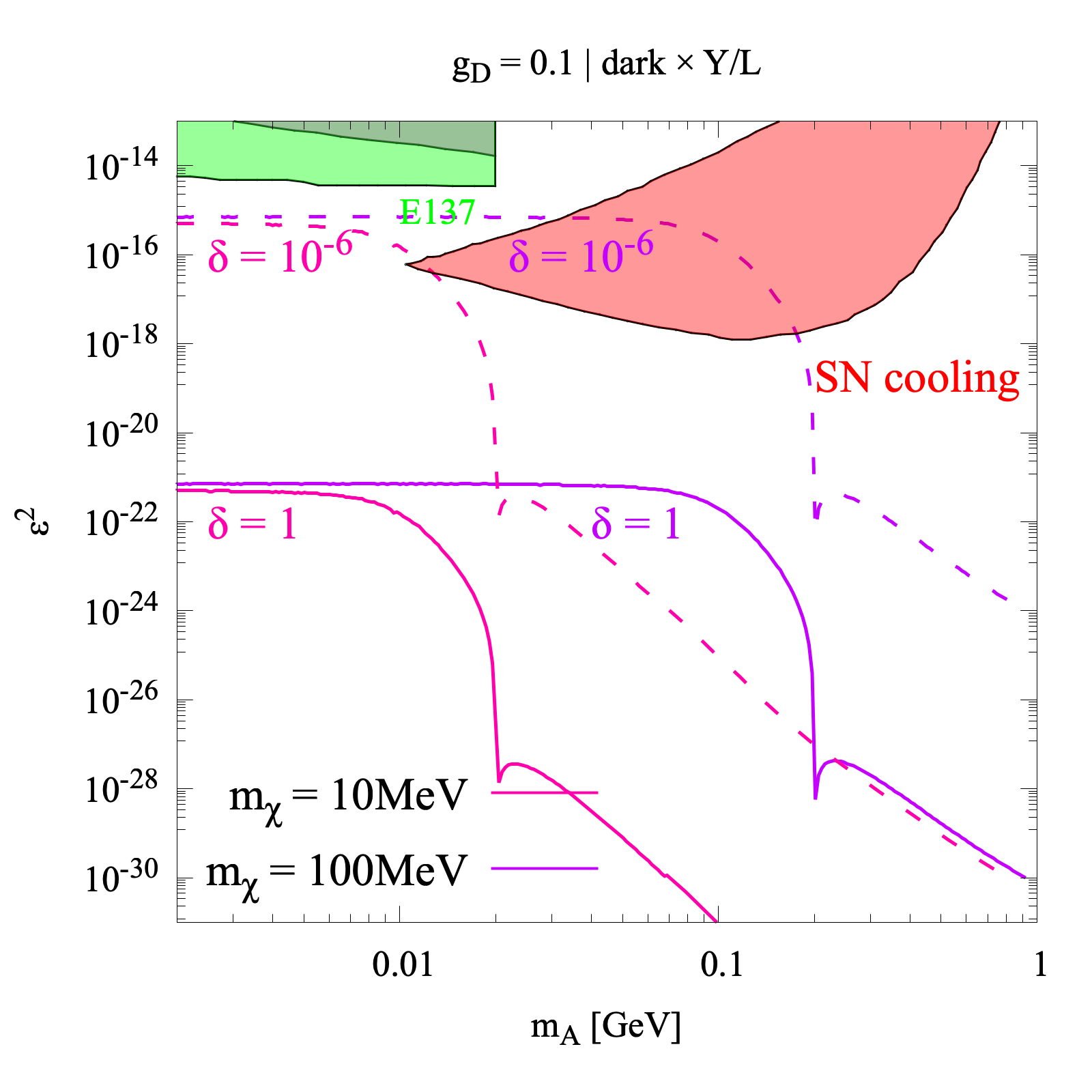} }}%
    \caption{Freeze-in predictions for coupling constant values $g_D = 1$, $g_D = 0.1$ and DM mass $m_\chi = 10$ MeV, $m_\chi = 100$ MeV; the top panels show the predictions for a dark $\times$ electromagnetic model, while the bottom panels those for a dark $\times$ Y/L model, both including a $\delta$ parameter to account for additional, non-freeze-in DM production.}
    \label{f.i.10/100MeVeps(mA)}%
\end{figure}

In Fig.~\ref{f.i.10/100MeVeps(mA)} we analyze DM candidates with masses $m_\chi = 10$ MeV (magenta lines) and $m_\chi = 100$ MeV (violet lines) for coupling constants $g_D = 1$ and $g_D = 0.1$ in a normal freeze-in production (continuous lines) as well as in a paradigm where a more complex dark sector may be present (dashed lines). In particular, for the latter we  assume that not all the DM contributing to $\Omega h^2$ results from freeze-in, but only a fraction $\delta$ of it, % a thermal production (such as freeze-in), then our predictions would just refer to a small percentage of the entire dark sector. Hence, we introduced a $\delta$ parameter to account for our ignorance in this respect, for which
%\begin{equation}
%    \delta = \dfrac{\Omega h^2_\text{thermal}}{0.12}
%\end{equation}
with the remaining $(1-\delta \Omega h^2)$ produced by a different mechanism, such as e.g. post-inflationary decay of a scalar field.

We find that only if $\delta\ll 1$ is freeze-in ever close to laboratory and astrophysical constraints such as those from SN cooling.
%Only thermal predictions are characterized by being out of bounds in terms of experimental sensitivity, being the closer bound we currently own the one of SN-cooling. To be more precise, 
As a reminder, SN events may produce sufficiently light particles like axions or DPs, in our case possibly emitted through channels $p + p \to p + p + A'$ and $p + n \to p + n + A'$ via both bremsstrahlung and pion emission. The emission of new physics light degrees of freedom generically alters the supernova energy loss rate. %, which can be in turn be expressed as luminosity in the emitted light particle. 
The maximum energy loss allowed by observations of SN1987A \cite{raffelt96} reads
\begin{equation}
    \epsilon = \dfrac{L}{M} \sim 10^{19}\dfrac{\text{erg}}{\text{g s}},
\end{equation}
where $M$ is the supernova mass and $L$ its luminosity. This yields a lower limit to $\epsilon$; however, for large values of $\epsilon$ ``trapping'' is possible \cite{dreiner14}, due to the fact that sufficiently strongly interacting DPs may not be able to escape the supernova before decaying again. Trapping then gives an upper limit on $\epsilon$. %Unfortunately these constraints are escaped by a MeV DM candidate produced by a freeze-in mechanism, unlike freeze-out and equilibrium mechanism explained before. \\
As noted above, if not all of the DM is produced via freeze-in, and thus $\delta < 1$, then  SN cooling and beam dumps experiments may play an important role and provide significant constraints, especially in the case of  dark $\times $ electromagnetic for  $g_D \lesssim 0.1$.
% End production  -------------------

% Begin future searches
\section{Future Searches}\label{sec:future}
In the scenario discussed here, WIMP-like DM candidates and portal DPs lie around the MeV scale. The direct detection of MeV dark matter is challenging, since the recoil energy is well below the threshold sensitivity of most current detectors. In the region $m_{A'} > 1$ MeV, probes of the model under consideration include experiments at colliders and beam dumps. In both cases, a resonance is searched for over a smooth background, with a prompt or slightly displaced vertex with respect to the beam interaction point, in case of a collider, while greatly displaced in the case of beam dump experiments. 

As far as colliders experiments are concerned, we have already discussed  annihilation processes such as $e^+ +e^- \to \gamma+ A'$ testable with  experiments such as BaBar \cite{collabBaBar}, but also Bremsstrahlung ($e^- + Z \to e^- + Z + A'$) and pion decays ($\pi^0 \to \gamma A'$) at KLOE \cite{archilli12}. Beam dump experiments, instead, make use of protons or electrons beams with fixed targets to produce dark photons through Bremsstrahlung and meson production. Examples of these are E141, E137 at SLAC and E774 at Fermilab \cite{andreas12}. Improvements in these directions will be taken up by NA62 and NA64$(e)^{++}$ at SPS, CERN \cite{NA62, Gninenko:2300189}, FASER and FASER2 at LHC, CERN \cite{Feng_2018}, HPS at Jefferson Laboratory \cite{Adrian_2018}, SeaQuest at Fermilab \cite{Berlin_2018}, MAGIX detector for Bremsstrahlung productions \cite{MAGIX} and MESA accelerator \cite{MESA} (see \cite{Fabbrichesi_2021} for more). It is finally worth noticing the effort towards the measurement of millicharged particles below 10 MeV at CERN and SLAC with the proposed LDMX experiment \cite{ldmx, dasel, Akesson:2640784}.

Turning to indirect detection, the annihilation of MeV-scale dark matter might yield a signature at telescopes sensitive to the relevant energy range. Such signatures consist of often unmistakable features in the electromagnetic spectrum. Gamma rays from WIMPs in the 0.5 $-$ 250 MeV mass range would lie predominantly in the range $\mathcal{O}$(0.1 $-$ 100 MeV), which includes the $\pi^0\to \gamma\gamma$ decay peak centered at roughly 70 MeV. This energy domain was last searched for by EGRET \cite{EGRET} and COMPTEL \cite{COMPTEL}; Future telescopes include  proposed experiments such as GAMMA$-$400 \cite{GAMMA-400}, Advanced Compton Telescope (ACT) \cite{ACT}, Advanced Energetic Pair Telescope (AdEPT) \cite{AdEPT}, PANGU \cite{PANGU}, GRAMS \cite{GRAMS}, MAST \cite{MAST}, AMEGO \cite{AMEGO}, All-Sky-ASTROGAM \cite{All-Sky-ASTROGRAM} and GECCO (Galactic Explorer with a Coded Aperture Mask Compton Telescope), which encapsulates at once the principles of a Compton telescope and of a coded-aperture mask telescope. Performance of the latter in MeV dark matter detection has already been studied closely and we will refer to those studies in what follows \cite{subGeV1, subGeV2}.

Future CMB surveys will additionally offer constraining power; %, a complementary analysis of recent \textit{Planck} CMB data \cite{planck16, planck2020} made this latter measurement an extremely important and complementary avenue to constraint dark matter from a cosmological perspective \cite{Acharya_2019, Slatyer_2016,Slatyer_2017,Liu_2016}.
thermally produced dark matter can, in principle, annihilate (or decay) into electrons and photons, re-ionizing an amount of the neutral baryonic gas during the Dark Age. Free electrons may enlarge the CMB last scattering surface, allowing for measurable imprints on the temperature and polarization anisotropy spectra of CMB. The ``annihilation parameter'' $p_\text{ann} = f_\text{eff}\langle \sigma v\rangle / m_\chi$ encapsulates all the needed information, taking into account the efficiency of ionization injection through annihilation (or decay) with a redshift-independent (see \cite{slatyer16} for detailed arguments) parameter $f_\text{eff}$. Operating and upcoming experiments that will improve \textit{Planck} data are BICEP3/KECK Array \cite{keck} and South Pole Telescope-3G \cite{South-Pole-Telescope} in Antarctica, Advanced Atacama Cosmology Telescope Polarimeter (AdvACTPol) \cite{Advanced-Atacama}, Cosmology Large Angular Scale Surveyor(CLASS) \cite{CLASS}, Simons Array \cite{Simons-Array} and Simons Observatory \cite{Simons-Observatory}.

\begin{figure}[H]
\centering
    \includegraphics[width=11cm]{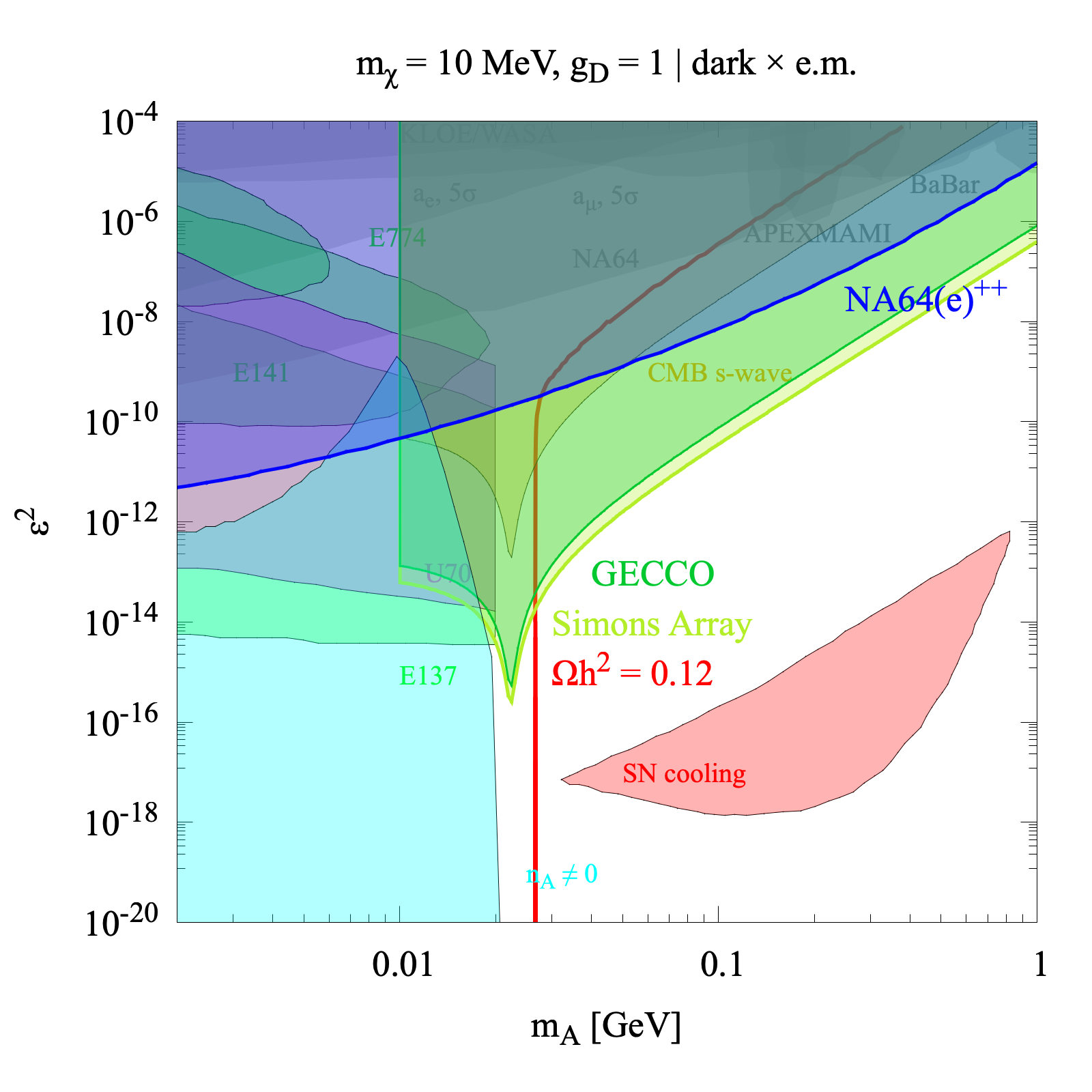}
    \caption{Predictions for the sensitivity of the future direct detection experiment NA64$^{++}$ \cite{Gninenko:2300189}, and for indirect searches for DM with the GECCO telescopes \cite{subGeV1,subGeV2} and with precision CMB measurements by the Simons Array \cite{Simons-Array}.}
    \label{fig:predictions}%
\end{figure}

We conclude by comparing the sensitivity reach of three future experimental avenues. In particular we considered NA64$^{++}$ projections for direct detection experiments \cite{Gninenko:2300189}, GECCO for indirect detection \cite{subGeV1,subGeV2} and the Simons Arrays for CMB precision measurements \cite{Simons-Array}. Predictions are shown in Fig. \ref{fig:predictions} with, respectively, blue, dark green and light green lines, for a representative slice of parameter space with $g_D = 1$, $m_\chi = 10$ MeV in the dark $\times$ electromagnetic theory.
NA64$^{++}$ becomes increasingly important in the high $A'$ mass region, where, however, $\Omega h^2$ predictions are already ruled out by \textit{Planck}. GECCO and the Simons Array will play a fundamental role in extending the current parameter space sensitivity down to smaller values of $\varepsilon^2$ (roughly by three orders of magnitudes) in the finely-tuned region approximately around $m_{A'} \sim 2 m_\chi$, where DM candidates are currently unconstrained.
% End future searches

% Begin conclusions
\section{Conclusions}\label{sec:conclusions}
Previous studies on dark photon portal dark matter models analyzed massless force carriers obtained after symmetry breaking from kinetic mixing operators involving the hypercharge boson \cite{Chu_2012}. The generalization to massive gauge fields was fully studied in Ref.~\cite{hambye19}, and visually portrayed with a suggestive ``Mesa'' phase diagram: in it, the authors studied different (and sometimes rather  contrived) mechanisms for producing DM while varying the model parameters. The result is a ``phase diagram'' where the ``phases'' represent different production mechanisms. Moreover, in the literature, scalar mediator fields (like the Higgs portal) have also been entertained (upon which there are already stringent bounds \cite{Krnjaic_2016}) as well as fermions like the right handed neutrino in the hypothesis of non-thermal DM production (for a complete review, see e.g.  \cite{alexander16}). 

Here, we focused on a particular configuration within the mentioned ``Mesa'' diagram, where the DM is assumed to be in equilibrium, allowing  freeze-out  to take place. We discussed the presence of DPs in the thermal bath, treating their thermalization independently and deriving constraints on the resulting  Boltzmann equations. Moreover, we considered, to our knowledge for the first time, strongly interacting Dark Sectors, which yield novel results for thermal freeze-out.

We considered the most significant experimental constraints, and additionally considered  non-standard cosmological setups such as a late-time inflationary period  to broaden the relevant region of parameter space under consideration. 
We also investigated  freeze-in production, again taking into account the presence of DPs and their possibly strong interactions with DM. Again to entertain a broader range of possible parameters, we assumed that only part of the DM be produced by freeze-in.

%From a theoretical standpoint, we analyzed three different theories for studying a  MeV DM model, where a bridge between the SM and the Dark Sector is established by a vector mediator DP. In particular, among the numerous possibilities, we focused on a "Secluded" DM with no millicharge, whose visible manifestation may only due to the presence of the DP and we obtained it through kinetic mixing an initial neutral gauge boson with the abelian content of the SM gauge group. \\
%We studied the implications of having DPs in equilibrium within the primordial plasma to Boltzmann equations and derived a condition for their thermal decoupling, through which we've set a new bounded region in parameters space (both in the $\varepsilon^2(m_{A'})$-plane and $g_D(m_{A'})$-plane) in which one can't assume DP were totally disappeared at the moment of DM production.\\

Our main conclusions are  that the general setup accommodates the observed relic abundance for highly constrained configurations in parameter space:  $m_\chi \sim m_{A'}$ in the $\varepsilon^2(m_{A'})$-plane, for dark $\times$ electromagnetic, $\varepsilon^2 \lesssim 10^{-14}$ when $(m_{A'} = 26.6\text{ MeV}, g_D = 1)$, $(m_{A'} = 22.5\text{ MeV}, g_D = 10^{-1})$ whilst for dark $\times$ Y, $\varepsilon^2 \lesssim 10^{-18}$ when $(m_{A'} = 21.6\text{ MeV}, g_D = 1)$ and $\varepsilon^2 \lesssim 10^{-17}$ when $(m_{A'} = 22.1\text{ MeV}, g_D = 10^{-1})$. In the case of freeze-out scenario, two regions emerged: the first corresponds to when $m_{A'}<m_\chi$ and DM production is driven by the interaction $\chi\chi\to A'A'$ and manifestly appears to be independent by mixing parameter $\varepsilon^2$; a second one when $m_{A'}>m_\chi$ and the process $\chi\chi\to ff$ dominates. In particular, when $m_{A'}<m_\chi$ for both cases $m_\chi =10$ MeV and $m_\chi =100$ MeV, we get a sharp prediction for $g_D$ independent of $\varepsilon^2$. We then turned to considering a heavier candidate $m_\chi = 100$  MeV that allowed us also to introduce a new phenomenological parameter, a ``dilution parameter'' $\Delta$ stemming from a hypothetical late inflationary period, possibly driven by a classical scalar field. For this new class of models we derived the same predictions relating $\varepsilon^2(m_{A'})$ and $g_D(m_{A'})$ and highlighted some interesting differences between these models and those with a lighter DM mass; in particular we found a non trivial intersection between our density parameter contour plot and the $n_{A'} \neq 0$ region which excluded DM candidates around the singular behaviour of $\Omega h^2$. In summary, in the case of freeze-out production we find that an important subspace region corresponds to a ``strong'' dark force - $g_D \gtrsim 1$, where experimental bounds are generally weak, while small regions of interest are still present for small values of $g_D$, especially considering late time inflation.

In the freeze-in case, we considered both the presence and absence of DPs in connection with  DM  production and found out that equilibrium is almost immediately reached if the DPs are present. Assuming $n_{A'} = 0$ we solved simplified Boltzmann equations obtaining again contour plots in the $\varepsilon^2(m_{A'})$-plane, noting that the DM is mainly produced by electron-positron pairs with light quarks contributing at most $\sim 1\%$  to $\Omega h^2$. We found that SN cooling bounds are the closest to constrain model predictions, but are still unable to reach the preferred region of parameter space due to the extremely small kinetic mixing parameter.

Finally we reviewed direct, indirect and CMB measurements future experiments in a particular (and representative) slide of parameters space. We saw that indirect and CMB measures will play an important role in constraining our models in the $m_\chi \sim m_{A'}$ region, where the dependency on $\varepsilon^2$ disappears. Future MeV gamma-ray telescopes and precision CMB surveys will significantly extend the discovery potential of the MeV DM/DP models we considered here.

%The nature of the cosmological dark matter continues to be a pressing problem  in modern and contemporary physics, both theoretically and experimentally. The idea of a new particle, extending the SM of particle physics, was (formally) born in the mid '70 in parallel with the completion of the already mentioned SM. This means that for the first forty years the "missing mass" problem was perceived differently due to a different (and in some sense smaller) theoretical and experimental set of knowledge than the one we currently have. Forty years have passed since then and the first centenary since Zwicky measurement is getting closer. Our ongoing efforts toward this direction are more sophisticated than ever, for it may be said it'll be only a matter of time before unveiling DM once for all. \\
%The problem is that there's no way for us to know how much this time amounts to.
% End conclusions

% Begin acknowledgments ----------
\section*{Acknowledgments}
S.P. is partly supported by the U.S. Department of Energy grant number de-sc0010107. N.F. acknowledges support from: {\sl Department of Excellence} grant awarded by the Italian Ministry of Education, University and Research (MIUR); Research grant {\sl The Dark Universe: A Synergic Multimessenger Approach}, Grant No. 2017X7X85K funded by the Italian Ministry of Education, University and Research (MIUR); Research grant {\sl TAsP (Theoretical Astroparticle Physics)} funded by Istituto Nazionale di Fisica Nucleare (INFN).
% End acknowledgments ----------

%\newpage
% Begin appedix ------------------
\begin{appendices}\label{sec:app}
\section{Thermal averaged cross section}\label{sec:thermal_averaged_cross_section}
Fundamental ingredient in Boltzmann Equations \eqref{Boltzmann_equation} are thermal averaged cross sections whose general expression is
\begin{equation}\label{thermal_average}
    \langle \sigma v_{\text{M}}  \rangle = \dfrac{\displaystyle\int \mathop{d^3p_1}\mathop{d^3p_2} f(p_1) f(p_2) \sigma v_{\text{M}}}{\displaystyle\int \mathop{d^3p_1}\mathop{d^3p_2} f(p_1)f(p_2)}
\end{equation}
If we ignore quantum corrections to phase space distributions i.e. assume Maxwell-Boltzmann distributions $f(p) = \exp(-E/T)$, we may manipulate \eqref{thermal_average} up to the point where we're forced to specify $\sigma v_\text{M}$ \cite{gondologel90}.

\begin{itemize}
    \item For a $\text{a}_{(p_1,E_1)}+\text{a}_{(p_2,E_2)}\to\text{b}_{(k_1,\omega_1)}+\text{b}_{(k_2,\omega_2)}$ process (e.g. $\chi\chi\to ff$) we perform a change of variables $\phi$ such that
    \begin{equation*}
    \phi:\begin{cases}
    E_+ = E_1+E_2\\
    E_- = E_1-E_2\\
    s = 2m_1^2+2E_1E_2-2\lvert \vv{p_1}\rvert \lvert \vv{p_2}\rvert\cos\vartheta
    \end{cases}
    \end{equation*}
transforming volume element and boundaries as
\begin{equation*}
    \begin{split}
        \mathop{d^3p_1}\mathop{d^3p_2} = 8 \pi^2 E_1p_1E_2p_2 \mathop{dE_1}\mathop{dE_2}\mathop{d \cos\vartheta} \stackrel{ \phi}{=} 2 \pi^2 E_1E_2 \mathop{dE_+}\mathop{dE_-}\mathop{ds} \\
        \begin{cases}
    E_1\geq m_1^2\\
    E_2\geq m_1^2\\
    \lvert \cos\vartheta \rvert \leq 1
    \end{cases} \stackrel{ \phi}{\longrightarrow}
    \begin{cases}
    s\geq4m_1^2\\
    E_+\geq \sqrt{s}\\
    \lvert E_- \rvert \leq \sqrt{1-\dfrac{4m_1^2}{s}}\sqrt{E_+^2-s}
    \end{cases}
    \end{split}
\end{equation*}
obtaining
\begin{equation}
     \langle \sigma v_{\text{M}_{\text{a}+\text{a}\to \text{b}+\text{b}}}  \rangle =\dfrac{2 \pi^2 T g_\text{a}^2}{(2\pi)^6 (n_\text{a}^{\text{eq}})^2}\int_{4m_1^2}^{\infty}\mathop{ds} \sqrt{s} K_1(\sqrt{s}/T) \left( s-4m_1^2 \right)\sigma_{\text{a}\text{a}\to\text{b}\text{b}}
\end{equation}
\item For a $\text{a}_{(p_1,E_1)}+\text{b}_{(k_1,\omega_1)}\to\text{a}_{(p_2,E_2)}+\text{b}_{(k_2,\omega_2)}$ process (e.g. $\chi A'\to \chi A'$), change of variables $\phi$ is in the form
\begin{equation*}
    \phi:\begin{cases}
    E_+ = E_1+\omega_1\\
    E_- = E_1-\omega_1\\
    s = m_1^2+m_2^2+2E_1\omega_1-2\lvert \vv{p_1}\rvert \lvert \vv{k_1}\rvert\cos\vartheta
    \end{cases}
    \end{equation*}
hence
\begin{equation*}
    \begin{split}
        \mathop{d^3p_1}\mathop{d^3k_1} = 8 \pi^2 E_1p_1\omega_1k_1 \mathop{dE_1}\mathop{d\omega_1}\mathop{d \cos\vartheta} \stackrel{ \phi}{=} 2 \pi^2 E_1\omega_1 \mathop{dE_+}\mathop{dE_-}\mathop{ds} 
        \end{split}
\end{equation*}
\begin{equation*}
    \begin{split}
        \begin{cases}
    E_1\geq m_1^2\\
    \omega_1\geq m_2^2\\
    \lvert \cos\vartheta \rvert \leq 1
    \end{cases} \stackrel{ \phi}{\longrightarrow} 
    \begin{cases}
    s\geq (m_1+m_2)^2\\
    E_+\geq \sqrt{s}\\
    \lvert E_- - E_+ \rvert \leq \sqrt{E_+^2-s}\sqrt{1-\dfrac{2(m_1^2+m_{2}^2)}{s}-\dfrac{2m_1^2m_{2}^2-m_1^4-m_{2}^4}{s^2}}
    \end{cases}
    \end{split}
\end{equation*}
and 
\begin{equation}
\begin{split}
     \langle \sigma v_{\text{M}_{\text{a}+\text{b}\to \text{a}+\text{b}}}  \rangle = & \dfrac{2 \pi^2 T g_\text{a}g_\text{b}}{(2\pi)^6 (n_\text{a}^{\text{eq}})(n_\text{b}^{\text{eq}})}\int_{(m_1+m_2)^2}^{\infty}\mathop{ds} \sqrt{s} K_1(\sqrt{s}/T) \times \\
     & \times \left( s-2(m_1^2+m_2^2) - \dfrac{2m_1^2m_2^2-m_1^4-m_2^4}{s} \right)\sigma_{\text{a}\text{b}\to\text{a}\text{b}}
     \end{split}
\end{equation}
\end{itemize}
We now have to turn to $\sigma$, recalling that
\begin{equation*}
    \begin{split}
        \dfrac{\mathop{d\sigma}_{\text{a}\text{a}\to\text{b}\text{b}}}{\mathop{d \text{cos}\vartheta}} & = \dfrac{1}{32\pi s}\dfrac{\sqrt{1-\dfrac{4m_2^2}{s}}}{\sqrt{1-\dfrac{4m_1^2}{s}}}\lvert \overline{\mathcal{M}_{\text{a}\text{a}\to\text{b}\text{b}}}
        \rvert^2 \\
        \dfrac{\mathop{d\sigma}_{\text{a}\text{b}\to \text{a}\text{b}}}{\mathop{d \text{cos}\vartheta}} & =\dfrac{1}{32\pi s}\lvert \overline{\mathcal{M}_{\text{a}\text{b}\to \text{a}\text{b}}}\rvert^2
    \end{split}
\end{equation*}
where we assume azimuthal symmetry and integrate over $\text{cos}\vartheta \in [-1,1]$ for distinguishable final states, while over $\text{cos}\vartheta \in [0,1]$ for indistinguishable final states.

\subsection{\textit{s}-channel (electromagnetic)}
\begin{figure}[H]
\centering
    \includegraphics[width=7cm]{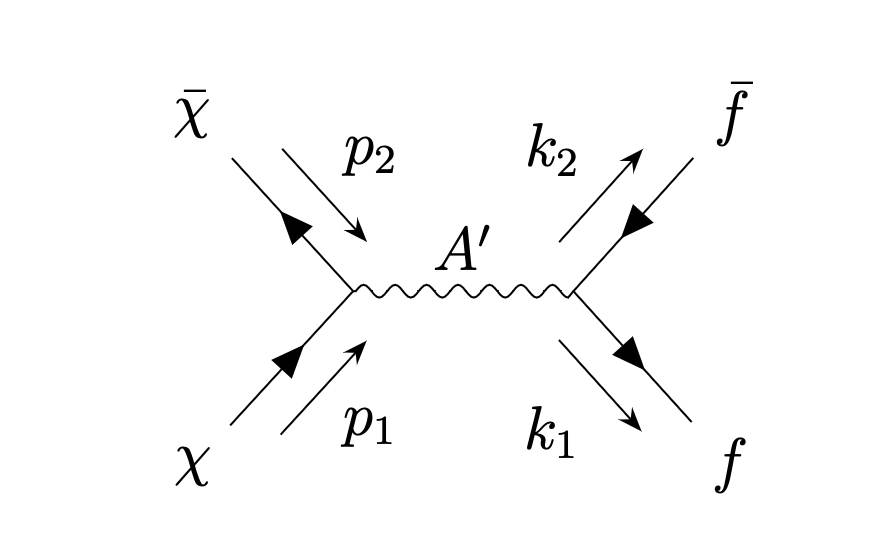}
\end{figure}
The amplitude in this case is just
\begin{equation*}
    \mathcal{M} = \mathcal{M}_{\chi\chi\to A'\to ff}
\end{equation*}
hence
\begin{equation*}
\begin{split}
\lvert\overline{\mathcal{M}}\rvert^2 = & \; 4\left(\varepsilon e q_f g_D\right)^2\Bigg[1+\dfrac{4m_\chi^2}{s}+\dfrac{4m_f^2}{s} + \text{cos}^2\vartheta\left(1-\dfrac{4m_\chi^2}{s}\right)\left(1-\dfrac{4m_f^2}{s}\right)\Bigg] \times \\
& \times \dfrac{s^2}{\left(s^2-m_{A'}^2\right)^2+m_{A'}^2\Gamma_{A'}^2} \\
\sigma_{\chi\chi\to ff} = & \; \dfrac{(e q_f g_D \varepsilon)^2}{12 \pi}\left(1+\dfrac{2m_f^2}{s}\right)\left(1+\dfrac{2m_\chi^2}{s}\right)\dfrac{\sqrt{1-\dfrac{4m_f^2}{s}}}{\sqrt{1-\dfrac{4m_\chi^2}{s}}}\dfrac{s}{(s-m_{A'}^2)^2+m_{A'}^2\Gamma_{A'}^2}
\end{split}
\end{equation*}

\subsection{\textit{s}-channel (Y/L)}
\begin{figure}[H]
\centering
    \includegraphics[width=12cm]{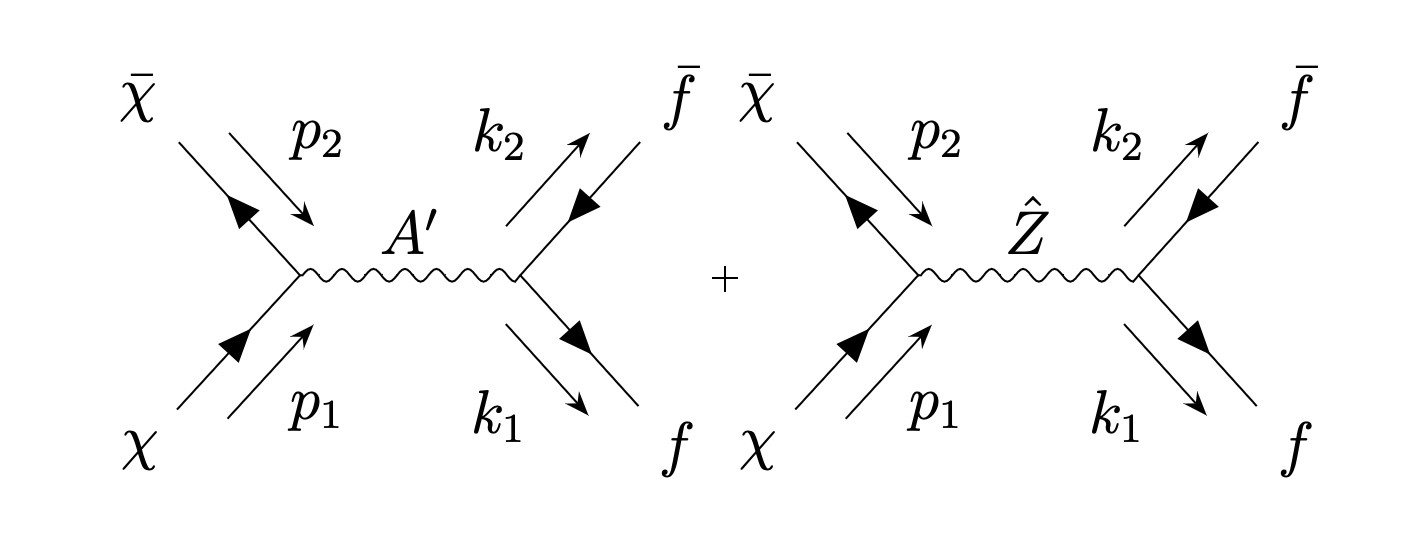}
\end{figure}
Now we got to consider the interference between the two process:
\begin{equation*}
    \mathcal{M} = \mathcal{M}_{\chi\chi\to A'\to ff} + \mathcal{M}_{\chi\chi\to \hat{Z} \to ff}
\end{equation*}
hence
\begin{equation*}
\begin{split}
\lvert\overline{\mathcal{M}}\rvert^2 = & \; \dfrac{\left( s^2+4s(m_f^2+m_\chi^2)\right)+\text{cos}^2\vartheta(s-4m_f^2)(s-4m_\chi^2)}{[(s-m_{A'}^2)^2+m_{A'}^2\Gamma_{A'}^2][(s-m_{\hat{Z}}^2)^2+m_{\hat{Z}}^2\Gamma_{\hat{Z}}^2]} \times\\
& \times [(g_f^{A'})^2(g_\chi^{A'})^2((s-m_{\hat{Z}}^2)^2+m_{\hat{Z}}^2\Gamma_{\hat{Z}}^2)-2g_f^{A'}g_\chi^{A'}g_f^{\hat{Z}}g_\chi^{\hat{Z}} \times \\
& \times ((s-m_{A'}^2)(s-m_{\hat{Z}}^2)-m_{A'}m_{\hat{Z}}\Gamma_{A'}\Gamma_{\hat{Z}}) + (g_f^{\hat{Z}})^2(g_\chi^{\hat{Z}})^2((s-m_{A'}^2)^2+m_{A'}^2\Gamma_{A'}^2)] \\
\sigma_{\chi\chi\to ff} = & \dfrac{s}{6 [(s-m_{A'}^2)^2+m_{A'}^2\Gamma_{A'}^2][(s-m_{\hat{Z}}^2)^2+m_{\hat{Z}}^2\Gamma_{\hat{Z}}^2]}\dfrac{\sqrt{1-\dfrac{4m_f^2}{s}}}{\sqrt{1-\dfrac{4m_\chi^2}{s}}} \times \\
& \times \left(1+\dfrac{2m_f^2}{s}\right)\left(1+\dfrac{2m_\chi^2}{s}\right)\{(g_f^{A'})^2(g_\chi^{A'})^2[(s-m_{\hat{Z}}^2)^2+m_{\hat{Z}}^2\Gamma_{\hat{Z}}^2] + 2g_f^{A'}g_\chi^{A'}g_f^{\hat{Z}}g_\chi^{\hat{Z}} \times \\
&\times [(s-m_{A'}^2)(s-m_{\hat{Z}}^2)-m_{A'}m_{\hat{Z}}\Gamma_{A'}\Gamma_{\hat{Z}}] + (g_f^{\hat{Z}})^2(g_\chi^{\hat{Z}})^2[(s-m_{A'}^2)^2+m_{A'}^2\Gamma_{A'}^2] \}
\end{split}
\end{equation*}

\subsection{DM-DP scattering}\label{compton}
\begin{figure}[H]
\centering
    \includegraphics[width=12cm]{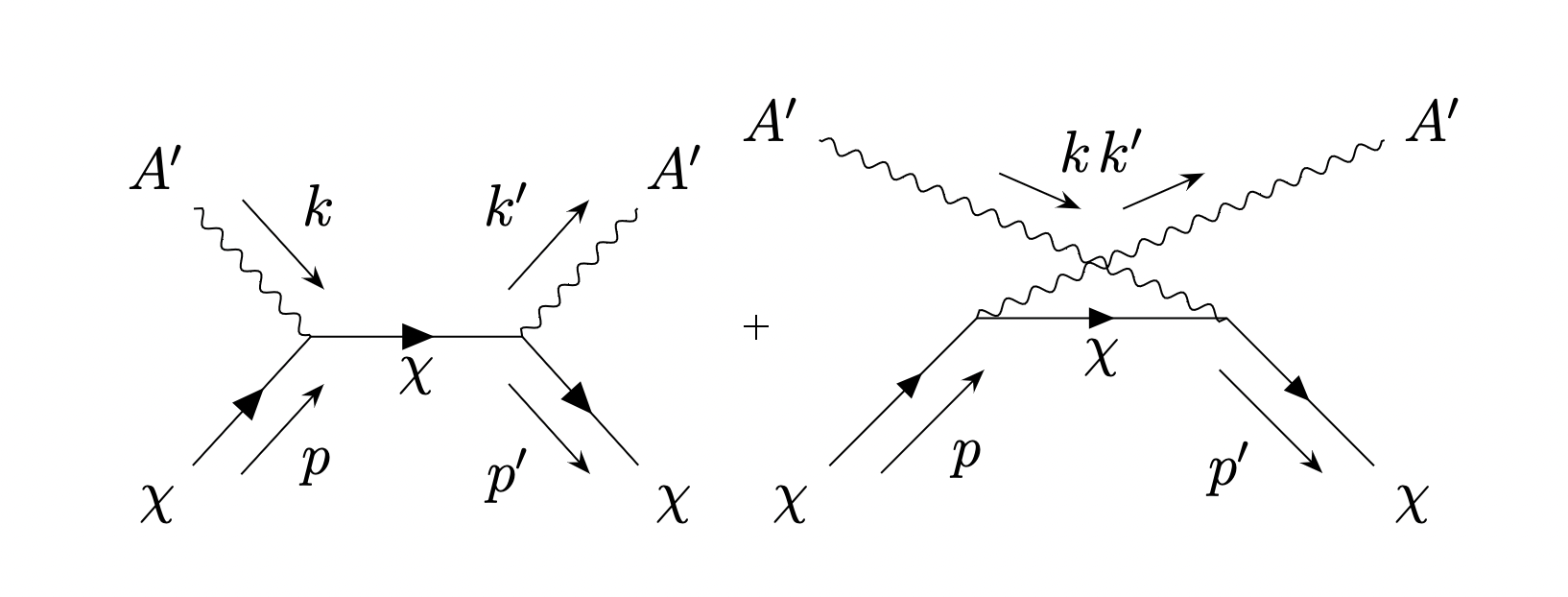}
\end{figure}
\begin{equation*}\label{squaredamplitudeXAXA}
\begin{split}
    \lvert \overline{\mathcal{M}} \rvert^2 = \dfrac{g_D^2}{6}\left(-\eta_{\nu\beta}+\dfrac{k_\nu k_\beta}{m_{A'}^2}\right)\left(-\eta_{\mu\alpha}+\dfrac{k'_\mu k'_\alpha}{m_{A'}^2}\right)\Bigg[ \dfrac{\textbf{I}}{\left(m_{A'}^2+2pk\right)^2} + \\ \dfrac{\textbf{II}}{\left(m_{A'}^2+2pk\right)\left(m_{A'}^2-2pk'\right)} +  \dfrac{\textbf{III}}{\left(m_{A'}^2-2pk'\right)\left(m_{A'}^2+2pk\right)} +  \dfrac{\textbf{IV}}{\left(m_{A'}^2-2pk'\right)^2}\Bigg]
\end{split}
\end{equation*}
where
\begin{fleqn}[\parindent]
\begin{equation*}
    \begin{split}
    \textbf{I} \equiv \; & \text{tr}\big[(\slashed{p'}+m_\chi)(\gamma^\alpha\slashed{k}\gamma^\beta+2\gamma^\alpha p^\beta)(\slashed{p}+m_\chi)(\gamma^\nu \slashed{k}\gamma^\mu + 2\gamma^\mu p^\nu)\big] \\
    \textbf{II} \equiv \; & \text{tr}\big[(\slashed{p'}+m_\chi)(\gamma^\alpha\slashed{k}\gamma^\beta+2\gamma^\alpha p^\beta)(\slashed{p}+m_\chi)(-\gamma^\mu \slashed{k'}\gamma^\nu+2 \gamma^\nu p^\mu)\big] \\
    \textbf{III} \equiv \; & \text{tr}\big[(\slashed{p'}+m_\chi)(-\gamma^\beta\slashed{k'}\gamma^\alpha+2\gamma^\beta p^\alpha)(\slashed{p}+m_\chi)(\gamma^\nu \slashed{k}\gamma^\mu + 2\gamma^\mu p^\nu)\big] \\
    \textbf{IV} \equiv \; & \text{tr}\big[(\slashed{p'}+m_\chi)(-\gamma^\beta\slashed{k'}\gamma^\alpha+2\gamma^\beta p^\alpha)(\slashed{p}+m_\chi)(-\gamma^\mu \slashed{k'}\gamma^\nu+2 \gamma^\nu p^\mu)\big]
    \end{split}
\end{equation*}
\end{fleqn}
Using \textit{FeynCalc} \cite{feyncalc} for contractions we get
\begin{equation*}
\begin{split}
\begin{aligned}
    \sigma_{\chi A'\to\chi A'}  = &  \dfrac{g_D^4}{24\pi s(s-m_\chi^2)^2}\Biggl\{ \dfrac{2s(s-m_\chi^2)(s^2-3m_\chi^4-m_\chi^2(6s-4m_{A'}^2)+8m_{A'}^4-4m_{A'}^2s)}{m_\chi^4-2m_\chi^2(m_{A'}^2+s)+(s-m_{A'}^2)^2} \times \\[10pt]
    & \times \text{log}\bigg[\dfrac{s(s-m_\chi^2-2m_{A'}^2)}{sm_\chi^2-(m_\chi-m_{A'})^2}\bigg]  + \\[10pt]
    & \dfrac{ s^4(m_\chi^2+m_{A'}^2)(13m_\chi^2+m_{A'}^2)+m_\chi^2(m_\chi^2-m_{A'}^2)^4(m_\chi^2+2m_{A'}^2)+m_\chi^2 s^5 - }{ s(m_\chi^2+2m_{A'}^2-s)(m_\chi^4-m_\chi^2(s+2m_{A'}^2)+m_{A'}^4)} \\[10pt]
    & \underline{- s^3(30m_\chi^6 + 24m_\chi^4m_{A'}^2 + 13 m_\chi^2m_{A'}^4+4m_{A'}^6) - s(m_\chi^2-m_{A'}^2)^2(3m_\chi^6-2m_\chi^4m_{A'}^2 + }  \\[10pt] & \underline{- 8m_\chi^2m_{A'}^4-2m_{A'}^6) + s^2(18m_\chi^8+4m_\chi^6m_{A'}^2-19m_\chi^4m_{A'}^4-8m_\chi^2m_{A'}^6+11m_{A'}^8)} \Biggr\}
    \end{aligned}
    \end{split}
\end{equation*}

\subsection{\textit{t} and \textit{u} channels}
\begin{figure}[H]
\centering
    \includegraphics[width=9cm]{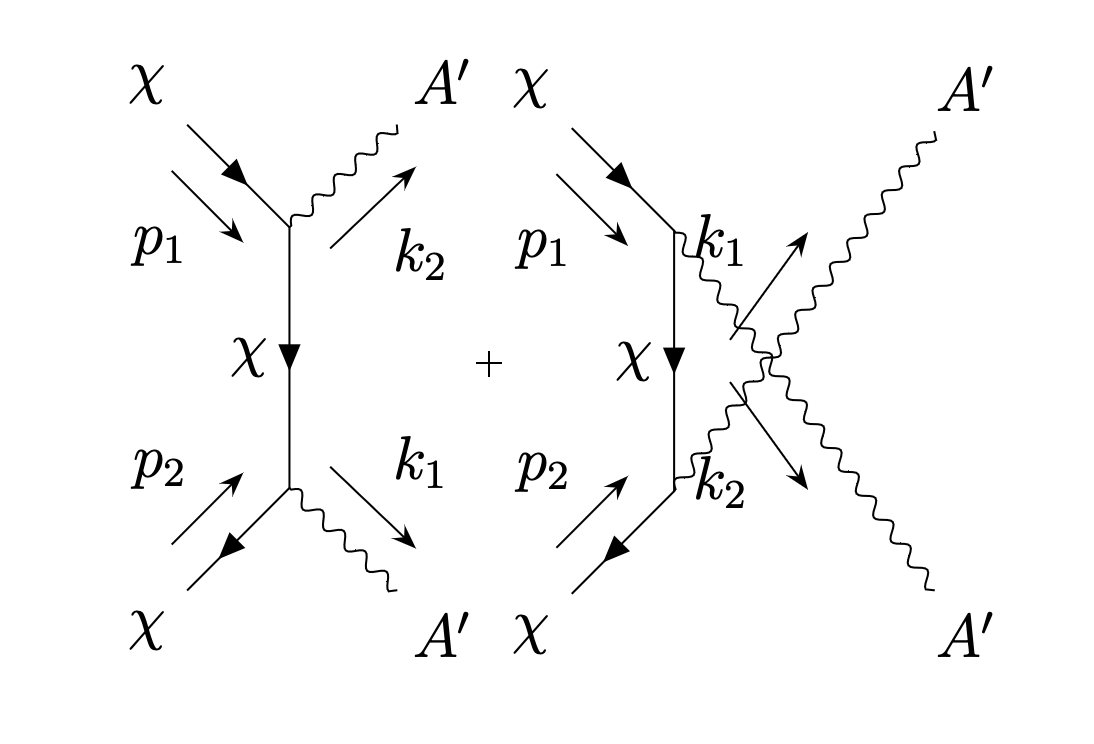}
\end{figure}
By means of crossing symmetry on \eqref{squaredamplitudeXAXA}:
\begin{equation}\label{crossingsymmetry}
p \to p_1, \;\;\;\; p' \to -p_2 \;\;\;\; k \to -k_1 \;\;\;\; k' \to k_2
\end{equation}
being careful of substituting
\begin{itemize}
    \item $ t = m_\chi^2+m_{A'}^2-\dfrac{s}{2}\left(1-\sqrt{1-\dfrac{4m_\chi^2}{s}}\sqrt{1-\dfrac{4m_{A'}^2}{s}}\text{cos}\vartheta \right)$ \\[2mm] in place of $t = (k-k')^2 = 2(\text{cos}\vartheta-1)(\omega^2-m_{A'}^2)$, 
    \item adjusting for the degrees of freedom we're averaging with,
    \item multiplying for $(-1)$ to fix the sign altered by the transformation (\ref{crossingsymmetry}).
\end{itemize}
we get:
\begin{equation*}\label{crosssectionXXAA}
    \begin{split}
        \sigma_{\chi\chi\to A'A'} = & \dfrac{g_D^4}{32\pi s(s-4m_\chi^2)^2}\dfrac{\sqrt{1-\dfrac{4m_{A'}^2}{s}}}{\sqrt{1-\dfrac{4m_\chi^2}{s}}}\Biggl\{\dfrac{2(s-m_\chi^2)}{\sqrt{s-4m_\chi^2}\sqrt{s-4m_{A'}^2}} \times \\
       & \times \big[-3m_\chi^4+m_\chi^2(4m_{A'}^2-6s)+8m_{A'}^4-4m_{A'}^2s+s^2\big] \times \\[2mm]
       & \times \text{log}\Bigg[\dfrac{(s-2m_{A'}^2)^2(s-2m_{A'}^2-\sqrt{s-4m_\chi^2}\sqrt{s-4m_{A'}^2})}{4\big[m_{A'}^4+m_\chi^2(s-4m_{A'}^2)\big](s-2m_{A'}^2+\sqrt{s-4m_\chi^2}\sqrt{s-4m_{A'}^2})}\Bigg]  + \\[10pt]
       & - 2(s-m_\chi^2)^2(2m_\chi^2+m_{A'}^2)^2 \Bigg[ \dfrac{4}{\sqrt{s-4m_\chi^2}\sqrt{s-4m_{A'}^2}(s-2m_{A'}^2)} + \\[10pt]
       & - \dfrac{(s-2m_{A'}^2)}{\sqrt{s-4m_\chi^2}\sqrt{s-4m_{A'}^2}\big[m_{A'}^4+m_\chi^2(s-4m_{A'}^2)\big]} + \dfrac{1}{\big[m_{A'}^4+m_\chi^2(s-4m_{A'}^2)\big]}\Bigg] - \\
       & - \dfrac{1}{2}\sqrt{s-4m_\chi^2}\sqrt{s-4m_{A'}^2}(s-m_\chi^2)-m_\chi^2(2m_{A'}^2+7s)-2m_{A'}^4-6m_{A'}^2s-s^2\Biggr\}
    \end{split}
\end{equation*}

\section{Electromagnetic mixing}\label{em_calculations}
Starting from lagrangian density given in \eqref{darkcrossemlagrangian}, rotate into the mass eigenstates:
\begin{equation}
    \begin{pmatrix}
    \tilde{A}'_\mu \\
    \tilde{A}_\mu
    \end{pmatrix}
    = \begin{pmatrix}
    \dfrac{1}{\sqrt{1-\varepsilon^2}} & 0 \\
    \dfrac{\varepsilon}{\sqrt{1-\varepsilon^2}} & 1
    \end{pmatrix}
    \begin{pmatrix}
    \text{c} & -\text{s} \\
    \text{s} & \text{c}
    \end{pmatrix}
    \begin{pmatrix}
    A'_\mu \\
    A_\mu
    \end{pmatrix}
\end{equation}
where c, s $= \text{cos}\vartheta$, $\text{sin}\vartheta$, we diagonalize the kinetic terms, while by setting $c\to1$, $s\to 0$ we obtain the following currents:
\begin{equation}
    \mathcal{L}_{\text{int}} \stackrel{ \varepsilon ^2 \ll 1}{\sim} g_D J'_\mu A'^\mu + e J_\mu \left( \varepsilon A'^\mu + A^\mu\right)
\end{equation}
Here $A_\mu$ is the classical SM photon, while $A'_\mu$ is the brand new DP, whose mass acquiring mechanism won't be addressed here, but may be obtained gauge-invariantly through a new Higgs sector or a Stueckelberg lagrangian \cite{ruegg04}.\\
We may notice that this model comprises of a SM photon and a DP coupling directly to DM through $g_D$ and to SM charged fermions through $\varepsilon e$: in other words, DM is totally blind to known interactions, except for gravity, and it can only be seen by means of the new force carrier DP. Notice we used the fact that $\lvert \varepsilon \rvert^2 \ll 1$ which is known to be true for instance from the "milli-charged" DM phenomenology \cite{Fabbrichesi_2021}, which is by the way obtainable within the same construction depicted above setting s$\to-\varepsilon$, c$\to\sqrt{1-\varepsilon^2}$:
\begin{equation}
\begin{split}
    \mathcal{L}_{int} & = \left(\dfrac{g_D}{\sqrt{1-\varepsilon^2}}\varepsilon J'_\mu+\dfrac{e}{\sqrt{1-\varepsilon^2}}\varepsilon J_\mu\right)A^\mu + g_D J'_\mu A'^\mu \\[2mm]
    & \stackrel{\lvert \varepsilon \rvert^2 \ll 1}{\sim} (g_D\varepsilon J'_\mu + e \varepsilon J_\mu)A^\mu + g_D J'_\mu A'^\mu
    \end{split}
\end{equation}
Notice how in this case, we're dealing with a electrically charged DM which directly interacts with the SM whilst the DP plays no role (at tree level, at least) and it gets secluded to the sole DS. 

\section{Hypercharge mixing}\label{Y/L_calculations}
Here we start from lagrangian \eqref{lagrangian_mixing_Yxdark}, first by recalling all the present covariant derivatives 
\begin{equation*}
    \begin{split}
        \mathcal{D}_\mu \Phi & = \left( \partial_\mu - i \dfrac{1}{2}g_2W^3_\mu \sigma^3 - i \dfrac{1}{2}g_1 B_\mu \right)\Phi \\
        \mathcal{D}_\mu L & = \left( \partial_\mu - i \dfrac{1}{2}g_2W^3_\mu \sigma^3 - i \dfrac{1}{2}g_1 B_\mu \right)L \\
        \mathcal{D}_\mu \ell_\text{R} & = \left( \partial_\mu +ig_1 B_\mu \right)\ell_\text{R} \\
        \mathcal{D}_\mu Q & = \left( \partial_\mu - i \dfrac{1}{2}g_2 \sigma^3 W^3_\mu - i \dfrac{1}{6}g_1B_\mu \right)Q \\
        \mathcal{D}_\mu u_\text{R} & = \left( \partial_\mu - i \dfrac{2}{3}g_1 B_\mu \right)u_\text{R} \\
        \mathcal{D}_\mu d_\text{R} & = \left( \partial_\mu + i \dfrac{1}{3}g_1 B_\mu \right)d_\text{R} \\
        \mathcal{D}_\mu \chi & = \left( \partial_\mu + i g_D a'_\mu \right)\chi
    \end{split}
\end{equation*}
excluding charged boson within SU(2)$_\text{L}$ for simplicity.\\
We perform the first transformation to diagonalize the kinetic sector:
\begin{equation*}
    \begin{pmatrix}
    B_\mu\\
    W^3_\mu\\
    a'_\mu
    \end{pmatrix}
    = G_\text{Y}(\varepsilon)
    \begin{pmatrix}
    \Tilde{B}_\mu\\
    \Tilde{W^3}_\mu\\
    \Tilde{a'}_\mu
    \end{pmatrix}
\end{equation*}
with \begin{equation*}
G_\text{Y}(\varepsilon) = 
    \begin{pmatrix}
    1 & 0 & -\dfrac{\varepsilon}{\sqrt{1-\varepsilon^2}}\\
    0 & 1 & 0 \\
    0 & 0 & \dfrac{1}{\sqrt{1-\varepsilon^2}}
    \end{pmatrix}
\end{equation*}
Let now the Higgs acquire a VEV (ignoring $\phi^+$,$\phi^-$ and $\phi^0$):
\begin{equation*}
    \Phi \to \dfrac{1}{\sqrt{2}}\begin{pmatrix}
    v + H \\
    0
    \end{pmatrix}
\end{equation*}
and substitute the former transformation in the mass term of \eqref{lagrangian_mixing_Yxdark}:
\begin{equation*}
    \mathcal{L}_\text{mass} = \dfrac{1}{2}
    \begin{pmatrix}
    \Tilde{B}_\mu & \Tilde{W^3}_\mu & \Tilde{a}'_\mu
    \end{pmatrix}
    \dfrac{v^2}{4}
    \begin{pmatrix}
    g_1^2 & - g_1g_2 & -g_1^2\varepsilon \\
    - g_1g_2 & g_2^2 & g_1g_2\varepsilon \\
    -g_1^2\varepsilon & g_1g_2\varepsilon & g_1^2\varepsilon^2+\dfrac{4m_{a'}^2(1+\varepsilon^2)}{v^2}
    \end{pmatrix}
    \begin{pmatrix}
    \Tilde{B}_\mu\\
    \Tilde{W^3}_\mu\\
    \Tilde{a}'_\mu
    \end{pmatrix}
\end{equation*}
Finally define the mass eigenstates through orthogonal transformation P such that:
\begin{equation*}
    \begin{pmatrix}
    \Tilde{B}_\mu\\
    \Tilde{W^3}_\mu\\
    \Tilde{a}'_\mu
    \end{pmatrix}
    = \text{P}^{-1}
    \begin{pmatrix}
    A_\mu\\
   \hat{Z}_\mu\\
    A'_\mu
    \end{pmatrix}
\end{equation*}
where we defined P to be:
\begin{equation*}
\text{P}(\xi,\vartheta_\text{W}) = R_1(\xi)R_2(\vartheta_\text{W}) =
    \begin{pmatrix}
    1 & 0 & 0 \\
    0 & \cos\xi & \sin\xi \\
    0 & -\sin\xi & \cos\xi
    \end{pmatrix}
    \begin{pmatrix}
    \cos\vartheta_{\text{W}} & \sin\vartheta_{\text{W}} & 0 \\
    -\sin\vartheta_{\text{W}} & \cos\vartheta_{\text{W}} & 0 \\
    0 & 0 & 1
    \end{pmatrix}
\end{equation*}
where $\xi = (1/2)\arctan\left(\dfrac{2\,\varepsilon \,\sin\vartheta_\text{W}}{1-\delta}\right)$, $\delta = m_{a'}^2/m_{Z}^2$, $m_Z \simeq$ 91.2 GeV and cos$\vartheta_\text{W} \simeq$ 0.88. Notice that we distinguished $\hat{Z}$ from $Z$ due to a $\mathcal{O}(\varepsilon^2)$ difference between their masses, discrepancy which is well known in the literature to be a signature of kinetic mixing effects.
From these transformations we can read out the eigenvalues from the diagonal mass matrix. Define $\Gamma \equiv 1/(1-\delta)$, $\gimel \equiv 1/2(\delta-1)^2$:
\begin{equation*}
    \begin{split}
        m_{A}^2 = \; & 0 \\
        m_{A'}^2 \sim \; & m_{a'}\{1+ \varepsilon^2[1-s_\text{W}^2(1+\delta + 3\delta^2)] \} \\
        m_{\hat{Z}}^2 \sim \; & m_{Z}^2(1+s_\text{W}^2\varepsilon^2\Gamma)\\
    \end{split}
\end{equation*}
and applying $G_\text{Y}(\varepsilon) \text{P}(\xi,\vartheta_\text{W})^{-1}$ to the currents we find the interacting part of our lagrangian
\begin{equation}
\begin{split}
    \mathcal{L} = & -\dfrac{1}{4}F_{\mu\nu}F^{\mu\nu} -\dfrac{1}{4}\hat{Z}_{\mu\nu}\hat{Z}^{\mu\nu} - \dfrac{1}{4}F'_{\mu\nu}F'^{\mu\nu} + \dfrac{1}{2}m_{\hat{Z}}^2\hat{Z}_\mu \hat{Z}^\mu + \dfrac{1}{2}m_{A'}^2 A'_\mu A'^\mu + \\
    & + i \Bar{L}(g_L^{A}\slashed{A} + g_L^{A'}\slashed{A'} + g_L^{\hat{Z}}\slashed{\hat{Z}})L + i \Bar{\ell}_\text{R}(g_{\ell_\text{R}}^{A}\slashed{A} + g_{\ell_\text{R}}^{A'}\slashed{A'} + g_{\ell_\text{R}}^{\hat{Z}}\slashed{\hat{Z}}){\ell_\text{R}} + \\
    & + \Bar{Q}(g_L^{A}\slashed{A} + g_Q^{A'}\slashed{A'} + g_Q^{\hat{Z}}\slashed{\hat{Z}})Q + \Bar{u}_\text{R}(g_{\ell_\text{R}}^{A}\slashed{A} + g_{u_\text{R}}^{A'}\slashed{A'} + g_{u_\text{R}}^{\hat{Z}}\slashed{\hat{Z}}){u_\text{R}} + \\
    & + \Bar{d}_\text{R}(g_{\ell_\text{R}}^{A}\slashed{A} + g_{d_\text{R}}^{A'}\slashed{A'} + g_{d_\text{R}}^{\hat{Z}}\slashed{\hat{Z}}){d_\text{R}} + g_{H\hat{Z}\hat{Z}}H\hat{Z}_\mu\hat{Z}^\mu + g_{HH\hat{Z}\hat{Z}}HH\hat{Z}_\mu\hat{Z}^\mu + \\
    & + g_{HA'A'}HA'_\mu A'^\mu + g_{HHA'A'}HHA'_\mu A'^\mu +
    g_{H\hat{Z}A'}H\hat{Z}_\mu A'^\mu + g_{HH\hat{Z}A'}HH\hat{Z}_\mu A'^\mu
\end{split}
\end{equation}
with couplings,
\begin{equation*}
    \begin{split}
        g_{\nu_L}^{A'} & = -\dfrac{e \varepsilon}{2 c_\text{W}}(1-\Gamma) \\
        g_{\ell_L}^{A'} & = -\dfrac{e\varepsilon}{2c_\text{W}}(1-\Gamma + 2\Gamma c_\text{W}^2) \\
        g_{\ell_R}^{A'} & = - \dfrac{e\varepsilon}{c_\text{W}}(1-s_\text{W}^2\Gamma) \\
        g_{u_L}^{A'} & =  \dfrac{e\varepsilon}{2c_\text{W}}\bigg[ \dfrac{1}{3}+\Gamma\left(c_\text{W}^2-\dfrac{s_\text{W}^2}{3}\right)\bigg]\\
        g_{d_L}^{A'} & = \dfrac{e\varepsilon}{2c_\text{W}}(s_\text{W}^2\Gamma-1)\\
        g_{u_R}^{A'} & = -\dfrac{2}{3}\dfrac{e\varepsilon}{c_\text{W}}(s_\text{W}^2\Gamma-1)\\
        g_{d_R}^{A'} & = \dfrac{1}{3}\dfrac{e\varepsilon}{c_\text{W}}(s_\text{W}^2\Gamma-1) \\
        g_\chi^{A'} & = g_D\bigg[ 1+\varepsilon^2 \left( \dfrac{1}{2}-s_{\text{W}}^2\gimel\right)\bigg] \\
        \\
        g_{\nu_L}^{\hat{Z}} & = -\dfrac{e}{2c_\text{W}s_\text{W}}+\dfrac{e\varepsilon^2\tan\vartheta_\text{W}}{2}(\gimel - \Gamma)\\
        g_{\ell_L}^{\hat{Z}} & = -\dfrac{e}{2c_\text{W}s_\text{W}}(s_\text{W}^2-c_\text{W}^2)+\dfrac{e\varepsilon^2}{2}[\tan\vartheta_\text{W}(\gimel -\Gamma) - 2s_\text{W}c_\text{W}]\\
        g_{\ell_R}^{\hat{Z}} & = -e\tan\vartheta_\text{W}+ \dfrac{e\varepsilon^2}{c_\text{W}}[s_\text{W}(\gimel - \Gamma) - c_\text{W}^2\gimel]\\
        g_{u_L}^{\hat{Z}} & = \dfrac{e}{2}\left(\dfrac{1}{3}\tan\vartheta_\text{W}-\cot\vartheta_\text{W}\right)-\dfrac{e\varepsilon^2}{2}\bigg[ \dfrac{1}{3}\tan\vartheta_\text{W}(\gimel - \Gamma) - \dfrac{4}{3}s_\text{W}c_\text{W}\gimel\bigg]\\
        g_{d_L}^{\hat{Z}} & = \dfrac{e}{2}\left(\dfrac{1}{3}\tan\vartheta_\text{W}+\cot\vartheta_\text{W}\right)-\dfrac{e\varepsilon^2}{2}\bigg[\dfrac{2}{3}c_\text{W}s_\text{W}\gimel + \dfrac{1}{3}\tan\vartheta_\text{W}(\gimel-\Gamma)\bigg] \\
        g_{u_R}^{\hat{Z}} & = \dfrac{2}{3}e\tan\vartheta_\text{W} - \dfrac{2}{3}e\varepsilon^2[\tan\vartheta_\text{W}(\gimel -\Gamma)-s_\text{W}c_\text{W}\gimel]\\
        g_{d_R}^{\hat{Z}} & = -\dfrac{1}{3}e\tan\vartheta_\text{W} + \dfrac{1}{3}e\varepsilon^2[\tan\vartheta_\text{W}(\gimel-\Gamma)-s_\text{W}c_\text{W}\gimel] \\
        g_{\chi}^{\hat{Z}} & = g_D\varepsilon s_\text{W}\Gamma \\
        \\
        g_{HA'A'} & = \dfrac{e^2\varepsilon^2}{4c_\text{W}^2}(\Gamma-1)^2v \\
        g_{HHA'A'} & = \dfrac{e^2\varepsilon^2}{8c_\text{W}^2}(\Gamma-1)^2\\
        g_{HA'\hat{Z}} & = -\dfrac{e^2\varepsilon}{2s_\text{W}c_\text{W}}(\Gamma-1)v \\
        g_{HHA'\hat{Z}} & = -\dfrac{e^2\varepsilon}{4s_\text{W}c_\text{W}}(\Gamma-1) \\
        g_{H\hat{Z}\hat{Z}} & = \dfrac{e^2}{4s_\text{W}^2c_\text{W}^2}v + \dfrac{e^2\varepsilon^2}{2c_\text{W}^2}(\Gamma-1)v \\
        g_{HH\hat{Z}\hat{Z}} & = \dfrac{e^2}{8c_\text{W}^2s_\text{W}^2} + \dfrac{e^2\varepsilon^2}{4c_\text{W}^2}(\Gamma-1)
    \end{split}
\end{equation*}

\section{Isospin mixing}
The mathematics for this model is almost the same one used in [\ref{Y/L_calculations}]. We just need to substitute $G_\text{Y}$ with
\begin{equation}
G_\text{L}(\varepsilon) = 
    \begin{pmatrix}
    1 & 0 & 0\\
    0 & 1 & -\dfrac{\varepsilon}{\sqrt{1-\varepsilon^2}} \\
    0 & 0 & \dfrac{1}{\sqrt{1-\varepsilon^2}}
    \end{pmatrix}
\end{equation}
After diagonalizing the mass matrix and applying $G_\text{L}(\varepsilon) \text{P}(\xi,\vartheta_\text{W})^{-1}$ to the currents terms we get the followings eigenvalues and couplings:
\begin{equation*}
    \begin{split}
        m_A^2 & = 0 \\
        m_{A'}^2 & \sim m_{a'}^2\biggl\{1+\varepsilon^2\bigg[\dfrac{1}{2\delta}(3+4c_\text{W}s_\text{W}-2s_\text{W}^2)+(1+\delta)(2+2c_\text{W}s_\text{W}-s_\text{W}^2) \bigg] \biggl\} \\
        m_{\hat{Z}}^2 & \sim m_Z^2[1-\varepsilon^2(2c_\text{W}s_\text{W}+s_\text{W}^2)\Gamma] 
    \end{split}
\end{equation*}
\begin{equation*}
    \begin{split}
        g_{\nu_L}^{A'} & = \dfrac{e \varepsilon}{2}\left(\dfrac{1}{s_\text{W}}+\dfrac{\Gamma}{c_\text{W}}\right) \\
        g_{\ell_L}^{A'} & = -\dfrac{e\varepsilon}{2s_\text{W}}\left(1-\dfrac{s_\text{W}}{c_\text{W}}(2s_\text{W}^2-1)\Gamma\right) \\
        g_{\ell_R}^{A'} & =e\varepsilon  \dfrac{s_\text{W}^2}{c_\text{W}}\Gamma \\
        g_{u_L}^{A'} & =  \dfrac{e\varepsilon}{2}\bigg[ \dfrac{1}{s_\text{W}}+s_\text{W}\left(\cot{\vartheta_\text{W}}-\dfrac{1}{3}\tan{\vartheta_\text{W}}\right)\Gamma\bigg]\\
        g_{d_L}^{A'} & = -\dfrac{e\varepsilon}{2}\bigg[\dfrac{1}{s_\text{W}}+\dfrac{1}{c_\text{W}}\left(c_\text{W}^2+\dfrac{1}{3}s_\text{W}^2\right)\Gamma \bigg]\\
        g_{u_R}^{A'} & = -\dfrac{2}{3}e\varepsilon\dfrac{s_\text{W}^2}{c_\text{W}}\Gamma\\
        g_{d_R}^{A'} & = \dfrac{1}{3}e\varepsilon\dfrac{s_\text{W}^2}{c_\text{W}}\Gamma \\
        g_\chi^{A'} & = g_D\bigg[ 1+\varepsilon^2 \left( \dfrac{1}{2}-s_{\text{W}}^2\gimel\right)\bigg] \\
        \\
        g_{\nu_L}^{\hat{Z}} & = -\dfrac{e}{2c_\text{W}s_\text{W}}+\dfrac{e\varepsilon^2\tan\vartheta_\text{W}}{2}(\tan{\vartheta_\text{W}}\gimel - \Gamma)\\
        g_{\ell_L}^{\hat{Z}} & = -\dfrac{e}{2c_\text{W}s_\text{W}}(s_\text{W}^2-c_\text{W}^2) + \dfrac{e\varepsilon^2}{2}[\tan{\vartheta_\text{W}}(s_\text{W}^2-c_\text{W}^2)\gimel-\Gamma]\\
        g_{\ell_R}^{\hat{Z}} & = -e\tan\vartheta_\text{W}+e\varepsilon^2 \dfrac{s_\text{W}^3}{c_\text{W}}\gimel\\
        g_{u_L}^{\hat{Z}} & = \dfrac{e}{2}\left(\dfrac{1}{3}\tan\vartheta_\text{W}-\cot\vartheta_\text{W}\right)+\dfrac{e\varepsilon^2}{2}\bigg[\tan{\vartheta_\text{W}}\left(c_\text{W}^2-\dfrac{1}{3}s_\text{W}^2\right)\gimel + \Gamma \bigg]\\
        g_{d_L}^{\hat{Z}} & = \dfrac{e}{2}\left(\dfrac{1}{3}\tan\vartheta_\text{W}+\cot\vartheta_\text{W}\right)-\dfrac{e\varepsilon^2}{2}\bigg[\tan{\vartheta_\text{W}}\left(c_\text{W}^2+\dfrac{1}{3}s_\text{W}^2\right)\gimel + \Gamma\bigg] \\
        g_{u_R}^{\hat{Z}} & = \dfrac{2}{3}e\tan\vartheta_\text{W} - \dfrac{2}{3}e\varepsilon^2\dfrac{s_\text{W}^3}{c_\text{W}}\gimel\\
        g_{d_R}^{\hat{Z}} & = -\dfrac{1}{3}e\tan\vartheta_\text{W} + \dfrac{1}{3}e\varepsilon^2\dfrac{s_\text{W}^3}{c_\text{W}}\gimel \\
        g_{\chi}^{\hat{Z}} & = g_D\varepsilon s_\text{W}\Gamma \\
        \\
        g_{HA'A'} & = \dfrac{e^2\varepsilon^2}{4c_\text{W}^2s_\text{W}^2}(c_\text{W}+s_\text{W}\Gamma)^2v \\
        g_{HHA'A'} & = \dfrac{e^2\varepsilon^2}{8c_\text{W}^2s_\text{W}^2}(c_\text{W}+s_\text{W}\Gamma)^2\\
        g_{HA'\hat{Z}} & = -\dfrac{e^2\varepsilon}{2s_\text{W}^2c_\text{W}^2}(c_\text{W}+s_\text{W}\Gamma)v \\
        g_{HHA'\hat{Z}} & = -\dfrac{e^2\varepsilon}{4s_\text{W}^2c_\text{W}^2}(c_\text{W}+s_\text{W}\Gamma) \\
        g_{H\hat{Z}\hat{Z}} & = \dfrac{e^2}{4s_\text{W}^2c_\text{W}^2}v - \dfrac{e^2\varepsilon^2}{2c_\text{W}^2s_\text{W}}(c_\text{W}\Gamma-s_\text{W}\gimel)v 
        \end{split}
        \end{equation*}
        \newpage
        \begin{equation*}
            \begin{split}
        g_{HH\hat{Z}\hat{Z}} = \dfrac{e^2}{8c_\text{W}^2s_\text{W}^2} - \dfrac{e^2\varepsilon^2}{4c_\text{W}^2s_\text{W}}(c_\text{W}\Gamma-s_\text{W}\gimel)
    \end{split}
\end{equation*}
\end{appendices}
% End appedix ----------------------

\bibliography{references.bib}

\providecommand{\latin}[1]{#1}
\makeatletter
\providecommand{\doi}
  {\begingroup\let\do\@makeother\dospecials
  \catcode`\{=1 \catcode`\}=2 \doi@aux}
\providecommand{\doi@aux}[1]{\endgroup\texttt{#1}}
\makeatother
\providecommand*\mcitethebibliography{\thebibliography}
\csname @ifundefined\endcsname{endmcitethebibliography}
  {\let\endmcitethebibliography\endthebibliography}{}
\begin{mcitethebibliography}{62}
\providecommand*\natexlab[1]{#1}
\providecommand*\mciteSetBstSublistMode[1]{}
\providecommand*\mciteSetBstMaxWidthForm[2]{}
\providecommand*\mciteBstWouldAddEndPuncttrue
  {\def\EndOfBibitem{\unskip.}}
\providecommand*\mciteBstWouldAddEndPunctfalse
  {\let\EndOfBibitem\relax}
\providecommand*\mciteSetBstMidEndSepPunct[3]{}
\providecommand*\mciteSetBstSublistLabelBeginEnd[3]{}
\providecommand*\EndOfBibitem{}
\mciteSetBstSublistMode{f}
\mciteSetBstMaxWidthForm{subitem}{(\alph{mcitesubitemcount})}
\mciteSetBstSublistLabelBeginEnd
  {\mcitemaxwidthsubitemform\space}
  {\relax}
  {\relax}

\bibitem[Aghanim and et~al.(2020)Aghanim, and et~al.]{2020planck}
Aghanim,~N.; et~al., \emph{Astronomy \& Astrophysics} \textbf{2020},
  \emph{641}, A6\relax
\mciteBstWouldAddEndPuncttrue
\mciteSetBstMidEndSepPunct{\mcitedefaultmidpunct}
{\mcitedefaultendpunct}{\mcitedefaultseppunct}\relax
\EndOfBibitem
\bibitem[’t Hooft()]{thooft71}
’t Hooft,~G. Renormalizable Lagrangians for Massive Yang-mills Fields\relax
\mciteBstWouldAddEndPuncttrue
\mciteSetBstMidEndSepPunct{\mcitedefaultmidpunct}
{\mcitedefaultendpunct}{\mcitedefaultseppunct}\relax
\EndOfBibitem
\bibitem[Workman \latin{et~al.}(2022)Workman, \latin{et~al.} others]{PDGreview}
Workman,~R.~L., \latin{et~al.}  \emph{PTEP} \textbf{2022}, \emph{2022},
  083C01\relax
\mciteBstWouldAddEndPuncttrue
\mciteSetBstMidEndSepPunct{\mcitedefaultmidpunct}
{\mcitedefaultendpunct}{\mcitedefaultseppunct}\relax
\EndOfBibitem
\bibitem[Springel \latin{et~al.}(2008)Springel, White, Frenk, Navarro, Jenkins,
  Vogelsberger, Wang, Ludlow, and Helmi]{Springel:2008zz}
Springel,~V.; White,~S. D.~M.; Frenk,~C.~S.; Navarro,~J.~F.; Jenkins,~A.;
  Vogelsberger,~M.; Wang,~J.; Ludlow,~A.; Helmi,~A. \emph{Nature}
  \textbf{2008}, \emph{456N7218}, 73--80\relax
\mciteBstWouldAddEndPuncttrue
\mciteSetBstMidEndSepPunct{\mcitedefaultmidpunct}
{\mcitedefaultendpunct}{\mcitedefaultseppunct}\relax
\EndOfBibitem
\bibitem[Arcadi \latin{et~al.}(2018)Arcadi, Dutra, Ghosh, Lindner, Mambrini,
  Pierre, Profumo, and Queiroz]{arcadi}
Arcadi,~G.; Dutra,~M.; Ghosh,~P.; Lindner,~M.; Mambrini,~Y.; Pierre,~M.;
  Profumo,~S.; Queiroz,~F.~S. \emph{Eur. Phys. J. C} \textbf{2018}, \emph{78},
  203\relax
\mciteBstWouldAddEndPuncttrue
\mciteSetBstMidEndSepPunct{\mcitedefaultmidpunct}
{\mcitedefaultendpunct}{\mcitedefaultseppunct}\relax
\EndOfBibitem
\bibitem[Jaeckel and Ringwald(2010)Jaeckel, and Ringwald]{jaeckel10}
Jaeckel,~J.; Ringwald,~A. \emph{Annual Review of Nuclear and Particle Science}
  \textbf{2010}, \emph{60}, 405--437\relax
\mciteBstWouldAddEndPuncttrue
\mciteSetBstMidEndSepPunct{\mcitedefaultmidpunct}
{\mcitedefaultendpunct}{\mcitedefaultseppunct}\relax
\EndOfBibitem
\bibitem[Essig(2013)]{essig13}
Essig,~R. e.~a. Dark Sectors and New, Light, Weakly-Coupled Particles. 2013;
  \url{https://arxiv.org/abs/1311.0029}\relax
\mciteBstWouldAddEndPuncttrue
\mciteSetBstMidEndSepPunct{\mcitedefaultmidpunct}
{\mcitedefaultendpunct}{\mcitedefaultseppunct}\relax
\EndOfBibitem
\bibitem[Alexander(2016)]{alexander16}
Alexander,~J. e.~a. Dark Sectors 2016 Workshop: Community Report. 2016;
  \url{https://arxiv.org/abs/1608.08632}\relax
\mciteBstWouldAddEndPuncttrue
\mciteSetBstMidEndSepPunct{\mcitedefaultmidpunct}
{\mcitedefaultendpunct}{\mcitedefaultseppunct}\relax
\EndOfBibitem
\bibitem[Fabbrichesi \latin{et~al.}(2021)Fabbrichesi, Gabrielli, and
  Lanfranchi]{Fabbrichesi_2021}
Fabbrichesi,~M.; Gabrielli,~E.; Lanfranchi,~G. \emph{The Physics of the Dark
  Photon}; Springer International Publishing, 2021\relax
\mciteBstWouldAddEndPuncttrue
\mciteSetBstMidEndSepPunct{\mcitedefaultmidpunct}
{\mcitedefaultendpunct}{\mcitedefaultseppunct}\relax
\EndOfBibitem
\bibitem[Asaka \latin{et~al.}(2007)Asaka, Ishiwata, and Moroi]{Asaka_2007}
Asaka,~T.; Ishiwata,~K.; Moroi,~T. \emph{Physical Review D} \textbf{2007},
  \emph{75}\relax
\mciteBstWouldAddEndPuncttrue
\mciteSetBstMidEndSepPunct{\mcitedefaultmidpunct}
{\mcitedefaultendpunct}{\mcitedefaultseppunct}\relax
\EndOfBibitem
\bibitem[Gopalakrishna \latin{et~al.}(2006)Gopalakrishna, de~Gouv{\^{e}}a, and
  Porod]{Gopalakrishna_2006}
Gopalakrishna,~S.; de~Gouv{\^{e}}a,~A.; Porod,~W. \emph{Journal of Cosmology
  and Astroparticle Physics} \textbf{2006}, \emph{2006}, 005--005\relax
\mciteBstWouldAddEndPuncttrue
\mciteSetBstMidEndSepPunct{\mcitedefaultmidpunct}
{\mcitedefaultendpunct}{\mcitedefaultseppunct}\relax
\EndOfBibitem
\bibitem[Hall \latin{et~al.}(2010)Hall, Jedamzik, March-Russell, and
  West]{Hall_2010}
Hall,~L.~J.; Jedamzik,~K.; March-Russell,~J.; West,~S.~M. \emph{Journal of High
  Energy Physics} \textbf{2010}, \emph{2010}\relax
\mciteBstWouldAddEndPuncttrue
\mciteSetBstMidEndSepPunct{\mcitedefaultmidpunct}
{\mcitedefaultendpunct}{\mcitedefaultseppunct}\relax
\EndOfBibitem
\bibitem[Dvorkin \latin{et~al.}(2021)Dvorkin, Lin, and Schutz]{Dvorkin_2021}
Dvorkin,~C.; Lin,~T.; Schutz,~K. \emph{Physical Review Letters} \textbf{2021},
  \emph{127}\relax
\mciteBstWouldAddEndPuncttrue
\mciteSetBstMidEndSepPunct{\mcitedefaultmidpunct}
{\mcitedefaultendpunct}{\mcitedefaultseppunct}\relax
\EndOfBibitem
\bibitem[Gninenko(2018)]{Gninenko:2300189}
Gninenko,~S. \emph{{Addendum to the NA64 Proposal: Search for the $A'\to
  invisible$ and $X\to e^+e^-$ decays in 2021}}; 2018\relax
\mciteBstWouldAddEndPuncttrue
\mciteSetBstMidEndSepPunct{\mcitedefaultmidpunct}
{\mcitedefaultendpunct}{\mcitedefaultseppunct}\relax
\EndOfBibitem
\bibitem[Coogan \latin{et~al.}(2021)Coogan, Moiseev, Morrison, and
  Profumo]{subGeV1}
Coogan,~A.; Moiseev,~A.; Morrison,~L.; Profumo,~S. Hunting for Dark Matter and
  New Physics with (a) GECCO. 2021;
  \url{https://arxiv.org/abs/2101.10370}\relax
\mciteBstWouldAddEndPuncttrue
\mciteSetBstMidEndSepPunct{\mcitedefaultmidpunct}
{\mcitedefaultendpunct}{\mcitedefaultseppunct}\relax
\EndOfBibitem
\bibitem[Essinger-Hileman \latin{et~al.}(2014)Essinger-Hileman, Ali, Amiri,
  Appel, Araujo, Bennett, Boone, Chan, Cho, Chuss, Colazo, Crowe, Denis,
  Dünner, Eimer, Gothe, Halpern, Harrington, Hilton, Hinshaw, Huang, Irwin,
  Jones, Karakla, Kogut, Larson, Limon, Lowry, Marriage, Mehrle, Miller,
  Miller, Moseley, Novak, Reintsema, Rostem, Stevenson, Towner, U-Yen, Wagner,
  Watts, Wollack, Xu, and Zeng]{subGeV2}
Essinger-Hileman,~T. \latin{et~al.}  {CLASS}: the cosmology large angular scale
  surveyor. {SPIE} Proceedings. 2014\relax
\mciteBstWouldAddEndPuncttrue
\mciteSetBstMidEndSepPunct{\mcitedefaultmidpunct}
{\mcitedefaultendpunct}{\mcitedefaultseppunct}\relax
\EndOfBibitem
\bibitem[Suzuki \latin{et~al.}(2016)Suzuki, Ade, Akiba, Aleman, Arnold,
  Baccigalupi, Barch, Barron, Bender, Boettger, Borrill, Chapman, Chinone,
  Cukierman, Dobbs, Ducout, Dunner, Elleflot, Errard, Fabbian, Feeney, Feng,
  Fujino, Fuller, Gilbert, Goeckner-Wald, Groh, Haan, Hall, Halverson, Hamada,
  Hasegawa, Hattori, Hazumi, Hill, Holzapfel, Hori, Howe, Inoue, Irie, Jaehnig,
  Jaffe, Jeong, Katayama, Kaufman, Kazemzadeh, Keating, Kermish, Keskitalo,
  Kisner, Kusaka, Jeune, Lee, Leon, Linder, Lowry, Matsuda, Matsumura, Miller,
  Mizukami, Montgomery, Navaroli, Nishino, Peloton, Poletti, Puglisi, Rebeiz,
  Raum, Reichardt, Richards, Ross, Rotermund, Segawa, Sherwin, Shirley,
  Siritanasak, Stebor, Stompor, Suzuki, Tajima, Takada, Takakura, Takatori,
  Tikhomirov, Tomaru, Westbrook, Whitehorn, Yamashita, Zahn, and
  Zahn]{Simons-Array}
Suzuki,~A. \latin{et~al.}  \emph{Journal of Low Temperature Physics}
  \textbf{2016}, \emph{184}, 805--810\relax
\mciteBstWouldAddEndPuncttrue
\mciteSetBstMidEndSepPunct{\mcitedefaultmidpunct}
{\mcitedefaultendpunct}{\mcitedefaultseppunct}\relax
\EndOfBibitem
\bibitem[Kolb and Turner(1990)Kolb, and Turner]{kolbturner}
Kolb,~E.~W.; Turner,~M.~S. \emph{{The early universe}}; Frontiers in physics;
  Westview Press: Boulder, CO, 1990\relax
\mciteBstWouldAddEndPuncttrue
\mciteSetBstMidEndSepPunct{\mcitedefaultmidpunct}
{\mcitedefaultendpunct}{\mcitedefaultseppunct}\relax
\EndOfBibitem
\bibitem[{Planck Collaboration} \latin{et~al.}(2016){Planck Collaboration},
  {Ade, P. A. R.}, {Aghanim, N.}, {Arnaud, M.}, {Ashdown, M.}, {Aumont, J.},
  {Baccigalupi, C.}, {Banday, A. J.}, {Barreiro, R. B.}, {Bartlett, J. G.},
  {Bartolo, N.}, {Battaner, E.}, {Battye, R.}, {Benabed, K.}, {Beno\^{\i}t,
  A.}, {Benoit-L\'evy, A.}, {Bernard, J.-P.}, {Bersanelli, M.}, {Bielewicz,
  P.}, {Bock, J. J.}, {Bonaldi, A.}, {Bonavera, L.}, {Bond, J. R.}, {Borrill,
  J.}, {Bouchet, F. R.}, {Boulanger, F.}, {Bucher, M.}, {Burigana, C.},
  {Butler, R. C.}, {Calabrese, E.}, {Cardoso, J.-F.}, {Catalano, A.},
  {Challinor, A.}, {Chamballu, A.}, {Chary, R.-R.}, {Chiang, H. C.}, {Chluba,
  J.}, {Christensen, P. R.}, {Church, S.}, {Clements, D. L.}, {Colombi, S.},
  {Colombo, L. P. L.}, {Combet, C.}, {Coulais, A.}, {Crill, B. P.}, {Curto,
  A.}, {Cuttaia, F.}, {Danese, L.}, {Davies, R. D.}, {Davis, R. J.}, {de
  Bernardis, P.}, {de Rosa, A.}, {de Zotti, G.}, {Delabrouille, J.}, {D\'esert,
  F.-X.}, {Di Valentino, E.}, {Dickinson, C.}, {Diego, J. M.}, {Dolag, K.},
  {Dole, H.}, {Donzelli, S.}, {Dor\'e, O.}, {Douspis, M.}, {Ducout, A.},
  {Dunkley, J.}, {Dupac, X.}, {Efstathiou, G.}, {Elsner, F.}, {En\ss{}lin, T.
  A.}, {Eriksen, H. K.}, {Farhang, M.}, {Fergusson, J.}, {Finelli, F.}, {Forni,
  O.}, {Frailis, M.}, {Fraisse, A. A.}, {Franceschi, E.}, {Frejsel, A.},
  {Galeotta, S.}, {Galli, S.}, {Ganga, K.}, {Gauthier, C.}, {Gerbino, M.},
  {Ghosh, T.}, {Giard, M.}, {Giraud-H\'eraud, Y.}, {Giusarma, E.}, {Gjerl\o{}w,
  E.}, {Gonz\'alez-Nuevo, J.}, {G\'orski, K. M.}, {Gratton, S.}, {Gregorio,
  A.}, {Gruppuso, A.}, {Gudmundsson, J. E.}, {Hamann, J.}, {Hansen, F. K.},
  {Hanson, D.}, {Harrison, D. L.}, {Helou, G.}, {Henrot-Versill\'e, S.},
  {Hern\'andez-Monteagudo, C.}, {Herranz, D.}, {Hildebrandt, S. R.}, {Hivon,
  E.}, {Hobson, M.}, {Holmes, W. A.}, {Hornstrup, A.}, {Hovest, W.}, {Huang,
  Z.}, {Huffenberger, K. M.}, {Hurier, G.}, {Jaffe, A. H.}, {Jaffe, T. R.},
  {Jones, W. C.}, {Juvela, M.}, {Keih\"anen, E.}, {Keskitalo, R.}, {Kisner, T.
  S.}, {Kneissl, R.}, {Knoche, J.}, {Knox, L.}, {Kunz, M.}, {Kurki-Suonio, H.},
  {Lagache, G.}, {L\"ahteenm\"aki, A.}, {Lamarre, J.-M.}, {Lasenby, A.},
  {Lattanzi, M.}, {Lawrence, C. R.}, {Leahy, J. P.}, {Leonardi, R.},
  {Lesgourgues, J.}, {Levrier, F.}, {Lewis, A.}, {Liguori, M.}, {Lilje, P. B.},
  {Linden-V\o{}rnle, M.}, {L\'opez-Caniego, M.}, {Lubin, P. M.},
  {Mac\'{\i}as-P\'erez, J. F.}, {Maggio, G.}, {Maino, D.}, {Mandolesi, N.},
  {Mangilli, A.}, {Marchini, A.}, {Maris, M.}, {Martin, P. G.}, {Martinelli,
  M.}, {Mart\'{\i}nez-Gonz\'alez, E.}, {Masi, S.}, {Matarrese, S.}, {McGehee,
  P.}, {Meinhold, P. R.}, {Melchiorri, A.}, {Melin, J.-B.}, {Mendes, L.},
  {Mennella, A.}, {Migliaccio, M.}, {Millea, M.}, {Mitra, S.},
  {Miville-Desch\^enes, M.-A.}, {Moneti, A.}, {Montier, L.}, {Morgante, G.},
  {Mortlock, D.}, {Moss, A.}, {Munshi, D.}, {Murphy, J. A.}, {Naselsky, P.},
  {Nati, F.}, {Natoli, P.}, {Netterfield, C. B.}, {N\o{}rgaard-Nielsen, H. U.},
  {Noviello, F.}, {Novikov, D.}, {Novikov, I.}, {Oxborrow, C. A.}, {Paci, F.},
  {Pagano, L.}, {Pajot, F.}, {Paladini, R.}, {Paoletti, D.}, {Partridge, B.},
  {Pasian, F.}, {Patanchon, G.}, {Pearson, T. J.}, {Perdereau, O.}, {Perotto,
  L.}, {Perrotta, F.}, {Pettorino, V.}, {Piacentini, F.}, {Piat, M.},
  {Pierpaoli, E.}, {Pietrobon, D.}, {Plaszczynski, S.}, {Pointecouteau, E.},
  {Polenta, G.}, {Popa, L.}, {Pratt, G. W.}, {Pr\'ezeau, G.}, {Prunet, S.},
  {Puget, J.-L.}, {Rachen, J. P.}, {Reach, W. T.}, {Rebolo, R.}, {Reinecke,
  M.}, {Remazeilles, M.}, {Renault, C.}, {Renzi, A.}, {Ristorcelli, I.},
  {Rocha, G.}, {Rosset, C.}, {Rossetti, M.}, {Roudier, G.}, {Rouill\'e
  d\'{}Orfeuil, B.}, {Rowan-Robinson, M.}, {Rubi\~no-Mart\'{\i}n, J. A.},
  {Rusholme, B.}, {Said, N.}, {Salvatelli, V.}, {Salvati, L.}, {Sandri, M.},
  {Santos, D.}, {Savelainen, M.}, {Savini, G.}, {Scott, D.}, {Seiffert, M. D.},
  {Serra, P.}, {Shellard, E. P. S.}, {Spencer, L. D.}, {Spinelli, M.},
  {Stolyarov, V.}, {Stompor, R.}, {Sudiwala, R.}, {Sunyaev, R.}, {Sutton, D.},
  {Suur-Uski, A.-S.}, {Sygnet, J.-F.}, {Tauber, J. A.}, {Terenzi, L.},
  {Toffolatti, L.}, {Tomasi, M.}, {Tristram, M.}, {Trombetti, T.}, {Tucci, M.},
  {Tuovinen, J.}, {T\"urler, M.}, {Umana, G.}, {Valenziano, L.}, {Valiviita,
  J.}, {Van Tent, F.}, {Vielva, P.}, {Villa, F.}, {Wade, L. A.}, {Wandelt, B.
  D.}, {Wehus, I. K.}, {White, M.}, {White, S. D. M.}, {Wilkinson, A.}, {Yvon,
  D.}, {Zacchei, A.}, and {Zonca, A.}]{planck16}
{Planck Collaboration}, \latin{et~al.}  \emph{A\&A} \textbf{2016}, \emph{594},
  A13\relax
\mciteBstWouldAddEndPuncttrue
\mciteSetBstMidEndSepPunct{\mcitedefaultmidpunct}
{\mcitedefaultendpunct}{\mcitedefaultseppunct}\relax
\EndOfBibitem
\bibitem[Gondolo and Gelmini(1991)Gondolo, and Gelmini]{gondologel90}
Gondolo,~P.; Gelmini,~G. \emph{Nucl. Phys. B} \textbf{1991}, \emph{360},
  145--179\relax
\mciteBstWouldAddEndPuncttrue
\mciteSetBstMidEndSepPunct{\mcitedefaultmidpunct}
{\mcitedefaultendpunct}{\mcitedefaultseppunct}\relax
\EndOfBibitem
\bibitem[Andreas \latin{et~al.}(2012)Andreas, Niebuhr, and Ringwald]{andreas12}
Andreas,~S.; Niebuhr,~C.; Ringwald,~A. \emph{Physical Review D} \textbf{2012},
  \emph{86}\relax
\mciteBstWouldAddEndPuncttrue
\mciteSetBstMidEndSepPunct{\mcitedefaultmidpunct}
{\mcitedefaultendpunct}{\mcitedefaultseppunct}\relax
\EndOfBibitem
\bibitem[Blümlein and Brunner(2011)Blümlein, and Brunner]{mlein11}
Blümlein,~J.; Brunner,~J. \emph{Physics Letters B} \textbf{2011}, \emph{701},
  155--159\relax
\mciteBstWouldAddEndPuncttrue
\mciteSetBstMidEndSepPunct{\mcitedefaultmidpunct}
{\mcitedefaultendpunct}{\mcitedefaultseppunct}\relax
\EndOfBibitem
\bibitem[Merkel \latin{et~al.}(2011)Merkel, Achenbach, Gayoso, Bernauer, Böhm,
  Bosnar, Debenjak, Denig, Distler, Esser, Fonvieille, Fri{\v{s}
  }{\v{c}}i{\'{c}}, Middleton, Müller, Nungesser, Pochodzalla, Rohrbeck,
  Majos, Schlimme, Schoth, {\v{S}}irca, and Weinriefer]{merkel11}
Merkel,~H. \latin{et~al.}  \emph{Physical Review Letters} \textbf{2011},
  \emph{106}\relax
\mciteBstWouldAddEndPuncttrue
\mciteSetBstMidEndSepPunct{\mcitedefaultmidpunct}
{\mcitedefaultendpunct}{\mcitedefaultseppunct}\relax
\EndOfBibitem
\bibitem[Abrahamyan \latin{et~al.}(2011)Abrahamyan, Ahmed, Allada, Anez,
  Averett, Barbieri, Bartlett, Beacham, Bono, Boyce, Brindza, Camsonne,
  Cranmer, Dalton, de~Jager, Donaghy, Essig, Field, Folts, Gasparian,
  Goeckner-Wald, Gomez, Graham, Hansen, Higinbotham, Holmstrom, Huang, Iqbal,
  Jaros, Jensen, Kelleher, Khandaker, LeRose, Lindgren, Liyanage, Long, Mammei,
  Markowitz, Maruyama, Maxwell, Mayilyan, McDonald, Michaels, Moffeit,
  Nelyubin, Odian, Oriunno, Partridge, Paolone, Piasetzky, Pomerantz, Qiang,
  Riordan, Roblin, Sawatzky, Schuster, Segal, Selvy, Shahinyan, Subedi,
  Sulkosky, Stepanyan, Toro, Walz, Wojtsekhowski, and Zhang]{abrahamyan11}
Abrahamyan,~S. \latin{et~al.}  \emph{Phys. Rev. Lett.} \textbf{2011},
  \emph{107}, 191804\relax
\mciteBstWouldAddEndPuncttrue
\mciteSetBstMidEndSepPunct{\mcitedefaultmidpunct}
{\mcitedefaultendpunct}{\mcitedefaultseppunct}\relax
\EndOfBibitem
\bibitem[Bjorken \latin{et~al.}(2009)Bjorken, Essig, Schuster, and
  Toro]{bjorken09}
Bjorken,~J.~D.; Essig,~R.; Schuster,~P.; Toro,~N. \emph{Physical Review D}
  \textbf{2009}, \emph{80}\relax
\mciteBstWouldAddEndPuncttrue
\mciteSetBstMidEndSepPunct{\mcitedefaultmidpunct}
{\mcitedefaultendpunct}{\mcitedefaultseppunct}\relax
\EndOfBibitem
\bibitem[{The BABAR Collaboration} and Aubert(2009){The BABAR Collaboration},
  and Aubert]{collabBaBar}
{The BABAR Collaboration},; Aubert,~B. Search for Dimuon Decays of a Light
  Scalar in Radiative Transitions Y(3S) -\&gt; gamma A0. 2009;
  \url{https://arxiv.org/abs/0902.2176}\relax
\mciteBstWouldAddEndPuncttrue
\mciteSetBstMidEndSepPunct{\mcitedefaultmidpunct}
{\mcitedefaultendpunct}{\mcitedefaultseppunct}\relax
\EndOfBibitem
\bibitem[Anastasi \latin{et~al.}(2016)Anastasi, Babusci, Bencivenni, Berlowski,
  Bloise, Bossi, Branchini, Budano, Balkest{\aa}hl, Cao, Ceradini, Ciambrone,
  Curciarello, Czerwi{\'{n}}ski, D{\textquotesingle}Agostini, Dan{\`{e}}, Leo,
  Lucia, Santis, Simone, Cicco, Domenico, Salvo, Domenici,
  D{\textquotesingle}Uffizi, Fantini, Felici, Fiore, Gajos, Gauzzi, Giardina,
  Giovannella, Graziani, Happacher, Heijkenskjöld, Andersson, Johansson,
  Kami{\'{n}}ska, Krzemien, Kupsc, Loffredo, Mandaglio, Martini, Mascolo,
  Messi, Miscetti, Morello, Moricciani, Moskal, Palladino, Papenbrock, Passeri,
  Patera, del Rio, Ranieri, Santangelo, Sarra, Schioppa, Silarski, Sirghi,
  Tortora, Venanzoni, Wi{\'{s}}licki, and Wolke]{archilli12}
Anastasi,~A. \latin{et~al.}  \emph{Physics Letters B} \textbf{2016},
  \emph{757}, 356--361\relax
\mciteBstWouldAddEndPuncttrue
\mciteSetBstMidEndSepPunct{\mcitedefaultmidpunct}
{\mcitedefaultendpunct}{\mcitedefaultseppunct}\relax
\EndOfBibitem
\bibitem[Slatyer(2016)]{slatyer16}
Slatyer,~T.~R. \emph{Physical Review D} \textbf{2016}, \emph{93}\relax
\mciteBstWouldAddEndPuncttrue
\mciteSetBstMidEndSepPunct{\mcitedefaultmidpunct}
{\mcitedefaultendpunct}{\mcitedefaultseppunct}\relax
\EndOfBibitem
\bibitem[Dreiner \latin{et~al.}(2014)Dreiner, Fortin, Hanhart, and
  Ubaldi]{dreiner14}
Dreiner,~H.~K.; Fortin,~J.-F.; Hanhart,~C.; Ubaldi,~L. \emph{Physical Review D}
  \textbf{2014}, \emph{89}\relax
\mciteBstWouldAddEndPuncttrue
\mciteSetBstMidEndSepPunct{\mcitedefaultmidpunct}
{\mcitedefaultendpunct}{\mcitedefaultseppunct}\relax
\EndOfBibitem
\bibitem[Davoudiasl \latin{et~al.}(2016)Davoudiasl, Hooper, and
  McDermott]{davoudiasl16}
Davoudiasl,~H.; Hooper,~D.; McDermott,~S.~D. \emph{Physical Review Letters}
  \textbf{2016}, \emph{116}\relax
\mciteBstWouldAddEndPuncttrue
\mciteSetBstMidEndSepPunct{\mcitedefaultmidpunct}
{\mcitedefaultendpunct}{\mcitedefaultseppunct}\relax
\EndOfBibitem
\bibitem[Bhattiprolu \latin{et~al.}(2022)Bhattiprolu, Elor, McGehee, and
  Pierce]{Bhattiprolu_22}
Bhattiprolu,~P.~N.; Elor,~G.; McGehee,~R.; Pierce,~A. Freezing-in hadrophilic
  dark matter at low reheating temperatures. 2022;
  \url{https://arxiv.org/abs/2210.15653}\relax
\mciteBstWouldAddEndPuncttrue
\mciteSetBstMidEndSepPunct{\mcitedefaultmidpunct}
{\mcitedefaultendpunct}{\mcitedefaultseppunct}\relax
\EndOfBibitem
\bibitem[Raffelt(1996)]{raffelt96}
Raffelt,~G.~G. \emph{{Stars as laboratories for fundamental physics}: {The
  astrophysics of neutrinos, axions, and other weakly interacting particles}};
  1996\relax
\mciteBstWouldAddEndPuncttrue
\mciteSetBstMidEndSepPunct{\mcitedefaultmidpunct}
{\mcitedefaultendpunct}{\mcitedefaultseppunct}\relax
\EndOfBibitem
\bibitem[Collaboration(2019)]{NA62}
Collaboration,~N. \emph{{ADDENDUM I TO P326 Continuation of the physics
  programme of the NA62 experiment}}; 2019\relax
\mciteBstWouldAddEndPuncttrue
\mciteSetBstMidEndSepPunct{\mcitedefaultmidpunct}
{\mcitedefaultendpunct}{\mcitedefaultseppunct}\relax
\EndOfBibitem
\bibitem[Feng \latin{et~al.}(2018)Feng, Galon, Kling, and
  Trojanowski]{Feng_2018}
Feng,~J.~L.; Galon,~I.; Kling,~F.; Trojanowski,~S. \emph{Physical Review D}
  \textbf{2018}, \emph{97}\relax
\mciteBstWouldAddEndPuncttrue
\mciteSetBstMidEndSepPunct{\mcitedefaultmidpunct}
{\mcitedefaultendpunct}{\mcitedefaultseppunct}\relax
\EndOfBibitem
\bibitem[Adrian \latin{et~al.}(2018)Adrian, Baltzell, Battaglieri,
  Bond{\'{\i}}, Boyarinov, Bueltmann, Burkert, Calvo, Carpinelli, Celentano,
  Charles, Colaneri, Cooper, Cuevas, D'Angelo, Dashyan, Napoli, Vita, Deur,
  Dupre, Egiyan, Elouadrhiri, Essig, Fadeyev, Field, Filippi, Freyberger,
  Gar{\c{c}}on, Gevorgyan, Girod, Graf, Graham, Griffioen, Grillo, Guidal,
  Herbst, Holtrop, Jaros, Kalicy, Khandaker, Kubarovsky, Leonora, Livingston,
  Maruyama, McCarty, McCormick, McKinnon, Moffeit, Moreno, Camacho, Nelson,
  Niccolai, Odian, Oriunno, Osipenko, Paremuzyan, Paul, Randazzo, Raydo, Reese,
  Rizzo, Schuster, Sharabian, Simi, Simonyan, Sipala, Sokhan, Solt, Stepanyan,
  Szumila-Vance, Toro, Uemura, Ungaro, Voskanyan, Weinstein, Wojtsekhowski, and
  and]{Adrian_2018}
Adrian,~P. \latin{et~al.}  \emph{Physical Review D} \textbf{2018},
  \emph{98}\relax
\mciteBstWouldAddEndPuncttrue
\mciteSetBstMidEndSepPunct{\mcitedefaultmidpunct}
{\mcitedefaultendpunct}{\mcitedefaultseppunct}\relax
\EndOfBibitem
\bibitem[Berlin \latin{et~al.}(2018)Berlin, Gori, Schuster, and
  Toro]{Berlin_2018}
Berlin,~A.; Gori,~S.; Schuster,~P.; Toro,~N. \emph{Physical Review D}
  \textbf{2018}, \emph{98}\relax
\mciteBstWouldAddEndPuncttrue
\mciteSetBstMidEndSepPunct{\mcitedefaultmidpunct}
{\mcitedefaultendpunct}{\mcitedefaultseppunct}\relax
\EndOfBibitem
\bibitem[Doria \latin{et~al.}(2018)Doria, Achenbach, Christmann, Denig,
  Guelker, and Merkel]{MAGIX}
Doria,~L.; Achenbach,~P.; Christmann,~M.; Denig,~A.; Guelker,~P.; Merkel,~H.
  Search for light dark matter with the MESA accelerator. 2018;
  \url{https://arxiv.org/abs/1809.07168}\relax
\mciteBstWouldAddEndPuncttrue
\mciteSetBstMidEndSepPunct{\mcitedefaultmidpunct}
{\mcitedefaultendpunct}{\mcitedefaultseppunct}\relax
\EndOfBibitem
\bibitem[Doria \latin{et~al.}(2019)Doria, Achenbach, Christmann, Denig, and
  Merkel]{MESA}
Doria,~L.; Achenbach,~P.; Christmann,~M.; Denig,~A.; Merkel,~H. Dark Matter at
  the Intensity Frontier: the new MESA electron accelerator facility. 2019;
  \url{https://arxiv.org/abs/1908.07921}\relax
\mciteBstWouldAddEndPuncttrue
\mciteSetBstMidEndSepPunct{\mcitedefaultmidpunct}
{\mcitedefaultendpunct}{\mcitedefaultseppunct}\relax
\EndOfBibitem
\bibitem[Åkesson \latin{et~al.}(2018)Åkesson, Berlin, Blinov, Colegrove,
  Collura, Dutta, Echenard, Hiltbrand, Hitlin, Incandela, Jaros, Johnson,
  Krnjaic, Mans, Maruyama, McCormick, Moreno, Nelson, Niendorf, Petersen,
  Pöttgen, Schuster, Toro, Tran, and Whitbeck]{ldmx}
Åkesson,~T. \latin{et~al.}  Light Dark Matter eXperiment (LDMX). 2018;
  \url{https://arxiv.org/abs/1808.05219}\relax
\mciteBstWouldAddEndPuncttrue
\mciteSetBstMidEndSepPunct{\mcitedefaultmidpunct}
{\mcitedefaultendpunct}{\mcitedefaultseppunct}\relax
\EndOfBibitem
\bibitem[Raubenheimer \latin{et~al.}(2018)Raubenheimer, Beukers, Fry, Hast,
  Markiewicz, Nosochkov, Phinney, Schuster, and Toro]{dasel}
Raubenheimer,~T.; Beukers,~A.; Fry,~A.; Hast,~C.; Markiewicz,~T.;
  Nosochkov,~Y.; Phinney,~N.; Schuster,~P.; Toro,~N. DASEL: Dark Sector
  Experiments at LCLS-II. 2018; \url{https://arxiv.org/abs/1801.07867}\relax
\mciteBstWouldAddEndPuncttrue
\mciteSetBstMidEndSepPunct{\mcitedefaultmidpunct}
{\mcitedefaultendpunct}{\mcitedefaultseppunct}\relax
\EndOfBibitem
\bibitem[Åkesson \latin{et~al.}(2018)Åkesson, Bossi, Boveia, Bryngemark,
  Brugger, Burrows, Carpinelli, Catalan, Catena, Ceccucci, Chappell, Colegrove,
  Collura, Conrad, Cornelis, Corsini, Danielsson, Dannheim, Doebert, Doglioni,
  Dukes, Dutheil, Dutta, Echenard, Evans, Fraser, Friedland, Gall, Gessner,
  Goddard, Group, Grudiev, Gschwendtner, Hedberg, Hiltbrand, Hitlin, Incandela,
  Jaros, Jensen, Johnson, Jones, Kozhuharov, Krnjaic, Lamont, Latina, Lefevre,
  Leonora, Linssen, Longhitano, Lundh, Lytken, Mans, Maruyama, McCormick,
  Mcmonagle, Montesinos, Moreno, Muggli, Mullier, Nelson, Niendorf, Osborne,
  Papaphilippou, Petersen, Pottgen, Prieto, Raggi, Randazzo, Read, Roloff,
  Rossi, Sailer, Schulte, Schuster, Sicking, Sipala, Stapnes, Syratchev, Toro,
  Tran, D'Urso, Valente, Whitbeck, and Wuensch]{Akesson:2640784}
Åkesson,~T. \latin{et~al.}  \emph{{Dark Sector Physics with a Primary Electron
  Beam Facility at CERN}}; 2018\relax
\mciteBstWouldAddEndPuncttrue
\mciteSetBstMidEndSepPunct{\mcitedefaultmidpunct}
{\mcitedefaultendpunct}{\mcitedefaultseppunct}\relax
\EndOfBibitem
\bibitem[{Hunter} \latin{et~al.}(1997){Hunter}, {Bertsch}, {Catelli}, {Dame},
  {Digel}, {Dingus}, {Esposito}, {Fichtel}, {Hartman}, {Kanbach}, {Kniffen},
  {Lin}, {Mayer-Hasselwander}, {Michelson}, {von Montigny}, {Mukherjee},
  {Nolan}, {Schneid}, {Sreekumar}, {Thaddeus}, and {Thompson}]{EGRET}
{Hunter},~S.~D. \latin{et~al.}  \textbf{1997}, \emph{481}, 205--240\relax
\mciteBstWouldAddEndPuncttrue
\mciteSetBstMidEndSepPunct{\mcitedefaultmidpunct}
{\mcitedefaultendpunct}{\mcitedefaultseppunct}\relax
\EndOfBibitem
\bibitem[Strong \latin{et~al.}(1998)Strong, Bloemen, Diehl, Hermsen, and
  Schoenfelder]{COMPTEL}
Strong,~A.~W.; Bloemen,~H.; Diehl,~R.; Hermsen,~W.; Schoenfelder,~V.
  \textbf{1998}, \relax
\mciteBstWouldAddEndPunctfalse
\mciteSetBstMidEndSepPunct{\mcitedefaultmidpunct}
{}{\mcitedefaultseppunct}\relax
\EndOfBibitem
\bibitem[Galper \latin{et~al.}(2014)Galper, \latin{et~al.} others]{GAMMA-400}
Galper,~A.~M., \latin{et~al.}  \textbf{2014}, \relax
\mciteBstWouldAddEndPunctfalse
\mciteSetBstMidEndSepPunct{\mcitedefaultmidpunct}
{}{\mcitedefaultseppunct}\relax
\EndOfBibitem
\bibitem[Boggs(2006)]{ACT}
Boggs,~S.~E. \emph{New Astronomy Reviews} \textbf{2006}, \emph{50}, 604--607,
  Astronomy with Radioactivities. V\relax
\mciteBstWouldAddEndPuncttrue
\mciteSetBstMidEndSepPunct{\mcitedefaultmidpunct}
{\mcitedefaultendpunct}{\mcitedefaultseppunct}\relax
\EndOfBibitem
\bibitem[Hunter \latin{et~al.}(2014)Hunter, Bloser, Depaola, Dion, DeNolfo,
  Hanu, Iparraguirre, Legere, Longo, McConnell, Nowicki, Ryan, Son, and
  Stecker]{AdEPT}
Hunter,~S.~D.; Bloser,~P.~F.; Depaola,~G.~O.; Dion,~M.~P.; DeNolfo,~G.~A.;
  Hanu,~A.; Iparraguirre,~M.; Legere,~J.; Longo,~F.; McConnell,~M.~L.;
  Nowicki,~S.~F.; Ryan,~J.~M.; Son,~S.; Stecker,~F.~W. \emph{Astroparticle
  Physics} \textbf{2014}, \emph{59}, 18--28\relax
\mciteBstWouldAddEndPuncttrue
\mciteSetBstMidEndSepPunct{\mcitedefaultmidpunct}
{\mcitedefaultendpunct}{\mcitedefaultseppunct}\relax
\EndOfBibitem
\bibitem[Wu \latin{et~al.}(2014)Wu, Su, Bravar, Chang, Fan, Pohl, and
  Walter]{PANGU}
Wu,~X.; Su,~M.; Bravar,~A.; Chang,~J.; Fan,~Y.; Pohl,~M.; Walter,~R. {PANGU}: A
  high resolution gamma-ray space telescope. {SPIE} Proceedings. 2014\relax
\mciteBstWouldAddEndPuncttrue
\mciteSetBstMidEndSepPunct{\mcitedefaultmidpunct}
{\mcitedefaultendpunct}{\mcitedefaultseppunct}\relax
\EndOfBibitem
\bibitem[Aramaki \latin{et~al.}(2020)Aramaki, Adrian, Karagiorgi, and
  Odaka]{GRAMS}
Aramaki,~T.; Adrian,~P. O.~H.; Karagiorgi,~G.; Odaka,~H. \emph{Astroparticle
  Physics} \textbf{2020}, \emph{114}, 107--114\relax
\mciteBstWouldAddEndPuncttrue
\mciteSetBstMidEndSepPunct{\mcitedefaultmidpunct}
{\mcitedefaultendpunct}{\mcitedefaultseppunct}\relax
\EndOfBibitem
\bibitem[Dzhatdoev and Podlesnyi(2019)Dzhatdoev, and Podlesnyi]{MAST}
Dzhatdoev,~T.; Podlesnyi,~E. \emph{Astroparticle Physics} \textbf{2019},
  \emph{112}, 1--7\relax
\mciteBstWouldAddEndPuncttrue
\mciteSetBstMidEndSepPunct{\mcitedefaultmidpunct}
{\mcitedefaultendpunct}{\mcitedefaultseppunct}\relax
\EndOfBibitem
\bibitem[Kierans(2020)]{AMEGO}
Kierans,~C.~A. \emph{Proc. SPIE Int. Soc. Opt. Eng.} \textbf{2020},
  \emph{11444}, 1144431\relax
\mciteBstWouldAddEndPuncttrue
\mciteSetBstMidEndSepPunct{\mcitedefaultmidpunct}
{\mcitedefaultendpunct}{\mcitedefaultseppunct}\relax
\EndOfBibitem
\bibitem[De~Angelis \latin{et~al.}(2020)De~Angelis, \latin{et~al.}
  others]{All-Sky-ASTROGRAM}
De~Angelis,~A., \latin{et~al.}  \emph{PoS} \textbf{2020}, \emph{ICRC2019},
  579\relax
\mciteBstWouldAddEndPuncttrue
\mciteSetBstMidEndSepPunct{\mcitedefaultmidpunct}
{\mcitedefaultendpunct}{\mcitedefaultseppunct}\relax
\EndOfBibitem
\bibitem[Grayson \latin{et~al.}(2016)Grayson, Ade, Ahmed, Alexander, Amiri,
  Barkats, Benton, Bischoff, Bock, Boenish, Bowens-Rubin, Buder, Bullock, Buza,
  Connors, Filippini, Fliescher, Halpern, Harrison, Hilton, Hristov, Hui,
  Irwin, Kang, Karkare, Karpel, Kefeli, Kernasovskiy, Kovac, Kuo, Leitch,
  Lueker, Megerian, Monticue, Namikawa, Netterfield, Nguyen,
  O{\textquotesingle}Brient, Ogburn, Pryke, Reintsema, Richter, Schwarz,
  Sorenson, Sheehy, Staniszewski, Steinbach, Teply, Thompson, Tolan, Tucker,
  Turner, Vieregg, Wandui, Weber, Wiebe, Willmert, Wu, and Yoon]{keck}
Grayson,~J.~A. \latin{et~al.}  {BICEP}3 performance overview and planned Keck
  Array upgrade. {SPIE} Proceedings. 2016\relax
\mciteBstWouldAddEndPuncttrue
\mciteSetBstMidEndSepPunct{\mcitedefaultmidpunct}
{\mcitedefaultendpunct}{\mcitedefaultseppunct}\relax
\EndOfBibitem
\bibitem[Anderson \latin{et~al.}(2018)Anderson, \latin{et~al.}
  others]{South-Pole-Telescope}
Anderson,~A.~J., \latin{et~al.}  \emph{J. Low Temp. Phys.} \textbf{2018},
  \emph{193}, 1057--1065\relax
\mciteBstWouldAddEndPuncttrue
\mciteSetBstMidEndSepPunct{\mcitedefaultmidpunct}
{\mcitedefaultendpunct}{\mcitedefaultseppunct}\relax
\EndOfBibitem
\bibitem[Henderson \latin{et~al.}(2016)Henderson, Allison, Austermann, Baildon,
  Battaglia, Beall, Becker, Bernardis, Bond, Calabrese, Choi, Coughlin,
  Crowley, Datta, Devlin, Duff, Dunkley, Dünner, van Engelen, Gallardo, Grace,
  Hasselfield, Hills, Hilton, Hincks, Hloẑek, Ho, Hubmayr, Huffenberger,
  Hughes, Irwin, Koopman, Kosowsky, Li, McMahon, Munson, Nati, Newburgh,
  Niemack, Niraula, Page, Pappas, Salatino, Schillaci, Schmitt, Sehgal,
  Sherwin, Sievers, Simon, Spergel, Staggs, Stevens, Thornton, Lanen,
  Vavagiakis, Ward, and Wollack]{Advanced-Atacama}
Henderson,~S.~W. \latin{et~al.}  \emph{Journal of Low Temperature Physics}
  \textbf{2016}, \emph{184}, 772--779\relax
\mciteBstWouldAddEndPuncttrue
\mciteSetBstMidEndSepPunct{\mcitedefaultmidpunct}
{\mcitedefaultendpunct}{\mcitedefaultseppunct}\relax
\EndOfBibitem
\bibitem[Essinger-Hileman \latin{et~al.}(2014)Essinger-Hileman, Ali, Amiri,
  Appel, Araujo, Bennett, Boone, Chan, Cho, Chuss, Colazo, Crowe, Denis,
  Dünner, Eimer, Gothe, Halpern, Harrington, Hilton, Hinshaw, Huang, Irwin,
  Jones, Karakla, Kogut, Larson, Limon, Lowry, Marriage, Mehrle, Miller,
  Miller, Moseley, Novak, Reintsema, Rostem, Stevenson, Towner, U-Yen, Wagner,
  Watts, Wollack, Xu, and Zeng]{CLASS}
Essinger-Hileman,~T. \latin{et~al.}  {CLASS}: the cosmology large angular scale
  surveyor. {SPIE} Proceedings. 2014\relax
\mciteBstWouldAddEndPuncttrue
\mciteSetBstMidEndSepPunct{\mcitedefaultmidpunct}
{\mcitedefaultendpunct}{\mcitedefaultseppunct}\relax
\EndOfBibitem
\bibitem[Ade \latin{et~al.}(2019)Ade, Aguirre, Ahmed, Aiola, Ali, Alonso,
  Alvarez, Arnold, Ashton, Austermann, Awan, Baccigalupi, Baildon, Barron,
  Battaglia, Battye, Baxter, Bazarko, Beall, Bean, Beck, Beckman, Beringue,
  Bianchini, Boada, Boettger, Bond, Borrill, Brown, Bruno, Bryan, Calabrese,
  Calafut, Calisse, Carron, Challinor, Chesmore, Chinone, Chluba, Cho, Choi,
  Coppi, Cothard, Coughlin, Crichton, Crowley, Crowley, Cukierman, D'Ewart,
  Dünner, de~Haan, Devlin, Dicker, Didier, Dobbs, Dober, Duell, Duff,
  Duivenvoorden, Dunkley, Dusatko, Errard, Fabbian, Feeney, Ferraro, Fluxà,
  Freese, Frisch, Frolov, Fuller, Fuzia, Galitzki, Gallardo, Ghersi, Gao,
  Gawiser, Gerbino, Gluscevic, Goeckner-Wald, Golec, Gordon, Gralla, Green,
  Grigorian, Groh, Groppi, Guan, Gudmundsson, Han, Hargrave, Hasegawa,
  Hasselfield, Hattori, Haynes, Hazumi, He, Healy, Henderson, Hervias-Caimapo,
  Hill, Hill, Hilton, Hilton, Hincks, Hinshaw, Hložek, Ho, Ho, Howe, Huang,
  Hubmayr, Huffenberger, Hughes, Ijjas, Ikape, Irwin, Jaffe, Jain, Jeong,
  Kaneko, Karpel, Katayama, Keating, Kernasovskiy, Keskitalo, Kisner, Kiuchi,
  Klein, Knowles, Koopman, Kosowsky, Krachmalnicoff, Kuenstner, Kuo, Kusaka,
  Lashner, Lee, Lee, Leon, Leung, Lewis, Li, Li, Limon, Linder,
  Lopez-Caraballo, Louis, Lowry, Lungu, Madhavacheril, Mak, Maldonado, Mani,
  Mates, Matsuda, Maurin, Mauskopf, May, McCallum, McKenney, McMahon, Meerburg,
  Meyers, Miller, Mirmelstein, Moodley, Munchmeyer, Munson, Naess, Nati,
  Navaroli, Newburgh, Nguyen, Niemack, Nishino, Orlowski-Scherer, Page,
  Partridge, Peloton, Perrotta, Piccirillo, Pisano, Poletti, Puddu, Puglisi,
  Raum, Reichardt, Remazeilles, Rephaeli, Riechers, Rojas, Roy, Sadeh, Sakurai,
  Salatino, Rao, Schaan, Schmittfull, Sehgal, Seibert, Seljak, Sherwin, Shimon,
  Sierra, Sievers, Sikhosana, Silva-Feaver, Simon, Sinclair, Siritanasak,
  Smith, Smith, Spergel, Staggs, Stein, Stevens, Stompor, Suzuki, Tajima,
  Takakura, Teply, Thomas, Thorne, Thornton, Trac, Tsai, Tucker, Ullom,
  Vagnozzi, van Engelen, Lanen, Winkle, Vavagiakis, Vergès, Vissers, Wagoner,
  Walker, Ward, Westbrook, Whitehorn, Williams, Williams, Wollack, Xu, Yu, Yu,
  Zago, Zhang, Zhu, and collaboration]{Simons-Observatory}
Ade,~P. \latin{et~al.}  \emph{Journal of Cosmology and Astroparticle Physics}
  \textbf{2019}, \emph{2019}, 056\relax
\mciteBstWouldAddEndPuncttrue
\mciteSetBstMidEndSepPunct{\mcitedefaultmidpunct}
{\mcitedefaultendpunct}{\mcitedefaultseppunct}\relax
\EndOfBibitem
\bibitem[Chu \latin{et~al.}(2012)Chu, Hambye, and Tytgat]{Chu_2012}
Chu,~X.; Hambye,~T.; Tytgat,~M.~H. \emph{Journal of Cosmology and Astroparticle
  Physics} \textbf{2012}, \emph{2012}, 034--034\relax
\mciteBstWouldAddEndPuncttrue
\mciteSetBstMidEndSepPunct{\mcitedefaultmidpunct}
{\mcitedefaultendpunct}{\mcitedefaultseppunct}\relax
\EndOfBibitem
\bibitem[Hambye \latin{et~al.}(2019)Hambye, Tytgat, Vandecasteele, and
  Vanderheyden]{hambye19}
Hambye,~T.; Tytgat,~M.~H.; Vandecasteele,~J.; Vanderheyden,~L. \emph{Physical
  Review D} \textbf{2019}, \emph{100}\relax
\mciteBstWouldAddEndPuncttrue
\mciteSetBstMidEndSepPunct{\mcitedefaultmidpunct}
{\mcitedefaultendpunct}{\mcitedefaultseppunct}\relax
\EndOfBibitem
\bibitem[Krnjaic(2016)]{Krnjaic_2016}
Krnjaic,~G. \emph{Physical Review D} \textbf{2016}, \emph{94}\relax
\mciteBstWouldAddEndPuncttrue
\mciteSetBstMidEndSepPunct{\mcitedefaultmidpunct}
{\mcitedefaultendpunct}{\mcitedefaultseppunct}\relax
\EndOfBibitem
\bibitem[Shtabovenko \latin{et~al.}(2020)Shtabovenko, Mertig, and
  Orellana]{feyncalc}
Shtabovenko,~V.; Mertig,~R.; Orellana,~F. \emph{Computer Physics
  Communications} \textbf{2020}, \emph{256}, 107478\relax
\mciteBstWouldAddEndPuncttrue
\mciteSetBstMidEndSepPunct{\mcitedefaultmidpunct}
{\mcitedefaultendpunct}{\mcitedefaultseppunct}\relax
\EndOfBibitem
\bibitem[Ruegg and Ruiz-Altaba(2004)Ruegg, and Ruiz-Altaba]{ruegg04}
Ruegg,~H.; Ruiz-Altaba,~M. \emph{International Journal of Modern Physics A}
  \textbf{2004}, \emph{19}, 3265--3347\relax
\mciteBstWouldAddEndPuncttrue
\mciteSetBstMidEndSepPunct{\mcitedefaultmidpunct}
{\mcitedefaultendpunct}{\mcitedefaultseppunct}\relax
\EndOfBibitem
\end{mcitethebibliography}

\end{document}